# PATRONES EN LA DISTRIBUCIÓN DE LA VEGETACIÓN EN ÁREAS DE PÁRAMO DE COLOMBIA: HETEROGENEIDAD Y DEPENDENCIA ESPACIAL

## Patterns in the distribution of vegetation in paramo areas: heterogeneity and spacial dependence

**Henry Arellano-P.**
*Programa de Posgrado, línea Biodiversidad* y *Conservación. Instituto de Ciencias Naturales, Facultad de Ciencias, Universidad Nacional de Colombia, Apartado 7495, Bogotá D. C., Colombia. henryarellc@gmail.com*

**J. Orlando Rangel-Ch.**
*Instituto de Ciencias Naturales, Facultad de Ciencias, Universidad Nacional de Colombia, Apartado 7495, Bogotá D. C., Colombia. jorangelc@unal.edu.co*

### RESUMEN

Se realizaron   análisis de heterogeneidad y dependencia espacial (auto-correlación) con base en  la caracterización y en la distribución de coberturas en diez áreas de páramo en las cordilleras Central y Oriental de Colombia, estas metodologías pertenecen al grupo de análisis exploratorios de datos espaciales (AEDE). De las áreas estudiadas muestran un buen estado de conservación y de conectividad, la alta montaña de la serranía de Perijá, los páramos del Parque Nacional Natural Los Nevados y gran parte del área  pertenecientes a los páramos de la jurisdicción de la corporación CORPOGUAVIO. La intervención de origen antrópico, el aumento del cultivo de papa, el cambio en el uso del suelo, principalmente en los páramos de la cordillera Oriental, han afectado la estructura y la distribución natural de los tipos de vegetación y han disminuido sensiblemente la conectividad entre bloques de área con lo cual se hace muy difícil cualquier proceso de re-vegetalización y recuperación. Esta situación se presenta en gran escala en los páramos La Rusia, Belén, Guantiva (hacia la vertiente Oriental), Telecom, Merchán, Tablazo, sector nororiental del páramo de Sumapaz y la vertiente Oriental de los páramos de Carmen de Carupa y sectores aledaños. Para los páramos de Guanacas y Las Delicias (área de influencia del volcán Puracé) la problemática está concentrada hacia el sector noroccidental. Los resultados obtenidos, se pueden aplicar y son útiles en los procesos de restauración y conservación, puesto que proporcionan información detallada acerca de la distribución de los bloques de áreas con los  tipos de vegetación, la conectividad y  además, pueden utilizarse como indicadores de la salud de los ecosistemas.

**Palabras clave.** Fragmentación, heterogeneidad espacial, conservación, páramos, biodiversidad.

**ABSTRACT**

Two methods of exploratory spatial data analysis (ESDA), analysis of spatial heterogeneity and dependence (auto-correlation), - - were applied to the cover patterns from ten paramo localities in the Central and Eastern cordilleras of Colombia. Among the localities studied, the high montane region of the Serrania de Perija, the paramo region of the Los Nevados National Park, and the paramo region under management of CORPOGUAVIO showed a good state of conservation and satisfactory level of connectivity among patches. Anthropic intervention, expansion of potato cultivation and other changes in land use mainly in the paramos of the Cordillera Oriental have alteredd the distribution patterns of natural vegetation types and resulted in significant losses of connectively. Under the circumstancesany attempt to restore the original conditions will be difficult and expensive. This problem is most acute in the paramo localities of La Rusia, Belén, Guantiva (towards the eastern slopes) Telecom, Merchan Tablazo, the northeastern sector of the paramo of Sumapaz and the eastern slopes of the mountains of Carmen de Carupa and surrounding areas. For the mountains of Guanacaste and Las Delicias (areas influenced by the volcano Puracé) the problem is more acute towards the northwestern sector. The results of the present study can be applied to ecological restoration and conservation as they provide detailed information about the distribution of vegetation types andhabitatconnectivity, both indicators of ecosystem health.

## INTRODUCCIÓN

El amplio conocimiento de la vegetación del páramo colombiano adquirido a través de una tradición investigativa que se inicia con los trabajos clásicos de Cuatrecasas (1934, 1958) y continua con los estudios de Cleef (1981), Rangel *et al.* (1982), Sturm & Rangel (1985), Monasterio (1980), Mora & Sturm (1994), Sturm (1998), Luteyn (1999), Rangel (2000, 2007) y Van der Hammen (2005, 2007) entre otros, han permitido abordar nuevos análisis que tienen como objetivos la especialización de la distribución de las unidades de vegetación e igualmente, detectar factores importantes que influyen en la salud y en el estado de conservación de los variados ecosistemas que se han caracterizado en el bioma paramuno.

A estas contribuciones se les adicionan los detallados estudios de campo actuales y el desarrollo de tecnología informática, con lo cual esta documentación básica adquiere un alcance práctico excepcional, por ejemplo en las propuestas de planes de manejo y uso sostenible de la oferta ambiental, en la detección de amenazas naturales y antrópicas y esencialmente en la elaboración de documentos que muestren realidades más acordes con la capacidad de ofrecer servicios ambientales por parte de los ecosistemas paramunos. Estas construcciones metodológicas son requeridas con urgencia por la problemática actual que se relaciona, con el avance de la frontera agropecuaria, la explotación excesiva de los recursos hídricos, la deficiente planeación del uso de la tierra y los cambios importantes en los ensambles bióticos que origina el conflicto armado y los cultivos ilícitos. A la par con estas afectaciones se ha despertado el interés en las entidades gubernamentales encargadas de la conservación por conocer nuevas alternativas que permitan mitigar la influencia antrópica directa sobre este importante capital natural de la nación.

En el marco general de la situación en cuanto a preservación de la región paramuna de Colombia, deben mencionarse los intentos por cristalizar una ley para la conservación de los páramos y la celebración de varios congresos y simposios nacionales e internacionales que han promovido la conservación de la biodiversidad del páramo en razón a los servicios ambientales fundamentales que presta. Sin embargo, estos esfuerzos disponen de reducidos insumos metodológicos para la implementación de recomendaciones acertadas, por ejemplo la cartografía detallada de la alta montaña es muy pobre, no se dispone de una adecuada delimitación catastral del territorio y en algunos casos, la división arbitraria del manejo de algunos sistemas paramunos impide su estudio integral al dividirlo por sectores administrativos (por ejemplo, la administración del páramo de Sumapaz está a cargo varias entidades, unidad de Parques Nacionales, secretaria de ambiente del distrito capital y la CAR).

A pesar de estos inconvenientes, hay esfuerzos importantes por generar información temática detallada que sirva como insumo para posteriores investigaciones, así como para el necesario monitoreo de estos sistemas bióticos. Los trabajos realizados entre otros, por el grupo de Investigación en Biodiversidad y Conservación del Instituto de Ciencias Naturales de la Universidad Nacional de Colombia como "Caracterización, zonificación ambiental e implementación de un Sistema de Información Geográfica del Territorio de los páramos Los Cristales, Castillejo, Cuchilla El Choque y nacimiento del río Bogotá"; "Estrategia corporativa para la caracterización con fines de manejo y conservación de páramos en el territorio CAR (énfasis en los páramos de Telecom y Merchán, El Tablón o Monquetiva, Sumapaz y Guerrero) (Convenio CAR-Universidad Nacional-Instituto de Ciencias Naturales)" y "Caracterización y plan de manejo de la alta montaña de la serranía de Perijá (Convenio CORPOCESAR-Universidad Nacional de Colombia", así lo demuestran.

Uno de los resultados de mayor impacto de estas investigaciones se relaciona con el estado de fragmentación y pérdida de la distintividad biológica en varias localidades paramunas de Colombia, razón por la cual conocer la discontinuidad en los parches de vegetación actual en parte ocasionados por la acción humana e igualmente por factores naturales, es de suma importancia para tener bases confiables que permitan planificar procesos adecuados de restauración y manejo de estos ecosistemas. Una de las herramientas para conocer la variación en el área cubierta por ecosistemas del páramo que se emplea en este documento proviene de la economía y forma parte del grupo de técnicas conocidas como análisis exploratorios de datos espaciales (AEDE). Entre éstas figuran los análisis de heterogeneidad y dependencia espacial que pueden ser utilizados con el fin de determinar los grados de conectividad y cercanía de los tipos de vegetación en un espacio geográfico, acción que se emprendió y cuyos resultados se muestran en esta contribución.

## ÁREA DE ESTUDIO

Las áreas de páramo estudiadas representan una gran variabilidad de patrones ubicados en las cordilleras Oriental y Central de Colombia (Figura 1). En la cordillera Oriental comprenden la zona de alta montaña de la serranía de Perijá (Cesar-La Guajira) en donde se analizaron 29139,13 hectáreas. Los límites exactos de la región en coordenadas

planas con datum Bogotá-Bogotá son Norte 1649601,24; Sur 1592831,34; Occidente 1110874,97; Oriente 1131786,4. El límite altitudinal inferior es 2800 metros y el superior 3436 metros. El área delimita la frontera con Venezuela al oriente y con la cota 2800 al oriente. Algunas generalidades de la zona se observan en la figura 2A. También en la cordillera Oriental, en los páramos de Guantiva, La Rusia y Belén se analizaron 103062,5 hectáreas. Los límites de la región son Norte 1195396,18; Sur 1139489,06; Occidente 1103785,76; Oriente 1136958,88. El límite altitudinal inferior es 2800 metros y el superior 4247 metros. El área se encuentra entre los departamentos de Santander y Boyacá (Figura 2B).

Algunas generalidades de la zona se observan en la figura 2B. En áreas cercanas a Bogotá, se examinó el páramo de Telecom (Figura 2C) en donde se analizaron 3637,67 hectáreas; los límites de la región son Norte 1128605,57; Sur 1119024,22; Occidente 1025495,08; Oriente 1033977,83; el límite altitudinal inferior está alrededor de 2900 metros y coincide con algunos límites prediales hacia el oriente; la altura más alta registrada es de 3358 metros. En el páramo de Merchán (Figura 2C) se analizaron 4668,67 hectáreas; los límites son Norte 1127664,15; Sur 1114678,48; Occidente 1041480,85; Oriente 1047076,43; el límite altitudinal inferior es de 2900 metros y el superior 3454,85 metros. Para el páramo de Santuario y sectores aledaños al municipio de Carmen de Carupa (Figura 2C) se analizaron 23865,72 hectáreas; los límites son Norte 1101244,48; Sur 1065584,44; Occidente 1008368,62; Oriente 1026645,16; el límite altitudinal inferior está alrededor de 2900 metros; la altura más alta registrada es de 3776 metros. En el páramo El Tablazo (Cundinamarca)

se analizaron 12510,72 hectáreas. Los límites de la región son Norte 10547553,16; Sur 1030890,18; Occidente 983693,32; Oriente 1000457,65. El límite altitudinal inferior es 2900 metros y el superior 3712 metros. Algunas generalidades de la zona se observan en la figura 2D. Para los páramos de la jurisdicción CORPOGUAVIO (Cundinamarca) se analizaron 33139,84 hectáreas. La zona de estudio comprende las áreas de Páramo en los municipios de Fómeque, Guasca, Gachetá, Ubalá, Gama, Junín y Gachalá, (exceptuando las áreas ubicadas dentro del Parque Nacional Natural Chingaza). Los límites de la región son Norte 1039686,87; Sur 978668,84; Occidente 1016373,48; Oriente 1070028,48. El límite altitudinal inferior es de 3000 metros y el superior 3722 metros (Figura 3A).

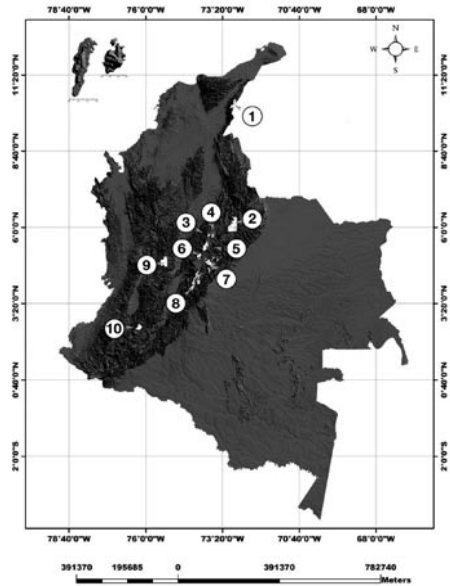

**Figura 1.** Localización de las áreas de páramo estudiadas en las cordilleras Oriental y Central de Colombia.

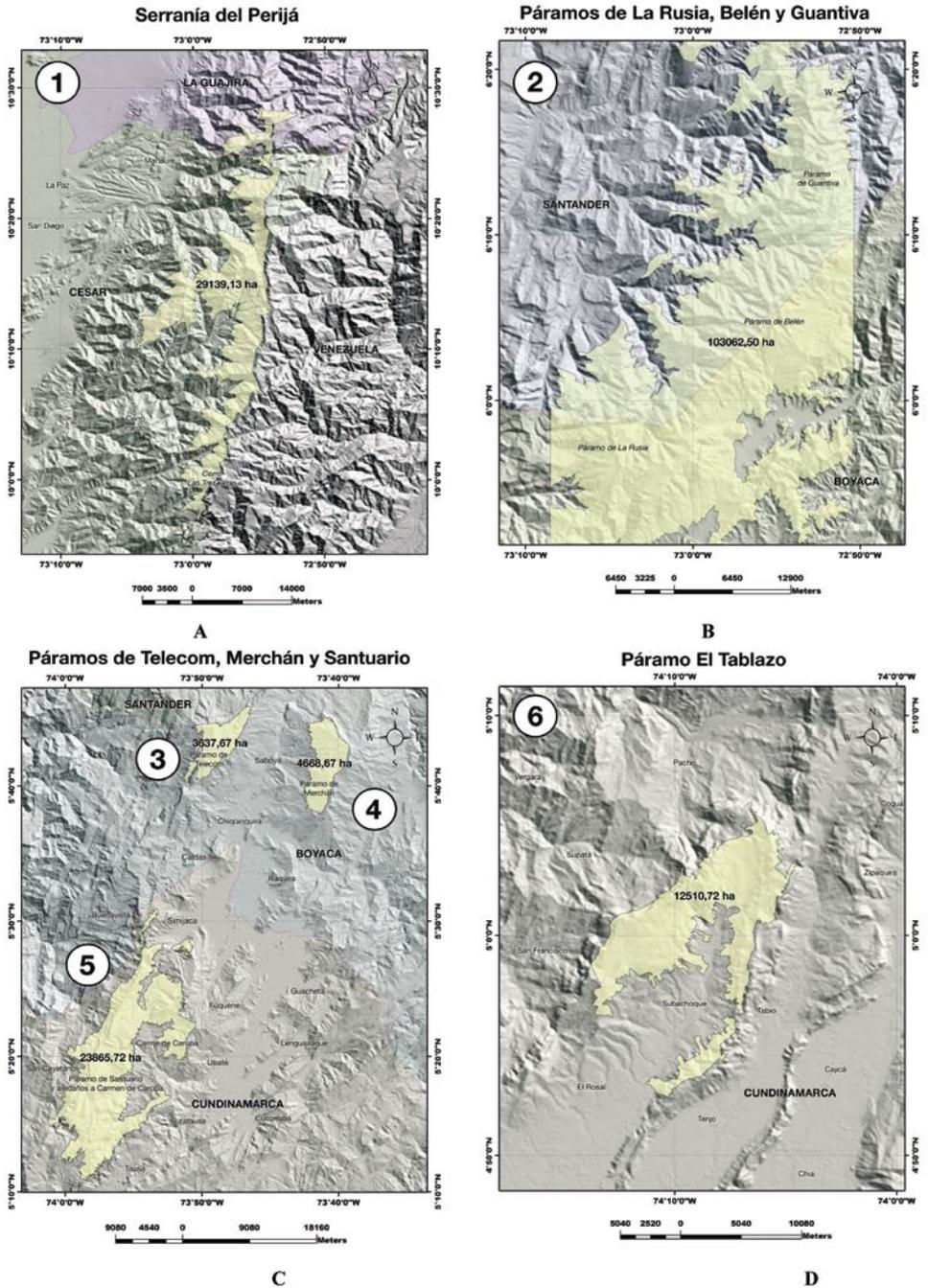

**Figura 2.** Cordillera Oriental: A- Alta montaña de la serranía de Perijá (29139,13 ha); B- Región perteneciente a los páramos la Rusia, Belén y Guantiva (103062,5 ha); C- Ubicación de los páramos cercanos a la sabana de Bogotá, Telecom (3637,67 ha), Merchán (4668,67 ha), Santuario y sectores paramunos del municipio Carmen de Carupa (23865,72 ha) y D-A- Ubicación del páramo El Tablazo (12510,72 ha).

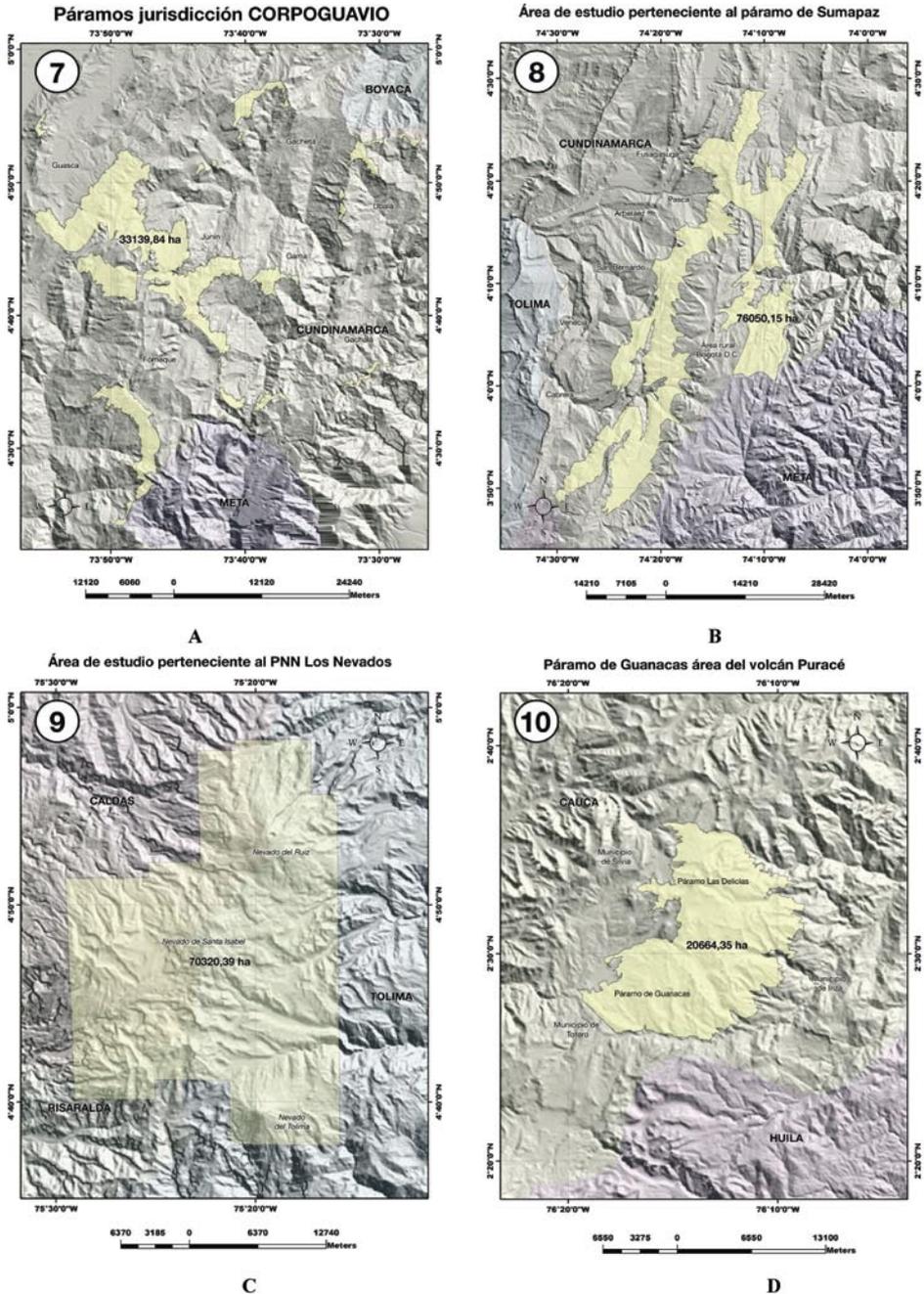

**Figura 3.** Cordillera Oriental: A- Localización de las áreas de páramo pertenecientes a la jurisdicción CORPOGUAVIO (33139,84 ha); B- Área de estudio perteneciente al páramo de Sumapaz (solo se incluyen 76050,15 ha de la jurisdicción CAR). Cordillera Central: C- Localización de la región paramuna definida por Kloosterman (1981) en el Parque Nacional Natural Los Nevados (70320,39 ha) y D-Distribución del área estudiada en los páramos de Guanacas y Las Delicias, cercanías del volcán Puracé (20664,35 ha).. Cartografía base, SRTM 90 metros, Nasa.

Para el área de estudio perteneciente al páramo de Sumapaz (jurisdicción CAR, Cundinamarca) se analizaron 76050,15 hectáreas. Los límites de la región son Norte 986795,1; Sur 911984,63; Occidente 953293,31; Oriente 997475,45. El límite altitudinal inferior está alrededor de 2900 metros y el superior alcanza 4015 metros (Figura 3B). En la cordillera Central se seleccionaron las áreas de páramo perteneciente al Parque Nacional Natural Los Nevados (Caldas, Risaralda y Tolima) y analizaron 70320,39 hectáreas. El área fue definida según la interpretación de F.H Kloosterman (1981). Los límites de la región son Norte 1041648,12; Sur 1004148,39; Occidente 843513,56; Oriente 868303,17. El límite altitudinal inferior está alrededor de 2800 metros y el superior alcanza 5252 metros (Figura 3C). En el macizo central, se estudió el área perteneciente a los páramos de Guanacas y Las Delicias (Cauca) se analizaron 20664,35 hectáreas. Los límites de la región son Norte 779389,57; Sur 760553,53; Occidente 1084475,84; Oriente 1102727,07. El límite altitudinal inferior está alrededor de 3000 metros y el superior alcanza 3767 metros (Figura 3D).

## METODOLOGÍA

Para la realización del trabajo se contó con la información ecológica y fitosociológica básica consolidada en los páramos de la Serranía de Perijá, El Parque Nacional Natural del Sumapaz, P.N.N. Los Nevados y Páramo de La Rusia. En las otras desde páramo, la caracterización es de tipo ecológico sobre la base de las especies dominantes se siguió la metodología básica reseñada en Rangel & Velásquez (1997) y las características y observaciones más importantes de campo, como tipo de cobertura general, estructura, matriz, patrones de distribución, escarpes, entre otras, se georreferenciaron mediante puntos gps en coordenadas geográficas y bajo el datum WGS84 . Cabe destacar la georreferenciación de elementos como vías, cultivos, áreas sin vegetación, canteras no pertenecientes a formaciones vegetales, con el fin de afinar aún más las interpretaciones de cobertura generadas.

La determinación del material recolectado se llevó a cabo por investigadores y colaboradores del Grupo de Investigación en Biodiversidad y Conservación del Instituto de Ciencias Naturales de la Universidad Nacional de Colombia. Con los mapas temáticos provenientes de la caracterización de coberturas se realizaron análisis de heterogeneidad y dependencia espacial (auto-correlación) con el fin de encontrar patrones de distribución en los parches de los tipos de vegetación, importantes para los procesos de restauración y conservación de estos ecosistemas. Aunque el estudio de heterogeneidad y de dependencia espacial puede ser abordados, mediante distintas formas y técnicas, cabe aclarar que el grado de confiabilidad depende además de la escala utilizada, de la calidad de las interpretaciones realizadas y es el reflejo de los levantamientos y anotaciones de campo. Teóricamente la técnica implica la cuantificación de la agregación o desagregación de patrones espaciales de cualquier tipo, que para el caso de unidades de cobertura indicarían el grado de homogeneidad y dependencia (para esta última hay que reinterpretar la metodología existente) entre los distintos tipos de vegetación y cobertura. Dentro de este proceso adaptativo de la metodología cabe aclarar que como el proceso de interpretación de coberturas busca mediante la generalización abstraer gran parte de la complejidad interior de los patrones, en este caso, el análisis no incluye la heterogeneidad interior propia de algunos tipos de cobertura (especialmente las entremezclas de vegetación) y cuyo análisis es objeto importante en otros tipos de estudio.

Cuando se comparan dos patrones (uno desconocido contra otro conocido) es necesario plantear hipótesis, de lo contrario,

la investigación realizada adquiere un carácter exploratorio en el cual la hipótesis es sustituida por una pregunta de investigación. En muchos casos el análisis geográfico se enmarca dentro de este segundo caso y los análisis presentados son de carácter descriptivo y exploratorio, ya que se desconoce como tal un patrón de distribución muestral lógico así como sus relaciones. Las metodologías existentes se encuentran dentro del grupo de los conocidos análisis exploratorios de datos espaciales (AEDE), las cuales fueron desarrollados dentro de la econometría espacial, una subdisciplina de la economía (Chasco, 2003). A continuación se presenta un resumen de los principales conceptos enmarcados dentro de estas metodologías y su respectiva adaptación al problema de la heterogeneidad y la dependencia de datos temáticos como los son los tipos de cobertura.

Para comprender la heterogeneidad desde el punto de vista de las metodologías AEDE es necesario comprender la naturaleza de los datos. Como ya se ha mencionado, los datos exploratorios no hacen parte de una metodología de captura sistemática de datos en una línea en el tiempo, si no que son parte de la información de un punto del mismo. Esta particularidad inherente a los datos espaciales puntuales de cualquier tipo (áreas de parches, perímetro, características de una región, cantidad de producción, votación, población) conlleva a tener una colección de información que en su mayoría proviene de múltiples mecanismos de generación y por esta razón, los hace heterogéneos mientras que en las series temporales los mecanismos de producción de datos son de carácter estable (distribución muestral) en cuanto al cambio, el crecimiento, la alteración o la constancia. Cabe aclarar que las series temporales de datos también presentan estas características y deben ser abordadas desde otras metodologías, mientras que las series temporales cartográficas comparten mecanismos múltiples de generación de datos, de tal manera que su análisis no se puede limitar a las diferencias entre las áreas cubiertas, ya que sus mecanismos son mucho más complejos.

La dependencia espacial según Moreno & Vayá (2000) es la relación funcional de fenómenos que se presentan en diferentes lugares geográficos debido a que una variable además de depender de muchos factores (conocidos o desconocidos), puede depender de valores del mismo tipo en regiones cercanas (la variable puede también influir en las otras regiones contiguas). La utilidad del análisis de dependencia es ampliamente documentada en la econometría espacial; sin embargo, este tipo de análisis al realizar las características de la distancia es de gran utilidad para comprender si existe una relación directa en la forma como se presenta un determinado patrón en el espacio. Las metodologías actuales, así como su enfoque, permiten evaluar una o varias variables que pueden estar distribuidas en el espacio de diferentes maneras. El análisis de dependencia utilizado sobre tipos de cobertura, sólo utiliza los valores de peso explicados más adelante con el fin de evaluar de manera general el grado de conectividad de las coberturas del mismo tipo. Cabe resaltar que aunque los objetivos de la autocorrelación no están diseñados para el análisis de las coberturas debido a que se toman como categorías, la utilidad resulta al generar las figuras de dispersión de una variable categórica con respecto a su peso estandarizado (dividir cada valor por la suma total) en una figura de Moran generalizada. Cabe aclarar que aunque en este tipo de análisis se genera un estadístico de dispersión (índice de Moran), éste pierde el sentido lógico, ya que al utilizar un identificador **ID** consecutivo para identificar cada patrón cobertura, los pesos de cada parche sólo se distribuyen en un eje de variación (el de los pesos), de ahí su utilidad en el análisis.

El paquete de análisis GeoDa 0.9.5-i (beta) es una herramienta de trabajo enfocada a realizar análisis exploratorios y confirmatorios de atributos espaciales de carácter red de datos, creado para el sistema operativo Windows por Anselin *et al*. (2004). Se caracteriza principalmente por generar un análisis basado en pesos de los datos espaciales provenientes de los polígonos de cartografía temática en formato "shapefile" de ArcGis y con la posibilidad de diferentes niveles de análisis dependiendo de los umbrales de cercanía o distancia que se quieran evaluar. Cabe resaltar que una de las ventajas más grandes del Software, es la de poder elegir un análisis de puntos o un análisis donde los bordes de los polígonos sean tenidos en cuenta. La necesidad de adaptar, según sea el caso, la interpretación de coberturas, es necesaria para la obtención de mejores resultados, así como la conversión a categorías numéricas de los símbolos correspondientes a los tipos de vegetación. Los pesos espaciales son el fundamento de los análisis de autocorrelación espacial y están basados en las distancias de los centroides de los polígonos análogos, en cuyo caso generaría un análisis de punto o en los bordes de los polígonos, en otros, el análisis estaría centrado en los problemas de contigüidad de los mismos.

Para tal fin, el sistema crea, según indicaciones previas, un archivo de texto basado en las observaciones y los atributos seleccionados con la extensión *.gal para los análisis que tienen en cuenta el borde de los polígonos (contigüidad) y *.gwt para los análisis basados en distancia entre los centroides. Según Moran (1948) los pesos encontrados, en general, deben responder a una matriz no aleatoria con valores entre uno (1) cuando dos regiones están adyacentes o contiguas o 0 cuando no lo están. No obstante, Anselin (1980), autor principal del software GeoDa asegura que los pesos responden mejor a una matriz inversa de distancias entre puntos al cuadrado, ya que la intensidad de la independencia entre dos regiones disminuye con la distancia que separa sus centros (Moreno & Vayá 2002).

## RESULTADOS

Los tipos de vegetación de las diez (10) áreas de páramo utilizadas en este manuscrito se presentan de las tablas 1 al 10. El orden presentado corresponde al identificador manejado dentro de los análisis de heterogeneidad y dependencia.

**Tabla 1.** Tipos de vegetación presentes en la alta montaña de la Serranía de Perijá. El orden corresponde al identificador utilizado a lo largo del presente trabajo.

| ALTA MONTAÑA DE LA Serranía de Perijá | | | |
|---|---|---|---|
| SIM_ID | SÍMBOLO | TIPO DE VEGETACIÓN | ESPECIES DOMINANTES |
| 64 | SV | Sin vegetación | Zona sin vegetación |
| 63 | SI | Sin información | Sin información |
| 62 | Hz-P/Och-Cef | Herbazales entremezclados con pajonales | *Orthrosanthus chimboracensis* y *Calamagrostis effusa* |
| 61 | Hz-P/Lst-Cef-Och | | *Lourteigia stoechadifolia, Calamagrostis effusa* y *Orthrosanthus chimboracensis* |
| 60 | Hz/Och-Vve | Herbazales | *Orthosanthus chimboracensis* y *Valeriana vetasana* |
| 59 | Ch/Sal-Cte | Chuscales | *Senecio albotectus* y *Chusquea tessellata* |
| 58 | Ch/Ani-Cte | | *Arcytophyllum nitidum* y *Chusquea tessellata* |
| 57 | Rfb-P-Mb-Baa1/Epe-Lst-Wpi | Rosetales frailejonales entremezclados con pajonales, matorrales bajos y bosques andinos conservados | *Espeletia perijaensis, Lourteigia stoechadifolia, Stevia lucida* y *Weinmannia pinnata* |
| 56 | Rfb-P/Epe-Cef | Frailejonales bajos entremezclados con pajonales | *Espeletia perijaensis* y *Calamagrostis effusa.* |
| 55 | Rfb-Mb-P/Epe-Lst-Cef-Cin | Frailejonales bajos entremezclados con matorrales bajos y pajonales | *Espeletia perijaensis, Lourteigia stoechadifolia, Stevia lucida, Calamagrostis effusa* y *Calamagrostis intermedia* |
| 54 | Rfb-Ma/Epe-Lst-Wpi | Frailejonales bajos entremezclados con matorrales altos | *Espeletia perijaensis, Lourteigia stoechadifolia, Stevia lucida* y *Calamagrostis effusa* |
| 53 | Rfb/Epe-Ale | Frailejonales bajos | *Espeletia perijaensis* y *Aa leucantha* |
| 52 | Rfa-P/Epe-Ldi-Cef | Frailejonales arborescentes entremezclados con pajonales | *Espeletia perijaensis, Libanothamnus divisoriensis* y *Calamagrostis effusa* |
| 51 | Rfa-Hz/Ldi-Epe-Och | Frailejonales arborescentes entremezclados con herbazales | *Libanothamnus divisoriensis, Espeletia perijaensis* y *Orthrosanthus chimboracensis* |
| 50 | Rfa/Ldi | Frailejonales arborescentes | *Libanothamnus divisoriensis* |
| 49 | P-Mb-Ma/Cef-Hph-Wpi | Pajonales entremezclados con matorrales bajos y altos | *Calamagrostis effusa, Hypericum magdalenicum* y *Weinmannia pinnata* |

## Continuación tabla 1.

| ALTA MONTAÑA DE LA Serranía de Perijá | | | |
|---|---|---|---|
| SIM_ID | SÍMBOLO | TIPO DE VEGETACIÓN | ESPECIES DOMINANTES |
| 48 | P-Mb-Baa1/Cef-Lst-Wpi | Pajonales entremezclados con matorrales bajos y bosques andinos atos conservados | *Calamagrostis effusa, Lourteigia stoechadifolia, Stevia lucida y Weinmannia pinnata* |
| 47 | P-Mb/Cef-Hba | Pajonales entremezclados con matorrales bajos | *Calamagrostis effusa, Hypericum baccharioides, Bejaria resinosa, Chaetolepis perijaensis y Arcythophyllum nitidum.* |
| 46 | P-Ma/Cin-Hfe | | *Calamagrostis intermedia y Hesperomeles ferruginea* |
| 45 | P-Ch/Cin-Cte | Pajonales entremezclados con chuscales | *Calamagrostis intermedia y Chusquea tessellata* |
| 44 | Mb-P-Ma/Hba-Cef-Gpu | Matorrales bajos entremezclados con pajonales | *Hypericum baccharoides, Calamagrostis effusa y Gaiadendron punctatum* |
| 43 | Mb-P/Lst-Cin-Cef | | *Lourteigia stoechadifolia, Stevia lucida, Calamagrostis intermedia y Calamagrostis effusa* |
| 42 | Mb-P/Hba-Cef | | *Hypericum baccharoides y Calamagrostis effusa* |
| 41 | Mb-P/Gbu-Hph-Cef | | *Gaylussacia buxifolia, Hypericum magdalenicum y Calamagrostis effusa* |
| 40 | Mb-P/Ani-Gbu-Cef | | *Arcytophyllum nitidum, Gaylussacia buxifolia y Calamagrostis effusa* |
| 39 | Mb/Lst-Hph | Matorrales bajos | *Lourteigia stoechadifolia e Hypericum magdalenicum* |
| 38 | Mb/Lst-Bae | | *Lourteigia stoechadifolia e Bejaria aestuans* |
| 37 | Mb/Hju-Pph | | *Hypericum juniperinum Bejaria sp., Arcytophyllum nitidum y Perissocoeleum phylloideum* |
| 36 | Mb/Ani-Hba | | *Arcytophyllum nitidum e Hypericum baccharioides* |
| 35 | Mb/Ani-Gbu-Cef | | *Arcytophyllum nitidum, Gaylusaccia buxifolia y Calamagrostis effusa* |
| 34 | Mb-Baa1/Hla-Pol | Matorrales bajos entremezclados con bosques andinos altos conservados | *Hypericum laricifolium y Podocarpus oleifolius* |
| 33 | Mb2/Paq-Och-Csp-Ati | Matorrales bajos no paramunos | *Pteridium aquilinum, Orthrosanthus chimboracensis, Cortaderia* spp. *y Ageratina tinifolia* |
| 32 | Ma-Hz-P/Gpu-Och-Cef | Matorrales altos entremezclados con herbazales y matorrales | *Gaiadendron puncatatum, Orthrosanthus chimboracensis y Calamagrostis effusa* |
| 31 | Ma2/Tme-Cmu | Matorrales altos | *Ternstroemia meridionalis, Clusia multiflora, Weinmannia pinnata y Prumnopitys montana.* |
| 30 | Ma/Wpi-Hba-Bae | | *Hypericum baccharoides, Bejaria aestuans y Weinmannia pinnata* |
| 29 | Ma/Gpu-Bgl | | *Gaiadendron puncatatum y Bejaria glauca.* |
| 28 | Baa2-Mb/Tme-Cmu-Hla | Bosque alto andino intervenido entremezclado con matorrales bajos | *Ternstroemia meridionalis, Clusia multiflora, Weinmannia pinnata e Hypericum laricifolium* |
| 27 | Baa2-Mb/Pol-Hla | | *Podocarpus oleifolius y Hypericum laricifolium.* |
| 26 | Baa2-Mb/Dga-Pol-Ani | | *Drimys granadensis, Podocarpus oleifolius, Clethra fimbriata y Arcytophyllum nitidum* |
| 25 | Baa2-Ma/Wpi-Hba-Bae | Bosque alto andino intervenido entremezclado con matorrales altos | *Hypericum baccharoides, Bejaria aestuans y Weinmannia pinnata* |
| 24 | Baa2-Ma/Tme-Cmu-Gpu | | *Ternstroemia meridionalis, Clusia multiflora, Weinmannia pinnata, Prumnopitys montana y Gaiadendron punctatum.* |
| 23 | Baa2-Hz-P/Tme-Cmu-Och-Cef | Bosque alto andino intervenido entremezclado con herbazales y pajonales | *Ternstroemia meridionalis, Clusia multiflora, Weinmannia pinnata, Prumnopitys montana y Orthrosanthus chimboracensis.* |
| 22 | Baa2-Hz/Hfe-Pmo-Och | Bosque alto andino intervenido entremezclado con herbazales | *Hesperomeles ferruginea, Prumnopitys montana, Myrcianthes* spp. *y Orthrosanthus chimboracensis* |
| 21 | Baa2/Wpi-Bco | Bosque alto andino intervenido | *Weinmannia pinnata, Brunellia integrifolia y Roupala pseudocordata* |
| 20 | Baa2/Tme-Cmu-Gpu | | *Ternstroemia meridionalis, Clusia multiflora, Weinmannia pinnata, Prumnopitys montana y Gaiadendron punctatum.* |
| 19 | Baa2/Tme-Cmu | | *Ternstroemia meridionalis, Clusia multiflora, Weinmannia pinnata y Prumnopitys montana.* |
| 18 | Baa2/Hfe-Pmo-Msp | | *Hesperomeles ferruginea, Prumnopitys montana y Myrcianthes* spp. |
| 17 | Baa1-Mb/Wpi-Ani-Gbu | Bosque alto andino conservado | *Weinmannia pinnata, Arcytophyllum nitidum y Gaylussacia buxifolia* |
| 16 | Baa1-Mb/Pol-Mli-Hfe | Bosque alto andino conservado entremezclado con matorrales bajos | *Podocarpus oleifolius, Miconia limitaris y Hesperomeles ferruginea* |
| 15 | Baa1-Mb/Pol-Hla | | *Podocarpus oleifolius e Hypericum laricifolium.* |
| 14 | Baa1-Mb/Hfe-Mli-Hla | | *Hesperomeles ferruginea, Miconia limitaris e Hypericum laricifolium* |
| 13 | Baa1-Ma-Mb/Pmo-Mli-Hfe-Hla | Bosque alto andino conservado entremezclado con matorrales altos y matorrales bajos | *Prumnopitys montana, Podocarpus oleifolius, Miconia limitaris, Hesperomeles ferruginea e Hypericum laricifolium.* |
| 12 | Baa1-Ma/Wpi-Hba-Bae | Bosque alto andino conservado entremezclado conmatorrales altos | *Hypericum baccharoides, Bejaria aestuans y Weinmannia pinnata* |
| 11 | Baa1-Ma/Tme-Cmu-Gpu | Bosque alto andino intervenido | *Ternstroemia meridionalis, Clusia multiflora, Weinmannia pinnata, Prumnopitys montana y Gaiadendron punctatum.* |
| 10 | Baa1-Ma/Pmo-Mli-Hfe | Bosque alto andino conservado entremezclado con matorrales altos | *Prumnopitys montana, Podocarpus oleifolius, Miconia limitaris y Hesperomeles ferruginea* |
| 9 | Baa1-Hz-P/Wpi-Cef-Och | Bosque alto andino conservado entremezclado con herbazales y pajonales | *Weinmannia pinnata, Brunellia integrifolia, Roupala pseudocordata, Calamagrostis effusa y Orthrosanthus chimboracensis* |
| 8 | Baa1-Hz-P/Tme-Cmu-Och-Cef | | *Ternstroemia meridionalis, Clusia multiflora, Weinmannia pinnata, Prumnopitys montana y Orthrosanthus chimboracensis* |

**Continuación tabla 1.**

| | | ALTA MONTAÑA DE LA Serranía de Perijá | |
|---|---|---|---|
| **SIM_ID** | **SÍMBOLO** | **TIPO DE VEGETACIÓN** | **ESPECIES DOMINANTES** |
| 7 | Baa1-Hz/Hfe-Pmo-Och | Bosque alto andino conservado entremezclado con herbazales | *Hesperomeles ferruginea, Prumnopitys montana, Myrcianthes* spp. y *Orthrosanthus chimboracensis* |
| 6 | Baa1/Wpi-Rro | | *Weinmania pinnata* y *Decussocarpus rospigliosii* |
| 5 | Baa1/Wpi-Bin | | *Weinmannia pinnata, Brunellia integrifolia* y *Roupala pseudocordata* |
| 4 | Baa1/Tme-Cmu | Bosque alto andino conservado | *Ternstroemia meridionalis, Clusia multiflora, Weinmannia pinnata* y *Prumnopitys montana* |
| 3 | Baa1/Pmo-Mli-Hfe | | *Prumnopitys montana, Podocarpus oleifolius, Miconia limitaris* y *Hesperomeles ferruginea* |
| 2 | Baa1/Pin-Adi | | *Prunus integrifolia* y *Acalypha diversifolia* |
| 1 | Baa1/Hfe-Mli | | *Hesperomeles ferruginea* y *Miconia limitaris* |

**Tabla 2.** Tipos de vegetación presentes en los páramos de La Rusia, Belén y Guantiva. El orden corresponde al identificador utilizado a lo largo del presente trabajo.

| | | PÁRAMOS DE LA RUSIA, BELÉN Y GUANTIVA | |
|---|---|---|---|
| **SIM_ID** | **SÍMBOLO** | **TIPO DE VEGETACIÓN** | **ESPECIES DOMINANTES** |
| 29 | U | Zonas urbanas destacables | |
| 28 | SV | Áreas sin vegetación | |
| 27 | A-Baa2/Wmi-Iku | Áreas agropecuarias con parches de bosque andino alto | |
| 26 | A | Áreas agropecuarias | |
| 25 | H | Cuerpos de agua | |
| 24 | Co/Ocl-Cco | Cortaderal | *Oreobolus cleefii, Cortaderia colombiana* y *Diplostephium colombianum.* |
| 23 | Ch/Cht | Chuscales | *Espeletia incana, Monnina salicifolia, Sphagnum sancto-josephense, Geranium sibbaldioides* y *Chusquea tessellata.* |
| 22 | Rf-P/Esp-Cef | Rosetales-frailejonales-pajonales | *Espeletia congestiflora, Calamagrostis effusa, Pentacalia vaccinioides, Paramiflos glandulosa* y *Arcytophyllum nitidum.* |
| 21 | Rf-Mb/Egu-Ani | Rosetal-frailejonal-matorral bajo | *Espeletiopsis guacharaca, Arcytophyllum nitidum* y *Bejaria resinosa* |
| 20 | Rf-Ch/Esp-Cht | Rosetales-frailejonales-chuscales | *Espeletia murilloi, Hypericum laricifolium, Espeletia incana, Chusquea tessellata* y *Calamagrostis effusa* |
| 19 | Rf/Eph | | *Espeletia phaneractis, Calamagrostis effusa* |
| 18 | Rf/Esp | | *Acaena cylindristachya* y *Espeletia boyacensis* |
| 17 | Rf/Blo-Emu | Rosetales-frailejonales | *Espeletia murilloi, Chusquea tessellata* y *Blechnum loxense* |
| 16 | Rf/Ale-Pgl | | *Achyrocline lehmannii, Clethra fimbriata* y *Paramiflos glandulosus* |
| 15 | P-Mb/Cef-Hla | Pajonales con matorrales | *Calamagrostis effusa, Hypericum laricifolium* y *Ageratina tinifolia* |
| 14 | P-Mb/Cef | | *Aragoa cleefii, Calamagrostis effusa* y *Castilleja fissifolia.* |
| 13 | P/Cin-Cbo | Pajonales | *Castilleja integrifolia, Bartsia santolinifolia* y *Calamagrostis bogotensis* |
| 12 | Mb/Hla | Matorrales bajos | *Hypericum laricifolium, Ageratina tinifolia, Escallonia myrtilloides* y *Geranium sibbaldioides* |
| 11 | Ma/Pqj | Matorrales altos | *Polylepis quadrijuga* |
| 10 | Ma/Msa | | *Miconia salicifolia* |
| 9 | Baa1-Mb/Wmi-Hla | Bosque andino alto entremezclado con vegetación de páramo | *Weinmannia microphylla, Ilex kunthiana* e *Hypericum laricifolium* |
| 8 | Baa1-Ch/Wmi-Ces | | *Weinmannia microphylla, Ilex kunthiana* y *Chusquea* spp. |
| 7 | Baa1-Ch/Pqj-Cht | | *Polylepis quadrijuga, Diplostephium tenuifolium* y *Chusquea tessellata* |
| 6 | Baa2-A/Wmi-Iku | Bosque andino alto intervenido o entremezclado con áreas agropecuarias | *Weinmannia microphylla, Ilex kunthiana* y *Miconia* sp. |
| 5 | Baa2/Wmi-Iku | | |
| 4 | Baa2/Qhu | | *Quercus humboldtii* |
| 3 | Baa1/Wmi-Iku | Bosque andino alto | *Weinmannia microphylla, Ilex kunthiana* y *Miconia* sp. |
| 2 | Baa1/Qhu | | *Quercus humboldtii* |
| 1 | Baa1/Pqj | | *Polylepis quadrijuga, Diplostephium tenuifolium* y *Escallonia myrtilloides* |

**Tabla 3.** Tipos de vegetación presentes en el páramo de Telecom. El orden corresponde al identificador utilizado a lo largo del presente trabajo.

| PÁRAMO DE TELECOM | | | |
|---|---|---|---|
| SIM_ID | SIMBOLO | TIPO DE VEGETACIÓN | ESPECIES DOMINANTES |
| 44 | SV | Áreas sin vegetación | |
| 43 | A-Mb/Chl | Áreas de uso agropecuario y matorrales bajos | Clethra fimbriata |
| 42 | A-Mb/Ani | | Arcytophyllum nitidum |
| 41 | A | Áreas de uso agropecuario | |
| 40 | Pl-Ma/Ppa-Clm | Plantaciones con matorrales altos | Pinus patula y Clusia multiflora |
| 39 | Pl/Ppa | Plantaciones | Pinus patula |
| 38 | Vr-Py/Pyn | Rosetales | Puya nitida |
| 37 | Vr-Mb/Mve | Vegetación riparia con matorrales bajos | Monticalia vernicosa |
| 36 | Vr-Mb/Hla | | Hypericum laricifolium |
| 35 | Vr-Mb/Ati | | Ageratina tinifolia |
| 34 | Vc-P-Rf/Cef | Vegetación casmófita con pajonales frailejonales | Calamagrostis effusa |
| 33 | Vc-P/Cef | | Calamagrostis effusa |
| 32 | Rf/Esp | Frailejonales | Espeletia sp. |
| 31 | Rf/Eph | | Espeletia phaneractis |
| 30 | P-Rf/Cef | Pajonales frailejonales | Calamagrostis effusa |
| 29 | Rf-Baa2/Wto | Frailejonales con relictos de bosque andino alto intervenido | Weinmannia tomentosa |
| 28 | Rf-Baa2/Eph-Bco | | Espeletia phaneractis y Brunellia colombiana |
| 27 | Rf-Baa2/Egr-Wto | | Espeletia grandiflora y Weinmannia tomentosa |
| 26 | Rf-Baa2/Bco | | Brunellia colombiana |
| 25 | Rf-Baa2/Cef-Wto | | Calamagrostis effusa y Weinmannia tomentosa |
| 24 | P-Rf/Cef-Esp | Pajonales frailejonales | Calamagrostis effusa y Espeletia sp. |
| 23 | P/Cef | Pajonales | Calamagrostis effusa |
| 22 | Mb-Baa2/Bj-Wto | Matorrales bajos con relictos de bosque andino alto intervenido | Bejaria resinosa y Weinmannia tomentosa |
| 21 | Mb/Mve | Matorrales bajos | Monticalia vernicosa |
| 20 | Mb/Emy | | Escallonia myrtilloides |
| 19 | Mb/Ani | | Ageratina tinifolia |
| 18 | Mb-Rf/Eph-Ani | Matorrales bajos y frailejonales | Arcytophyllum nitidum y Espeletia phaneractis |
| 17 | Mb-P-Rf/Cef-Ani | Matorrales bajos con pajonales | Arcytophyllum nitidum y Calamagrostis effusa |
| 16 | Mb/Chl | Matorrales bajos | Clethra fimbriata |
| 15 | Mb/Bj | | Bejaria resinosa |
| 14 | Ma/Wto | Matorrales altos | Weinmannia tomentosa |
| 13 | Ma/Chl | | Clethra fimbriata |
| 12 | Ma/Bco | | Brunellia colombiana |
| 11 | Bab1/Xsp-Dca | Bosque andino baja | Xylosma spiculiferum y Daphnopsis caracasana |
| 10 | Baa2-Pl/Wto | Bosque andino alto intervenido entremezclado con plantaciones forestales | Weinmannia tomentosa |
| 9 | Baa2-Mb/Wto-Ani | Bosque andino alto intervenido con matorrales bajos | Weinmannia tomentosa y Arcytophyllum nitidum |
| 8 | Baa2-Mb/Chl | Bosque andino alto intervenido y matorrales bajos | Clethra fimbriata |
| 7 | Baa2-A/Wto | Bosque andino alto intervenido y áreas agropecuarias | Weinmannia tomentosa y Weinmannia tomentosa |
| 6 | Baa2-A/Bco | | Brunellia colombiana |
| 5 | Baa2/Chl | Bosque andino alto intervenido | Clethra fimbriata |
| 4 | Baa1-Mb/Chl | Bosque andino alto con matorrales bajos | Clethra fimbriata |
| 3 | Baa1/Msa | Bosque andino alto | Miconia salicifolia |
| 2 | Baa1/Chl | | Clethra fimbriata |
| 1 | Baa1/Bco | | Brunellia colombiana |

**Tabla 4.** Tipos de vegetación presentes en el páramo de Merchán. El orden corresponde al identificador utilizado a lo largo del presente trabajo.

| PÁRAMO DE MERCHÁN | | | |
|---|---|---|---|
| SIM_ID | SÍMBOLO | TIPO DE VEGETACIÓN | ESPECIES DOMINANTES |
| 22 | Pl-Baa2/Ppa-Clm | Plantaciones con relictos de bosque andino alto intervenido | Pinus patula y Clusia multiflora |
| 21 | A-Rf/Eph | Áreas de uso agropecuario y rosetales frailejonales | Espeletia phaneractis |
| 20 | A-Ma/Chl | Áreas de uso agropecuario y matorrales altos | Clusia multiflora |
| 19 | A | Áreas de uso agropecuario | |
| 18 | Vr-Pl/Ati-Ppa | Vegetación riparia con planaciones forestales | Ageratina tinifolia y Pinus patula |
| 17 | Vr-Ma-Mb/Vst | Vegetación riparia con matorrales altos y bajos | Vallea stipularis |

## Continuación tabla 4.

| | | PÁRAMO DE MERCHÁN | |
|---|---|---|---|
| SIM_ID | SÍMBOLO | TIPO DE VEGETACIÓN | ESPECIES DOMINANTES |
| 16 | Vr-Ma/Emy | Vegetación riparia con matorrales altos | *Escallonia myrtilloides* |
| 15 | Vr-Ma/Ati | | *Ageratina tinifolia* |
| 14 | Rf/Eph | Frailejonales | *Espeletia phaneractis* |
| 13 | Rf/Eph-Chl | | *Espeletia phaneractis* y *Clethra fimbriata* |
| 12 | Rf-Baa2/Ati | Frailejonales con bosque andino alto intervenido | *Clethra fimbriata* |
| 11 | P/Cef | Pajonales | *Calamagrostis effusa* |
| 10 | Mb-Vr/Ati | Vegetación riparia de matorrales bajos | *Ageratina tinifolia* |
| 9 | Mb-Baa2/Chl | Matorrales bajos y bosque andino alto intervenido | *Clethra fimbriata* |
| 8 | Mb-Rf/Chl-Egr | Matorrales bajos con frailejonales | *Arcytophyllum nitidum* y *Espeletia grandiflora* |
| 7 | Mb/Hla | Matorrales bajos | *Hypericum laricifolium* |
| 6 | Mb/Ani-Chl | Matorrales bajos con relictos de matorrales | *Arcytophyllum nitidum* y *Clethra fimbriata* |
| 5 | Mb/Acl | Matorrales bajos | *Aragoa cleefii* |
| 4 | Mb/Bj | | *Bejaria resinosa* |
| 3 | Mb/Chl | | *Clethra fimbriata* |
| 2 | Baa2/Wto | Bosque andino alto intervenido | *Weinmannia tomentosa* |
| 1 | Baa2/Chl | | *Clethra fimbriata* |

**Tabla 5.** Tipos de vegetación presentes en los páramos de Carmen de Carupa y sectores aledaños. El orden corresponde al identificador utilizado a lo largo del presente trabajo.

| | | PÁRAMOS DE CARMEN DE CARUPA Y SECTORES ALEDAÑOS | |
|---|---|---|---|
| SIM_ID | SÍMBOLO | TIPO DE VEGETACIÓN | ESPECIES DOMINANTES |
| 44 | H | Cuerpos de agua | |
| 43 | A-Baa2/Wto | Áreas de uso agropecuario entremezcladas con bosques intervenidos | *Weinmannia tomentosa* |
| 42 | A | Áreas de uso agropecuario | |
| 41 | Vr-Py/Pyn | Rosetales | *Puya nitida* |
| 40 | Vp/B-Rf-Baa2/Esp-Wto | Vegetación de páramo (Rosetal-frailejonal) dominando sobre bosque altoandino | *Espeletia* sp. y *Weinmannia tomentosa* |
| 39 | Vp/B-Rf-Baa2/Egr-Wto | | *Espeletia grandiflora* y *Weinmannia tomentosa* |
| 38 | Vc-Mb/Ani-Cef | Vegetación casmófita con matorrales bajos | *Arcytophyllum nitidum* y *Calamagrostis effusa* |
| 37 | Rf-Mb/Pyn-Esp | Frailejonales con matorrales bajos y rosetales | *Espeletia* sp. y *Puya nitida* |
| 36 | Rf-Mb/Eph-Ani | Frailejonales con matorrales bajos | *Espeletia phaneractis* y *Arcytophyllum nitidum* |
| 35 | Rf-Mb/Eph | | *Espeletia phaneractis* |
| 34 | Rf-Baa1/Egr-Wto | Frailejonales con parches de bosque andino alto | *Weinmannia tomentosa* y *Espeletia grandiflora* |
| 33 | Rf/Eph | Frailejonales | *Espeletia phaneractis* |
| 32 | P-Mb/Cef-Ani | Pajonales con matorrales bajos | *Calamagrostis effusa* y *Arcytophyllum nitidum* |
| 31 | P-Mb/Cef | | *Calamagrostis effusa* |
| 30 | Pl/Ppa | Plantaciones | *Pinus patula* |
| 29 | P/Cef-Esp | Pajonales | *Calamagrostis effusa* y *Espeletia* sp. |
| 28 | P/Cef | | *Calamagrostis effusa* |
| 27 | Mb-Rf/Eph-Ani | Matorrales bajos y frailejonales | *Arcytophyllum nitidum* y *Espeletia phaneractis* |
| 26 | Mb-P/Cef | Matorrales bajos y pajonales | *Calamagrostis effusa* |
| 25 | Mb-P/Ani-Cef | Matorrales bajos con pajonales | *Arcytophyllum nitidum* y *Calamagrostis effusa* |
| 24 | Mb-Baa2/Bj-Wto | Matorrales bajos con relictos de bosque andino alto intervenido | *Bejaria resinosa* y *Weinmannia tomentosa* |
| 23 | Mb-Baa2/Ani-Wto | | *Arcytophyllum nitidum* y *Weinmannia tomentosa* |
| 22 | Mb/Mve | Matorrales bajos | *Monticalia vernicosa* |
| 21 | Mb/Hsp | | *Hypericum strictum* |
| 20 | Mb/Emy | | *Escallonia myrtilloides* |
| 19 | Mb/Bj | | *Bejaria resinosa* |
| 18 | Mb/Ani | | *Ageratina tinifolia* |
| 17 | Mb/Acl | | *Aragoa cleefii* |
| 16 | Ma1/Wto | | *Weinmannia tomentosa* |
| 15 | Ma/Ati | | *Ageratina tinifolia* |
| 14 | B/Vp-Baa2-P-Rf/Wto-Cef-Esp | Bosque alto andino intervenido y vegetación de páramos | *Weinmannia tomentosa* y *Calamagrostis effusa-Espeletia* sp. |
| 13 | Bab2/Xsp-Dca | Bosque andino bajo intervenido | *Xylosma spiculiferum* y *Daphnopsis caracasana* |
| 12 | Bab1/Xsp-Dca | Bosque andino baja | |
| 11 | Baa2-Pl/Wto-Ppa | Bosque andino alto intervenido entremezclado con plantaciones forestales | *Weinmannia tomentosa* y *Pinus patula* |
| 10 | Baa2-P/Wto-Cef | Bosque andino alto intervenido con pajonales | *Weinmannia tomentosa* y *Calamagrostis effusa* |

## Continuación tabla 5.

| | | PÁRAMOS DE CARMEN DE CARUPA Y SECTORES ALEDAÑOS | |
|---|---|---|---|
| SIM_ID | SÍMBOLO | TIPO DE VEGETACIÓN | ESPECIES DOMINANTES |
| 9 | Baa2-Ma/Wto | Bosque andino alto intervenido con matorrales altos | *Weinmannia tomentosa* |
| 8 | Baa2/Wto | Bosque andino alto intervenido | |
| 7 | Baa1-Rf/Wsp-Eph | Bosque andino alto y frailejonales | *Weinmannia* sp. y *Espeletia phaneractis* |
| 6 | Baa1-P/Wto-Cef | Bosque andino alto y pajonales | *Weinmannia tomentosa* y *Calamagrostis effusa* |
| 5 | Baa1-P/Wsp-Cef | | *Weinmannia* sp. y *Calamagrostis effusa* |
| 4 | Baa1-Mb/Wsp-Hju | Bosque andino alto y matorrales bajos | *Weinmannia* sp. y *Hypericum juniperinum* |
| 3 | Baa1-Mb/Wsp-Ani | | *Weinmannia* sp. y *Arcytophyllum nitidum* |
| 2 | Baa1-Ch/Wto-Cht | Bosque andino alto con chuscales | *Weinmannia tomentosa* y *Chusquea tessellata* |
| 1 | Baa1/Wto | Bosque andino alto | *Weinmannia tomentosa* |

**Tabla 6.** Tipos de vegetación presentes en el páramo El Tablazo. El orden corresponde al identificador utilizado a lo largo del presente trabajo.

| | | PÁRAMO DEL TABLAZO | |
|---|---|---|---|
| SIM_ID | SÍMBOLO | TIPO DE VEGETACIÓN | ESPECIES DOMINANTES |
| 45 | SVC | Áreas de explotación minera | |
| 44 | SV | Áreas sin vegetación | |
| 43 | Q | Áreas con quemas | |
| 42 | H | H | |
| 41 | Pl/Ppa | Plantaciones | *Pinus patula* |
| 40 | A-Baa2/Wto | Áreas de uso agropecuario entremezcladas con bosques intervenidos | *Weinmannia tomentosa* |
| 39 | A | Áreas de uso agropecuario | |
| 38 | Vr-Mb/Ani | vegetación riparia con matorrales bajos | *Arcytophyllum nitidum* |
| 37 | Vr-Ma-Mb/Vst | vegetación riparia con matorrales altos y bajos | *Vallea stipularis* |
| 36 | Vr-Ma/Ati | vegetación riparia con matorrales altos | *Ageratina tinifolia* |
| 35 | Rf-Mb/Eph-Ani | Frailejonales con matorrales bajos | *Espeletia phaneractis* y *Arcytophyllum nitidum* |
| 34 | Rf/Esp | Frailejonales | *Espeletia* sp. |
| 33 | Rf/Egr | | *Espeletia grandiflora* |
| 32 | P/Cef | Pajonales | *Calamagrostis effusa* |
| 31 | Mb/Vfl | Matorrales bajos | *Vaccinium floribundum* |
| 30 | Mb/Hla | | *Hypericum laricifolium* |
| 29 | Mb/Ani | | *Ageratina tinifolia* |
| 28 | Baa2/Ecy-Wto | Bosque andino alto intervenido | *Espeletia sp1* y *Weinmannia tomentosa* |
| 27 | Ch/Cht | Chuscales | *Chusquea tessellata* |
| 26 | P-Ma/Wto | Pajonales con matorrales altos | *Weinmannia tomentosa* |
| 25 | Mb-Vc/Bj | Matorrales bajos y vegetación casmófita | *Bejaria resinosa* |
| 24 | Mb-Rf/Ani-Egr | Matorrales bajos con frailejonales | *Arcytophyllum nitidum* y *Espeletia grandiflora* |
| 23 | Mb-Baa2/Bj-Wto | Matorrales bajos con relictos de bosque andino alto intervenido | *Bejaria resinosa* y *Weinmannia tomentosa* |
| 22 | Mb-P/Ani-Cef | Matorrales bajos con pajonales | *Arcytophyllum nitidum* y *Calamagrostis effusa* |
| 21 | Mb/Bj | Matorrales bajos | *Bejaria resinosa* |
| 20 | Mb/Acl | Matorrales bajos | *Aragoa cleefii* |
| 19 | Ma-Rf/Wto-Eph | | *Weinmannia tomentosa* y *Espeletia phaneractis* |
| 18 | Ma/Xsp | Matorrales altos cerrados | *Xylosma spiculiferum* |
| 17 | Ma/Wto-Ani | | *Weinmannia tomentosa* y *Arcytophyllum nitidum* |
| 16 | Ma/Wto-Acl | | *Weinmannia tomentosa* y *Aragoa Clefii* |
| 15 | Ma/Wto | Matorrales altos | *Weinmannia tomentosa* |
| 14 | Ma/Mru | | *Macleania rupestris* |
| 13 | Bab2-Pl/Xsp-Dca-Ppa | Bosque andino bajo intervenido y plantaciones forestales | *Xylosma spiculiferum* y *Daphnopsis-Pinus patula* |
| 12 | Bab2/Xsp-Dca | Bosque andino bajo intervenido | *Xylosma spiculiferum* y *Daphnopsis caracasana* |
| 11 | Bab1/Xsp-Dca | Bosque andino baja | |
| 10 | Baa2-Pl/Wto | Bosque andino alto intervenido entremezclado con plantaciones forestales | *Weinmannia tomentosa* |
| 9 | Baa2-Pl/Qhu | Bosque andino alto intervenido, áreas con quemas y plantaciones forestales | *Quercus humboldtii* |
| 8 | Baa2-Mb/Wto-Ani | Bosque andino alto intervenido con matorrales bajos | *Weinmannia tomentosa* y *Arcytophyllum nitidum* |
| 7 | Baa2/Wto | Bosque andino alto intervenido | *Weinmannia tomentosa* |
| 6 | Baa2/Vst | | *Vallea stipularis* |

## Continuación tabla 6.

| PÁRAMO DEL TABLAZO | | | |
|---|---|---|---|
| SIM_ID | SÍMBOLO | TIPO DE VEGETACIÓN | ESPECIES DOMINANTES |
| 5 | Baa2/Qhu | Bosque andino alto intervenido entremezclado con áreas con frecuentes quemas | *Quercus humboldtii* |
| 4 | Baa1/Wto | Bosque andino alto | *Weinmannia tomentosa* |
| 3 | Baa1/Qhu | Bosque andino alto entremezclado con áreas de frecuentes quemas | *Quercus humboldtii* |
| 2 | Baa1-Rf/Wto-Esp | Bosque andino alto y frailejonales | *Weinmannia tomentosa* y *Espeletia* sp. |
| 1 | Baa2-P-Rf/Wto-Cef-Esp | Bosque andino alto intervenido con pajonales y frailejonales | *Weinmannia tomentosa- Calamagrostis effusa* y *Espeletia* sp. |

**Tabla 7.** Tipos de vegetación presentes en el páramo El Tablazo. El orden corresponde al identificador utilizado a lo largo del presente trabajo.

| PÁRAMOS PERTENECIENTES A LA JUSISDICCIÓN CORPOGUAVIO | | | |
|---|---|---|---|
| SIM_ID | SÍMBOLO | TIPO DE VEGETACIÓN | ESPECIES DOMINANTES |
| 60 | X | Minería | |
| 59 | Q | Quemas | |
| 58 | Pl | Plantaciones Forestales | |
| 57 | Ap | Zonas agropecuarias | Especies de Poaceae |
| 56 | Ac | | Predominan Cultivos |
| 55 | H | Cuerpos de Agua | |
| 54 | E | Erial | |
| 53 | Vr | Vegetación Riparia | |
| 52 | Vt/Rhp | Vegetación deturberas | *Rhacocarpus purpurascens* |
| 51 | Vt/Ove | | *Oreobolus venezuelensis* |
| 50 | Rf-P/Es-Cef | Rosetales frailejonales entremezclados con pajoales | *Espeletia grandiflora* y *Calamagrostis effussa* |
| 49 | Rf-P/Egr | | *Espeletia grandiflora* |
| 48 | Rf/Egr | Rosetales frailejonales | |
| 47 | Rf/Ea | | *Espeletia argentea* |
| 46 | Pt/Pls-Crv-Els | | *Pleurozium schreberi, Crassula venezualensis* y *Eleocharis stenocarpa.* |
| 45 | Pt/Hol | | *Holcus lanatus* |
| 44 | Pt/Est | | *Eleocharis stenocarpa* |
| 43 | Pt/Cve | Vegetación de Pantano | *Crassula venezuelensis* |
| 42 | Pt/Cru | | *Cyperus* aff. *rufus* |
| 41 | Pt/Cli | | *Calamagrostis ligulata* |
| 40 | Pt/Cja | | *Carex jamesonii* var. *chordalis* |
| 39 | Pt/Cac | | *Carex acutata* |
| 38 | Ch-Rf/Cht-Egr | | *Chusquea tesellata* y *Espeletia grandiflora* |
| 37 | Ch-P/Cht-Cbo | Chuscales | *Chusquea tesellata* y *Calamagrostis bogotensis* |
| 36 | Ch/Cht | | *Chusquea tesellata* |
| 35 | P-Rf/Ch/Cef-Egr | | *Calamagrostis effusa* y *Espeletia grandiflora* |
| 34 | P-Rf/Cef-Egr | Pajonales entremezclados con frailejonales | |
| 33 | P-Rf/Cef-Ear | | *Calamagrostis effussa* y *Espeletia argentea* |
| 32 | P-Rf/Cef | | *Calamagrostis effusa* |
| 31 | P-Mb/Cef | Pajonales entremezclados con Matorrales | |
| 30 | P-Mb/An-Cef | | *Calamagostis effussa* y *Arcytophyllum nitidum* |
| 29 | P/Fdo | | *Festuca dolichophylla* |
| 28 | P/Cef | Pajonales | *Calamagrostis effusa* |
| 27 | P/Cbo | | *Calamagrostis bogotensis* |
| 26 | Mb-P/Cef | Matorrales bajos entremezcla con pajonales | *Calamagrostis effussa* |
| 25 | Mb-P/Ani-Cef | | *Arcytophyllum nitidum* y *Calamagrostis effussa* |
| 24 | Mb-Ch/Hgo-Cht | Matorrales bajos entremezclados con chuscales | *Hypericum goyanesii* y *Chusquea tessellata* |
| 23 | Mb/Vfl | | *Vaccinium floribundum* |
| 22 | Mb/Hgo | | *Hypericum goyanesii* |
| 21 | Mb/Gra | Matorrales bajos | *Gaultheria ramossisima* |
| 20 | Mb/Ani | | *Arcytophyllum nitidum* |
| 19 | Mb/Aab | | *Aragoa abietina* |
| 18 | Ma-P/Wmi-Cef | | *Weinmannia microphylla* y *Calamagrostis effussa* |
| 17 | Ma-P/Cef | Matorrales altos entremezclados con pajonales | *Calmagrostis effussa* |
| 16 | Ma-Ch/Msp-Cs | | *Miconia* sp. y *Chusquea spencei* |
| 15 | Ma-Ch/Bgl-Cht | | *Bucquetia glutinosa* y *Chusquea tessellata.* |

## Continuación tabla 7.

| | | PÁRAMOS PERTENECIENTES A LA JUSISDICCIÓN CORPOGUAVIO | |
|---|---|---|---|
| SIM_ID | SÍMBOLO | TIPO DE VEGETACIÓN | ESPECIES DOMINANTES |
| 14 | Ma/Wsp | Matorrales altos | *Weinmannnia* sp. 2 |
| 13 | Ma/Wmi | | *Weinmania microphylla* |
| 12 | Ma/Vfl | | *Vaccinium floribundum* |
| 11 | Ma/Ppp | | *Peperomia* sp. |
| 10 | Ma/Pga | | *Peperomia galioides* |
| 9 | Ma/Mni | | *Monticalia nitida* |
| 8 | Ma/Dri | | *Drimys granadensis* |
| 7 | Ma/Bgl | | *Bucquetia glutinosa* |
| 6 | Ma/Ati | | *Ageratina tinifolia* |
| 5 | Baa1-Ma/Wmi | Bosque Alto Andino conservado o semi conservado | *Weinmannia microphylla* |
| 4 | Baa1-Ch/Wmi-Chs | | *Weinmania microphylla* y *Chusquea spencei* |
| 3 | Baa1/Wsp | | *Weinmannia* sp. 2 |
| 2 | Baa1/Wmi | | *Weinmania microphylla* |
| 1 | Baa1/Dri | | *Drimys granadensis* |

**Tabla 8.** Tipos de vegetación presentes en el área de estudio perteneciente al páramo de Sumapaz. El orden corresponde al identificador utilizado a lo largo del presente trabajo.

| | | PÁRAMO DE SUMAPAZ | |
|---|---|---|---|
| SIM_ID | SÍMBOLO | TIPO DE VEGETACIÓN | ESPECIES DOMINANTES |
| 46 | U | Áreas urbanas | |
| 45 | SV-AQ | Áreas sin vegetación | |
| 44 | Si | Zonas sin información | |
| 43 | H | Sistemas hídricos visibles | |
| 42 | A-Pl-Egl | Áreas de uso agropecuario y plantaciones forestales | *Eucalyptus globulus* |
| 41 | A-Ma/Wro | Áreas de uso agropecuario y matorrales altos | *Weinmannia rollotii* |
| 40 | A-Ma/Clm | | *Clusia multiflora* |
| 39 | A | Áreas de uso agropecuario | |
| 38 | Vr-Py/Pyn | Rosetales | *Puya nitida* |
| 37 | Vr-Ma/Ati | matorrales altos | *Ageratina tinifolia* |
| 36 | Vp-Vr-P-Ma/Cef-Clm | Vegetación riparia, pajonales y matorrales | *Calamagrostis effusa - Clusia multiflora* |
| 35 | Vp/B-Rf-Baa2/Wro | Vegetación de páramo entremezclada con bosques intervenidos | *Weinmannia rollotii* |
| 34 | Vp/B-Rf-Baa2/Msa | | *Miconia salicifolia* |
| 33 | Vp/B-Rf-Baa2/Egr-Wto | | *Espeletia grandiflora - Weinmannia rollotii* |
| 32 | Vp/B-Rf-Baa2/Egr-Msa | Vegetación de páramo (Rosetal-frailejonal) dominando sobre bosque altoandino | *Espeletia grandiflora - Miconia salicifolia* |
| 31 | Vp/B-P-Rf-Baa2/Cef-Wto | Vegetación de páramo (Pajonal-rosetal-frailejonal) dominando sobre bosque altoandino | *Weinmannia tomentosa - Calamagrostis effusa* |
| 30 | Vp/B-Mb-Baa2/Aab-Clm | Vegetación de páramo (Matorrales bajos) dominando sobre bosque altoandino | *Aragoa abietina - Clusia multiflora* |
| 29 | Vp/B-Ch-Baa2/Cht-Wto | Vegetación de páramo (Chuscal) dominando sobre bosque altoandino | *Chusquea tessellata - Weinmannia tomentosa* |
| 28 | Vc-Mb/Aab | Vegetación casmófita principalmente matorrales bajos | *Aragoa abietina* |
| 27 | Ch/Cht | Chuscales | *Chusquea tessellata* |
| 26 | Rf/Egr | Rosetal frailejonal | *Espeletia grandiflora* |
| 25 | P-Rf/Cef-Esp | Pajonale con rosetales frailejonales | *Calamagrostis effusa - Espeletia sp* |
| 24 | Mb-P/Cef | Matorrales bajos y pajonales | *Calamagrostis effusa* |
| 23 | Mb-Ch/Cht | Matorrales bajos con chuscales | *Chusquea tessellata* |
| 22 | Mb/Vfl | Matorrales bajos | *Vaccinium floribundum* |
| 21 | Mb/Bj | | *Bejaria resinosa* |
| 20 | Mb/Ani | | *Arcytophyllum nitidum* |
| 19 | Ma/Wro | Matorrales altos | *Macleania rupestris* |
| 18 | Ma/Mru | | *Macleania rupestris* |
| 17 | B/Vp-Baa-Ch/Wto-Cht | Bosque alto andino y vegetación de páramos | *Weinmannia tomentosa - Chusquea tessellata* |
| 16 | B/Vp-Baa2-P-Rf/Wto-Cef-Esp | Bosque alto andino intervenido y vegetación de páramos | *Weinmannia tomentosa - Calamagrostis effusa-Espeletia* sp. |
| 15 | B/Vp-Baa2-Ch/Wto-Cht | | *Weinmannia tomentosa - Chusquea tessellata* |
| 14 | Bab1/Xsp-Dca | Bosque andino bajo | *Xylosma spiculiferum - Daphnopsis caracasana* |
| 13 | Baa2-P/Wto-Cef | Bosque andino alto intervenido con pajonales | *Weinmannia tomentosa - Calamagrostis effusa* |

## Continuación tabla 8.

| PÁRAMO DE SUMAPAZ | | | |
|---|---|---|---|
| SIM_ID | SÍMBOLO | TIPO DE VEGETACIÓN | ESPECIES DOMINANTES |
| 12 | Baa2-A/Wto | Bosque andino alto intervenido con áreas agropecuarias | *Weinmannia tomentosa - Weinmannia tomentosa* |
| 11 | Baa2-A/Wro | | *Weinmannia rollotii* |
| 10 | Baa2/Wto | Bosque andino alto intervenido | *Weinmannia tomentosa* |
| 9 | Baa2/Wro | | *Weinmannia rollotii* |
| 8 | Baa2/Msa | | *Miconia salicifolia* |
| 7 | Baa2/Clm | | *Clusia multiflora* |
| 6 | Baa1-P/Wto-Cef | | *Weinmannia tomentosa - Calamagrostis effusa* |
| 5 | Baa1-Baa2/Wro | Bosque andino alto conservado e intervenido | *Weinmannia rollotii* |
| 4 | Baa1-Baa2/Clm | | *Clusia multiflora* |
| 3 | Baa1/Wro | Bosque andino alto | *Weinmannia rollotii* |
| 2 | Baa1/Msa | | *Miconia salicifolia* |
| 1 | Baa1/Clm | | *Clusia multiflora* |

**Tabla 9.** Tipos de vegetación presentes en el área de estudio perteneciente al área de páramo presente ene el Parque Nacional Natural Los Nevados. El orden corresponde al identificador utilizado a lo largo del presente trabajo.

| PÁRAMO DEL PARQUE NACIONAL NATURAL LOS NEVADOS | | | |
|---|---|---|---|
| SIM_ID | SÍMBOLO | TIPO DE VEGETACIÓN | ESPECIES DOMINANTES |
| 36 | H | Cuerpos de agua | |
| 35 | N2 | Nieves perpetuas | |
| 34 | N1 | | |
| 33 | Hz/Bca-Lni | Herbazales | *Baccharis caespitosa-Lachemilla nivalis* |
| 32 | Cj/Dmu | Cojines | *Distichia muscoides* |
| 31 | Sd-Hz/Pge-Bca-Sve | Suelo desnudo y herbazales | *Pentacalia gelida-Baccharis caespitosa-Stereocaulon vesuvianum* |
| 30 | R/Sis-Lal | Rosetales | *Senecio isabeli-LuPinus alopecurioides* |
| 29 | Pv-P-Ps/Pri-Cre-Cef-Ahe | Cojines de plantas vasculares con pajonales y pastizales | *Plantago rigida-Calamagrostis recta-Calamagrostis effusa-Agrostis haenkeana* |
| 28 | Pv-P-Ps/Pri-Cef-Ahe | | *Plantago rigida-Calamagrostis effusa-Agrostis haenkeana* |
| 27 | Pv-P/Pri-Dmu-Cre-Cef | Cojines de plantas vasculares y pajonales | *Plantago rigida-Distichia muscoides-Calamagrostis recta-Calamagrostis effusa* |
| 26 | Pv-P/Pri-Dmu-Cre | | *Plantago rigida-Distichia muscoides-Calamagrostis recta* |
| 25 | Pv-P/Pri-Dmu-Cef | | *Plantago rigida-Distichia muscoides-Calamagrostis effusa* |
| 24 | Pv-P/Pri-Cre-Cef | | |
| 23 | Pv-P/Pri-Cre | | *Plantago rigida-Calamagrostis recta* |
| 22 | Pv/Pri | Cojines de plantas vasculares | *Plantago rigida* |
| 21 | Ps-Pv/Ahe-Pri | Pastizales y cojines de plantas vasculares | *Agrostis haenkeana-Plantago rigida* |
| 20 | Ps-P/Ahe-Cre-Cef | Pastizales y pajonales | *Agrostis haenkeana-Calamagrostis recta-Calamagrostis effusa* |
| 19 | Ps-P/Ahe-Cef | | *Agrostis haenkeana-Calamagrostis effusa* |
| 18 | Ps/Ahe | Pastizales | *Agrostis haenkeana* |
| 17 | P-Pv/Cre-Cef-Pri | Pajonales y cojines de plantas vasculares | |
| 16 | P-Ps-Pv/Cef-Ahe-Pri | Pajonales con pastizales y cojines de plantas vasculares | *Calamagrostis recta-Calamagrostis effusa-Plantago rigida* |
| 15 | P-Ps/Cef-Ahe | Pajonales y pastizales | *Calamagrostis effusa-Agrostis haenkeana* |
| 14 | Pd-Pv/Lor-Pri | Prados y cojines de plantas vasculares | *Lachemilla orbiculata-Plantago rigida* |
| 13 | Pd-Baa1-Pv/Lor-Hfe-Pri | Prados con parches de bosques y cojines de plantas vasculares | *Lachemilla orbiculata-Hesperomeles ferruginea-Plantago rigida* |
| 12 | Pd/Lor | Prados | *Lachemilla orbiculata* |
| 11 | P/Cre-Cef | Pajonales | *Calamagrostis recta-Calamagrostis effusa* |
| 10 | P/Cre | | *Calamagrostis recta* |
| 9 | Mb-Baa1/Pve-Cre-Pse | Matorrales bajos con enclaves de bosque achaparrado | *Pentacalia vernicosa-Calamagrostis recta-Polylepis sericea* |
| 8 | Mb/Pve | Matorrales bajos | *Pentacalia vernicosa* |
| 7 | Mb/Lco | | *Loricaria colombiana* |
| 6 | Mb/Hla | | *Hypericum laricifolium* |
| 5 | Ma-P/Emy-Cef | Matorrales altos | *Escallonia myrtilloides-Calamagrostis effusa* |
| 4 | Baa1-Pd-A/Wsp-Lor | Bosque andino alto entremezclado con prados | *Weinmannia* spp.-*Lachemilla orbiculata*-zonas alteradas |
| 3 | Baa1-Pd/Wsp-Lor | | *Weinmannia* spp.-*Lachemilla orbiculata* |
| 2 | Baa1-Pd/Hfe-Lor | | *Hesperomeles ferruginea-Lachemilla orbiculata* |
| 1 | Baa1/Hfe | Bosque andino alto | *Hesperomeles ferruginea* |

**Tabla 10.** Tipos de vegetación presentes en los páramos de Guanacas y Las Delicias, sectores aledaños al volcán Puracé. El orden corresponde al identificador utilizado a lo largo del presente trabajo.

| PÁRAMO DE GUANACAS SECTORES ALEDAÑOS AL VOCÁN PURACÉ | | | |
|---|---|---|---|
| SIM_ID | SÍMBOLO | TIPO DE VEGETACIÓN | ESPECIES DOMINANTES |
| 38 | A | Zonas agropecuarias | |
| 37 | H | Cuerpos de agua | |
| 36 | T/ Ope-Cyp | Turberas del complejo | *Oritrophium peruvianum* y *Distichia muscoides* |
| 35 | Ch-Rf-P/ Cht-Eha-Cef | Chuscales con matorrales frailejonales y pajonales | *Chusquea tessellata, Espeletia hartwegiana* y *Calamagrostis effusa* |
| 34 | Ch-Rf-Mb/ Cht-Eha-Dci | Chuscales con rosetales frailejonales y matorrales bajos | *Chusquea tessellata, Espeletia hartwegiana* y *Diplostephium* cf. *cinerascens* |
| 33 | Ch-Rf-Baa1/ Cht-Eha | Chuscales con Rosetales frailejonales entremezclados con bosque andino alto | |
| 32 | Ch-Rf-A/ Cht-Eha | Chuscales con rosetales frailejonales y áreas agropecuarias | *Chusquea tessellata* y *Espeletia hartwegiana* |
| 31 | Ch-Rf/ Cht-Eha | Chuscales con rosetales frailejonales | |
| 30 | Ch/ Cht-Dem | Chuscales | *Chusquea tessellata* y *Disterigma empetrifolium* |
| 29 | Rf-P-Mb/ Eha-Cef-Dci-Blo | Rosetales frailejonales con pajonales y matorrales bajos | *Espeletia hartwegiana, Calamagrostis effusa, Diplostephium* cf. *cinerascens* y *Blechnum loxense.* |
| 28 | Rf-P-Ma/ Eha-Cef-Bca | Rosetales frailejonales con pajonales y matorrales altos | *Espeletia hartwegiana, Calamagrostis effusa* y *Brunellia cayambensis* |
| 27 | Rf-P-Baa1/ Eha-Cef-Bca-Wpu | Rosetales frailejonales con pajonales entremezclados con bosque andino alto | *Espeletia hartwegiana, Calamagrostis effusa, Brunellia cayambensis* y *Weinmannia pubescens* |
| 26 | Rf-P/ Eha-Cef | Rosetales frailejonales con pajonales | *Espeletia hartwegiana* y *Calamagrostis effusa* |
| 25 | Rf-Ch/ Eha-Cht | Rosetales frailejonales con chuscales | *Espeletia hartwegiana* y *Chusquea tessellata* |
| 24 | P/Cef | Pajonales | *Calamagrostis effusa* |
| 23 | Mb-Ma/ Dci-Cht-Wbr-Msa | Matorrales bajos entremezclados con matorrales altos | *Diplostephium* cf. *cinerascens, Chusquea tessellatade, Weinmanna brachystachya* y *Miconia salicifolia* |
| 22 | Mb/ Hla | Matorrales bajos | *Hypericum laricifolium* |
| 21 | Mb/ Dci-Cht | | *Diplostephium* cf. *cinerascens* y *Chusquea tessellata* |
| 20 | Ma-Ps/ Cru-Dry-Emy-Asp | Matorrales densos (relictos de bosque) entremezclados con pastizales | *Clethra rugosa, Drimys granadensis, Escallonia myrtilloides* y *Agrostis* sp. 1 |
| 19 | Ma-Ch/ Bca-Wpu-Cht | Matorrales altos entremezclados con chuscales | *Brunellia cayambensis, Weinmannia pubescens* y *Chusquea tessellata* |
| 18 | Ma-Ch/ Ati-Cht | | *Ageratina tinifolia* y *Chusquea tessellata* |
| 17 | Ma-A/ Wbr-Msa | Matorrales densos (relictos de bosque) entremezclados con áreas agropecuarias | *Weinmannia brachystachya* y *Miconia salicifolia* |
| 16 | Ma-A/ Cru-Dry-Emy | | *Clethra rugosa, Drimys granadensis* y *Escallonia myrtilloides* |
| 15 | Ma-A/ Bca | | *Brunellia cayambensis* y *Weinmannia pubescens* |
| 14 | Ma/ Wbr-Msa | Matorrales densos (relictos de bosque) | *Weinmannia brachystachya* y *Miconia salicifolia* |
| 13 | Ma/ Cru-Dry-Emy | Matorrales densos (relictos de bosque) | *Clethra rugosa, Drimys granadensis* y *Escallonia myrtilloides* |
| 12 | Ma/Ast- Bca-Wpu | Matorrales densos (relictos de bosque) | Asteraceas con *Brunellia cayambensis* y *Weinmannia pubescens* |
| 11 | Ma/ Ati | Matorrales altos | *Ageratina tinifolia* |
| 10 | Baa1-Mb/ Bca-Wbr-Dci-Blo | Bosque andino alto entremezclado con matorrales bajos | *Brunellia cayambensis, Weinmannia pubescens, Diplostephium* cf. *cinerascens* y *Blechnum loxense.* |
| 9 | Baa1-Ch/ Cru-Dry-Emy-Cht | Bosque andino alto entremezclado con áreas agropecuarias | *Clethra rugosa, Drimys granadensis* y *Escallonia myrtilloides* |
| 8 | Baa1-Ch/ Bca-Wpu-Cht | Bosque andino alto entremezclado con chuscales | *Brunellia cayambensis, Weinmannia pubescens* y *Chusquea tessellata* |
| 7 | Baa2-A/ Wbr-Msa | Bosque andino alto entremezclado con áreas agropecuarias | *Weinmanna brachystachya* y *Miconia salicifolia* |
| 6 | Baa2-A/ Cru-Dry-Emy | | *Clethra rugosa, Drimys granadensis* y *Escallonia myrtilloides* |
| 5 | Baa2-A/ Bca-Wpu | Bosque andino alto | *Brunellia cayambensis* y *Weinmannia pubescens* |
| 4 | Baa1/ Wbr-Msa | | *Weinmannia brachystachya* y *Miconia salicifolia* |
| 3 | Baa2/ Cru-Dry-Emy | Bosque andino alto intervenido | *Clethra rugosa, Drimys granadensis* y *Escallonia myrtilloides* |
| 2 | Baa1/ Cru-Dry-Emy | Bosque andino alto | |
| 1 | Baa1/ Bca-Wpu | Bosque andino alto entremezclado con áreas agropecuarias | *Brunellia cayambensis* y *Weinmannia pubescens* |

## HETEROGENEIDAD ESPACIAL

### Páramos de la serranía de Perijá

Esta región de estudio presenta 64 tipos de cobertura distribuidos en cerca de 57 kilómetros en el eje **Y** (latitud) y 21 kilómetros en el eje **X** (longitud). En la Figura 4 se muestra la distribución de los patrones en una cuadrícula formada por el norte, centro y sur y por el occidente, centro longitudinal y oriente del área estudiada. Al interior de cada cuadrante se presenta una figura de tendencia formada por los valores de área en el eje **X** y el identificador de los tipos de vegetación y cobertura **SIM-VEG** en el eje Y. Para la correcta interpretación de la gráfica es necesario conocer la distribución de los tipos de cobertura del eje **Y** (Tabla 11). Tipos similares de cobertura se agruparon con el fin de reducir la complejidad en el análisis. Se observa que en el intervalo 0-20, se agrupan bosques conservados e intervenidos, el intervalo 20-40 engloba la transición de bosques intervenidos a matorrales altos y bajos (estos últimos pertenecientes a páramo); el intervalo 40-60 a los tipos de vegetación paramuna como pajonales, rosetales frailejonales, herbazales y chuscales y en el intervalo 60-80 dos tipos de herbazales junto a los tipos de cobertura sin información y sin vegetación. La expresión de tendencia puede tomar valores positivos cuando existe una concentración de áreas hacia los valores elevados del identificador de cobertura, valores cercanos a cero cuando la concentración de las áreas se distribuye de manera homogénea en un determinado intervalo o a lo largo de todos los cuadrantes y negativa cuando la tendencia es a concentrar valores de área hacia los intervalos inferiores o de tipos de vegetación boscosa. Cabe resaltar la importancia en la distribución de los puntos en la gráfica, ya que indican el punto de partida de la línea de tendencia, la cual puede estar ubicada dentro en cualquier intervalo.

**Tabla 11.** Eje de variación SIM_ID.

| EJE Y | |
|---|---|
| SIM_ID | NÚMERO |
| SV | 64 |
| SI | 63 |
| Hz-P/Och-Cef | 62 |
| Hz-P/Lst-Cef-Och | 61 |
| Hz/Och-Vve | 60 |
| Ch/Sal-Cte | 59 |
| Ch/Ani-Cte | 58 |
| Rfb-P-Mb-Baa1/Epe-Lst-Wpi | 57 |
| Rfb-P/Epe-Cef | 56 |
| Rfb-Mb-P/Epe-Lst-Cef-Cin | 55 |
| Rfb-Ma/Epe-Lst-Wpi | 54 |
| Rfb/Epe-Ale | 53 |
| Rfa-P/Epe-Ldi-Cef | 52 |
| Rfa-Hz/Ldi-Epe-Och | 51 |
| Rfa/Ldi | 50 |
| P-Mb-Ma/Cef-Hph-Wpi | 49 |
| P-Mb-Baa1/Cef-Lst-Wpi | 48 |
| P-Mb/Cef-Hba | 47 |
| P-Ma/Cin-Hfe | 46 |
| P-Ch/Cin-Cte | 45 |
| Mb-P-Ma/Hba-Cef-Gpu | 44 |
| Mb-P/Lst-Cin-Cef | 43 |
| Mb-P/Hba-Cef | 42 |
| Mb-P/Gbu-Hph-Cef | 41 |
| Mb-P/Ani-Gbu-Cef | 40 |
| Mb/Lst-Hph | 39 |
| Mb/Lst-Bae | 38 |
| Mb/Hju-Pph | 37 |
| Mb/Ani-Hba | 36 |
| Mb/Ani-Gbu-Cef | 35 |
| Mb-Baa1/Hla-Pol | 34 |
| Mb2/Paq-Och-Csp-Ati | 33 |
| Ma-Hz-P/Gpu-Och-Cef | 32 |
| Ma2/Tme-Cmu | 31 |
| Ma/Wpi-Hba-Bae | 30 |
| Ma/Gpu-Bgl | 29 |
| Baa2-Mb/Tme-Cmu-Hla | 28 |
| Baa2-Mb/Pol-Hla | 27 |
| Baa2-Mb/Dga-Pol-Ani | 26 |
| Baa2-Ma/Wpi-Hba-Bae | 25 |
| Baa2-Ma/Tme-Cmu-Gpu | 24 |
| Baa2-Hz-P/Tme-Cmu-Och-Cef | 23 |
| Baa2-Hz/Hfe-Pmo-Och | 22 |
| Baa2/Wpi-Bco | 21 |
| Baa2/Tme-Cmu-Gpu | 20 |
| Baa2/Tme-Cmu | 19 |
| Baa2/Hfe-Pmo-Msp | 18 |
| Baa1-Mb/Wpi-Ani-Gbu | 17 |
| Baa1-Mb/Pol-Mli-Hfe | 16 |
| Baa1-Mb/Pol-Hla | 15 |
| Baa1-Mb/Hfe-Mli-Hla | 14 |
| Baa1-Ma-Mb/Pmo-Mli-Hfe-Hla | 13 |
| Baa1-Ma/Wpi-Hba-Bae | 12 |
| Baa1-Ma/Tme-Cmu-Gpu | 11 |
| Baa1-Ma/Pmo-Mli-Hfe | 10 |
| Baa1-Hz-P/Wpi-Cef-Och | 9 |
| Baa1-Hz-P/Tme-Cmu-Och-Cef | 8 |
| Baa1-Hz/Hfe-Pmo-Och | 7 |
| Baa1/Wpi-Rro | 6 |
| Baa1/Wpi-Bin | 5 |
| Baa1/Tme-Cmu | 4 |
| Baa1/Pmo-Mli-Hfe | 3 |
| Baa1/Pin-Adi | 2 |
| Baa1/Hfe-Mli | 1 |

En la figura 4 **(1) Noroccidente**- la región carece de coberturas por encima de los 2800 m. **(2) Norte propiamente dicho**- existen coberturas en el intervalo 0-20 de bosques conservados e intervenidos; en el intervalo 40-60 de tipos de vegetación paramuna y en el intervalo 60-80 con la presencia de un herbazal. La pendiente negativa indica la concentración de áreas en los tipos de vegetación boscosa y la concentración de puntos se presenta en los tipos de bosques intervenidos. Las mayores áreas no superan 800 hectáreas. **(3) Nororiente**- existen coberturas en todos lo intervalos de identificador SIM-ID aunque el intervalo 60-80 solo presenta un tipo de herbazal (SIM-VEG 61). Las áreas más grandes las presentan los bosques conservados; sinembargo la región de chuscales muestra un registro que se encuentra alrededor de 800 hectáreas. Los bosques conservados de *Hesperomeles ferruginea* (SIM-VEG 1) presentan áreas por encima de cien hectáreas. **(4) Occidente** -existen registros en los tres primeros intervalos de cobertura aunque sus valores se encuentran concentrados en cinco tipos de cobertura en el intervalo 0-20 y dentro de estos hacia los bosques conservados; tres tipos de bosques conservados y uno intervenido presentan áreas superiores a 200 hectáreas que influye para que que la línea de tendencia tenga un valor negativo; en el intervalo 20-40 sólo se presenta una concentración de valores bajos de área hacia la región de los matorrales altos y apenas dos tipos de coberturas hacia la región de chuscales con un registro alrededor de 400 hectáreas. **(5) Centro**-esta región presenta una de las mayores densidades de coberturas con cuatro concentraciones de puntos distribuidas hacia el centro del intervalo 0-20 (bosques conservados entremezclados con herbazales y matorrales altos), en la transición del punto 20 (bosques intervenidos), en la transición del punto 40 (matorrales bajos entremezclados con pajonales) y en el centro del intervalo 40-60 (rosetales frailejonales arboresentes de

*Libanothamnus divisioriensis* y frailejonales de *Espeletia perijaensis*). Los dos herbazales ubicados en el intervalo 60-80 también se encuentran presentes. La línea de tendencia es más pronunciada y parte desde tipos de matorral bajo pertenecientes a vegetación paramuna, e indica una concentración de áreas en este tipo de vegetación. Los bosques conservados entremezclados con matorrales y herbazales presentan los mayores registros aproximadamente 600 hectáreas. **(6) Oriente**- hacia el sector oriental de la serranía de Perijá se presenta una dominancia de la vegetación paramuna con una concentración de más de 4700 hectáreas perteneciente a rosetales frailejonales entremezclados con matorrales bajos y pajonales. Otras combinaciones de rosetal frailejonal también son importantes sin embargo se encuentran distribuidas por debajo de 300 hectáreas de superficie. Este sector de la serranía también presenta un parche de bosque conservado de menos de 50 hectáreas de cobertura. **(7) Sur occidente**-presenta un tipo de cobertura dentro de los matorrales altos con cobertura inferior a 50 hectáreas. **(8) Sur**- esta región de la serranía presenta la mayor pendiente positiva, indica una concentración elevada de formaciones paramunas. Hacia el punto 60 (formaciónes de rosetales frailejonales y pastizales entremezclados con matorrales bajos y en algunos casos bosques no intervenidos) se registra una de las áreas más grandes, aproximadamente 500 hectáreas. Un registro similar al anterior se ubica hacia los matorrales bajos y a pesar de que estos sumados a los matorrales altos contienen individualmente áreas de al menos 300 hectáreas, la alta concentración de pequeñas zonas con frailejonales no los hace dominantes. **(9) Sur Oriente**- presenta una pendiente negativa y muy pocos registros de vegetación. Los mayores valores se registran alrededor de 300 hectáreas y hacen parte de los bosques identificados como conservados entremezclados con herbazales y de algunos chuscales.

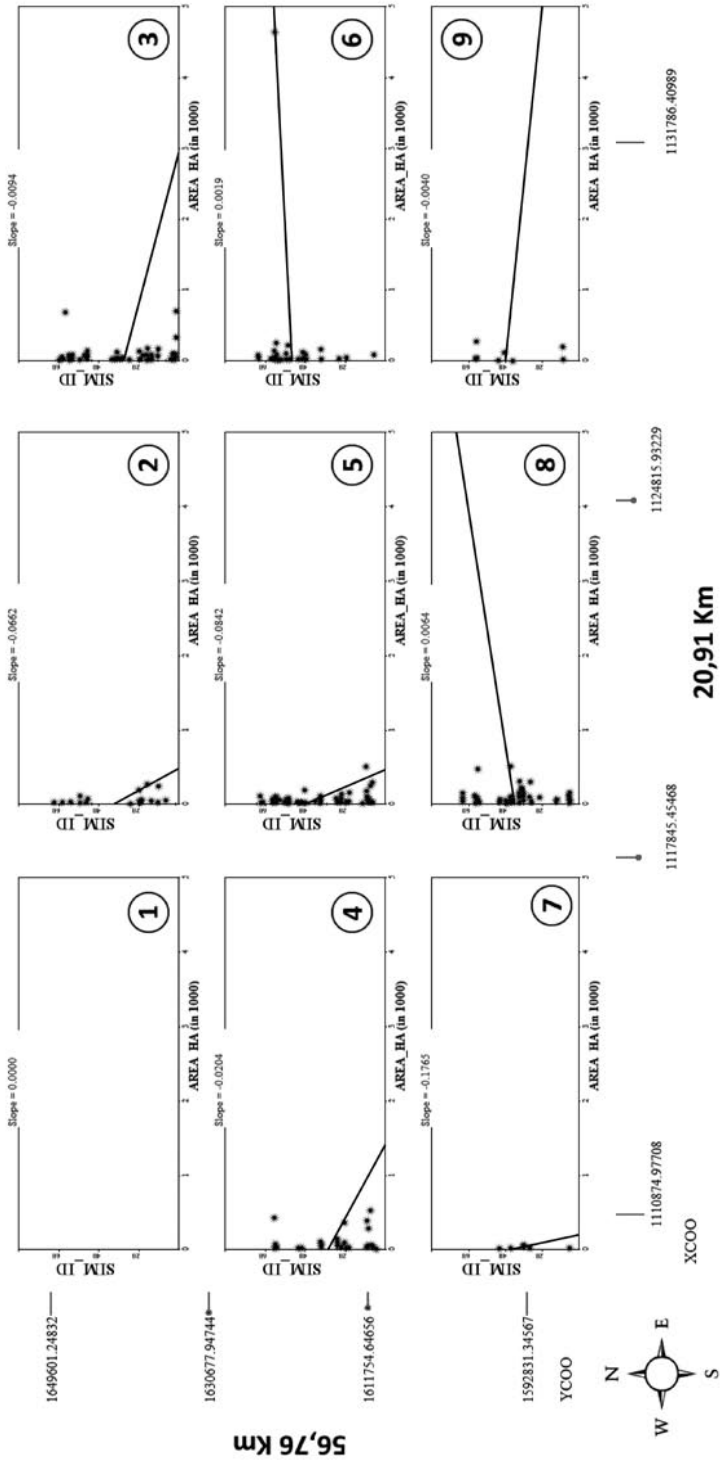

**Figura 4.** Distribución de la heterogeneidad en tipos de cobertura y área presentes en la alta montaña de la serranía de Perijá.

### Páramos: La Rusia, Belén y Guantiva

Los páramos de esta zona de estudio forman un continuo a lo largo de 55,9 kilómetros en el eje **Y** (latitud) y con una variación longitudinal 33,17 kilómetros (**X**). En esta región se diferenciaron 30 tipos generales de cobertura, los cuales se muestran en la tabla 12. El eje de variación SIM-ID o tipos de vegetación y cobertura presenta tres intervalos. De 0 a 10 se observan los bosques conservados sin entremezclas, los bosques alterados o intervenidos y los bosques conservados entremezclados con otros tipos de vegetación o cobertura. A diferencia de los patrones presentados en la serranía de Perijá, la intervención antrópica hace parte de estas coberturas en forma de áreas agropecuarias (A) como sucede con el tipo SIM-ID seis (6). El intervalo 10-20 contiene la transición entre matorrales altos hasta rosetales fraile­jonales entremezclados con chuscales y el intervalo 20-30 presenta dos tipos de rosetales, chuscales de *Chusquea tessellata*, vegetación de tipo colchón y otros tipos de cobertura como las áreas agropecuarias y las zonas urbanas. La heterogeneidad de la zona se resume en la figura 5. **(1) Noroccidente**- la región carece de coberturas por encima de los 2800 m. **(2) Norte**- debido al sentido nororiental de la cordillera, el sector norte presenta apenas seis tipos de cobertura. Las áreas más extensas son de bosques intervenidos de *Quercus humboldtii* y alcanzan alrededor de mil hectáreas. Otros tipos de bosque se encuentran en la región distribuidos en al menos 500 hectáreas. **(3) Nororiente**- para este sector de la región existen coberturas en todos los intervalos de identificador SIM-ID. Se presentan alrededor de cuatro mil hectáreas de bosque conservado (SIM-VEG 3) y 5000 hectáreas de bosque intervenido entremezclado con áreas agropecuarias. Los tipos de vegetación restante presentes en la zona no superan las mil hectáreas. **(4) Occidente** –hacia el occidente del área de estudio se encuentran doce tipos de cobertura. Al igual que en los casos anteriores las coberturas más extensas pertenecen a formaciones de tipo boscoso. Los bosques intervenidos entremezclados con áreas agropecuarias se presentan en aproximadamente 2000 hectáreas, mientras los bosques conservados de *Weinmannia microphylla* entremezclados con chuscales de *Chusquea tessellata* se presentan en más de 3000 hectáreas. **(5) Centro**-esta región presenta casi todos los tipos de cobertura diferenciados en el estudio. La tendencia general muestra un predominio de la formación de bosques conservados dominados por *Weinmannia microphylla*, *Ilex kunthiana* y *Miconia* sp. (SIM-ID 3). Es importante mencionar que aunque la región central en general presenta características de conservación ya se evidencia la llegada de la intervención antrópica con dominio de áreas agropecuarias entremezcladas con el tipo de cobertura anterior en aproximadamente 1000 hectáreas y su inversa (áreas con bosques conservados entremezclados con pequeñas áreas de producción agropecuaria) en al menos 2000 hectáreas. Los matorrales altos de *Polylepis quadrijuga* se concentran en la región aunque los valores de cobertura de sus parches no superan las 500 hectáreas. Otra cobertura con gran expresión en la zona son los pajonales de *Calamagrostis effusa* entremezclados con distintos tipos de matorrales bajos en alrededor de 1800 hectáreas. **(6) Oriente**- hacia el sector oriental se concentran los pajonales de *Calamagrostis effusa* entremezclados con distintos tipos de matorrales bajos cuyos valores individuales de cobertura no superan 1500 hectáreas. Los rosetales frailejonales entremezclados con pajonales también están presentes con coberturas que no superan las mil hectáreas.

**Tabla 12.** Eje de variación SIM_ID.

| EJE Y | |
|---|---|
| SIM_ID | NÚMERO |
| SI | 30 |
| U | 29 |
| SV | 28 |
| A-Baa2/Wmi-Iku | 27 |
| A | 26 |
| H | 25 |
| Co/Ocl-Cco | 24 |
| Ch/Cht | 23 |
| Rf-P/Esp-Cef | 22 |
| Rf-Mb/Egu-Ani | 21 |
| Rf-Ch/Esp-Cht | 20 |
| Rf/Eph | 19 |
| Rf/Esp | 18 |
| Rf/Blo-Emu | 17 |
| Rf/Alc-Pgl | 16 |
| P-Mb/Cef-Hla | 15 |
| P-Mb/Cef | 14 |
| P/Cin-Cbo | 13 |
| Mb/Hla | 12 |
| Ma/Pqj | 11 |
| Ma/Msa | 10 |
| Baa1-Mb/Wmi-Hla | 9 |
| Baa1-Ch/Wmi-Ces | 8 |
| Baa1-Ch/Pqj-Cht | 7 |
| Baa2-A/Wmi-Iku | 6 |
| Baa2/Wmi-Iku | 5 |
| Baa2/Qhu | 4 |
| Baa1/Wmi-Iku | 3 |
| Baa1/Qhu | 2 |
| Baa1/Pqj | 1 |

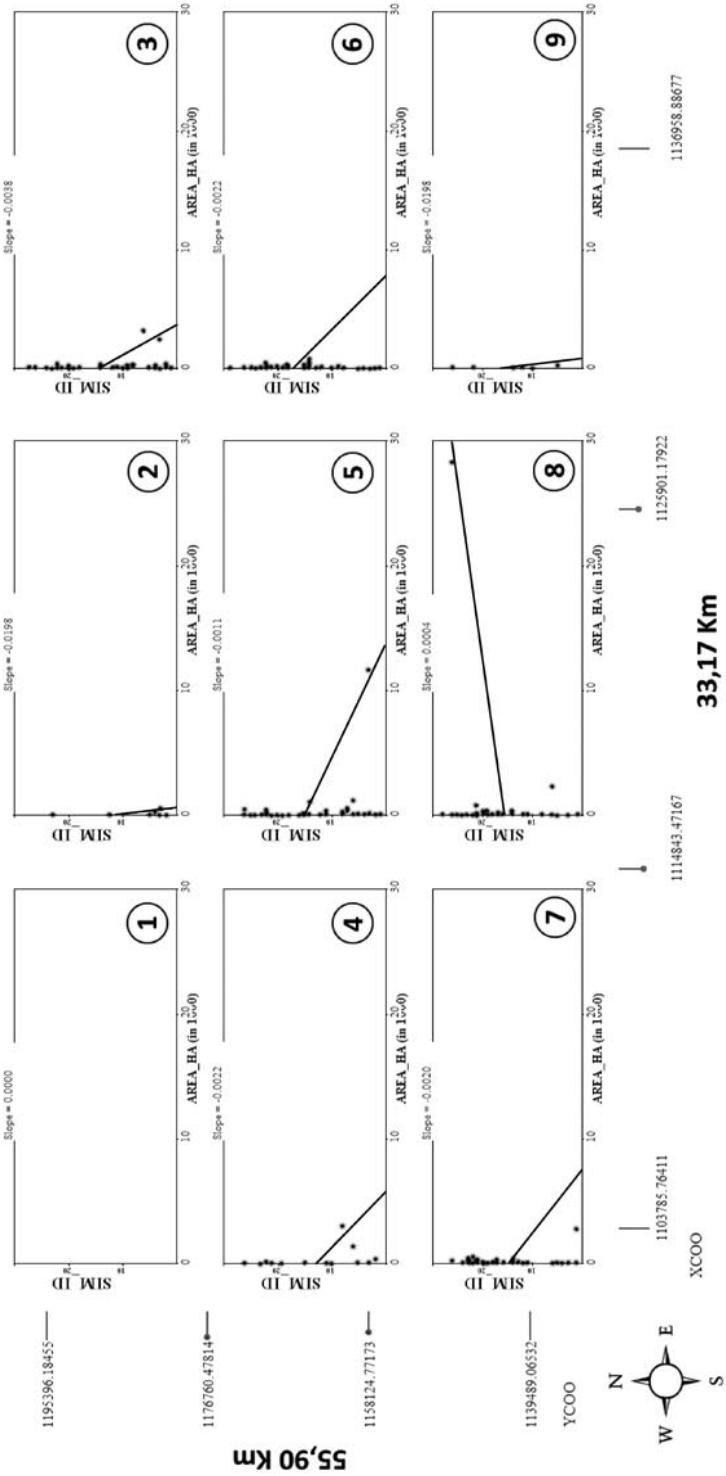

**Figura 5.** Distribución de la heterogeneidad en tipos de cobertura y área presentes en los páramos de La Rusia, Belén y Guantiva.

**(7) Sur occidente**-en este sector se presenta una disminución drástica de los tipos de vegetación boscosa a pesar de que se registra un área cercana a 3500 hectáreas de bosques conservados dominados por *Polylepis quadrijuga*. Los rosetales frailejonales dominados por *Espeletiopsis guacharaca*, *Arcytophyllum nitidum* y *Bejaria resinosa* entremezclados con matorrales bajos de *Arcytophyllum nitidum* se concentran con áreas individuales menores a 1000 hectáreas. **(8) Sur**- se presenta una gran extensión de vegetación boscosa intervenida entremezclada con áreas agropecuarias (alrededor de 4000 hectáreas), sinembargo este valor se opaca por la extensa región agropecuaria **(A)** que registra alrededor de 27000 hectáreas, la cual representa una gran porción de la intervención total registrada de 30600 hectáreas para toda el área de estudio. En esta región se presenta una concentración de vegetación de páramo de tipo rosetal frailejonal con parches que no superan 1500 hectáreas. **(9) Sur-Oriente**- presenta una pendiente negativa y muy pocos registros de vegetación. Los mayores valores se registran alrededor de 500 hectáreas y pertenecen a bosques intervenidos entremezclados con áreas agropecuarias.

**Páramo de Telecom**

Se evaluó la heterogeneidad de coberturas a lo largo de 9,58 kilómetros en el eje **Y** (latitud) y 8,48 kilómetros en el eje **X** (longitud). Debido a su ubicación y tamaño presenta una alta intervención la cual se evidencia en el elevado número de tipos de coberturas y en el grado de fragmentación de las mismas. El eje SIM-ID contiene 44 tipos de vegetación y están distribuidos en la figura seis (6) de la siguiente manera: el intervalo 0-10 representa bosques semi conservados y bosques alterados o intervenidos. Es importante destacar que en todos los bosques intervenidos hay mezclas con áreas de producción agropecuaria así como plantaciones forestales. El intervalo 10-20 presenta una formación de bosque andino bajo, la transición entre matorrales altos hasta matorrales bajos pertenecientes al páramo. El intervalo 20-30 contiene tipos de vegetación paramuna entre los que se encuentran pajonales y algunos rosetales frailejonales entremezclados con bosques alto andinos intervenidos. El intervalo 30-40 representa dos rosetales frailejonales y los tipos de vegetación casmófita y ribereña de la zona además de las coberturas con plantaciones forestales y el intervalo 40-50 todos los tipos de origen antrópico (tabla 13). La heterogeneidad de la zona se resume en la figura 6. **(1) Noroccidente**-predomina la cobertura agropecuaria en más de 150 hectáreas. Otros tipos de vegetación presente son los bosques intervenidos que se concentran en varios parches con áreas inferiores a 10 hectáreas y vegetación casmófita en pequeñas proporciones. **(2) Norte**- cinco coberturas presentan tamaños considerables con respecto a los otros tipos de vegetación. Las áreas agropecuarias dominan con tres fragmentos de los cuales uno presenta más de 250 hectáreas. Entre las coberturas naturales se destacan los rosetales frailejonales entremezclados con bosques intervenidos de *Weinmannia tomentosa* distribuidos en parches de al menos30 hectáreas. **(3) Nororiente**- esta región del área de estudio no presenta los tipos de cobertura del intervalo 20-30. La línea de tendencia indica que las mayores extensiones de la zona corresponden a plantaciones forestales y áreas agropecuarias en fragmentos que llegan cerca a 100 hectáreas. Los matorrales altos también encuentran significancia así como una zona semi conservada que alcanza más de 70 hectáreas. **(4) Occidente** –este sector del páramo de Telecom carece de los tipos de vegetación del intervalo 30-40 y la mayor extensión la cubren bosques muy intervenidos entremezclados con coberturas agropecuarias en alrededor de 120 hectáreas. **(5) Centro**- presenta una alta concentración de las coberturas de origen antrópico, aunque también se encuentran las mayores áreas cubiertas por frailejón en pequeños fragmentos. Al menos un fragmento alcanza 70 hectáreas. La zona también presenta la mayor diversificación de matorrales bajos con tamaños que no superan 20 hectáreas. Se evidencia un parche de bosques semi conservado en 50 hectáreas y bosques muy intervenidos entremezclados con áreas agropecuarias en cerca de 60 hectáreas.

**(6) Oriente**- presenta la mayor cobertura agropecuaria en alrededor de 570 hectáreas. Los tipos de vegetación de los intervalos 0-10 y 20-30 no se encuentran representados en la zona aunque se evidencian matorrales bajos dominados por *Bejaria resinosa* y *Clethra fimbriata* en fragmentos menores a 20 hectáreas. **(7) Sur occidente**-en este sector se expresan todos los tipos de cobertura generales encontrados. Los bosques semi conservados alcanzan una buena extensión en la zona (210 hectáreas aproximadamente) así como los matorrales altos de *Brunellia colombiana* (120 hectáreas) y los matorrales bajos de *Bejaria resinosa* entremezclados con bosques intervenidos dominados por *Weinmannia tomentosa* (alrededor de 100 hectáreas). **(8) Sur**-en esta región ningún patrón de vegetación o cobertura sobrepasa 50 hectáreas. La pendiente de la línea de tendencia indica una concentración de coberturas hacia los matorrales altos y los bosques semi conservados e intervenidos **(9) Sur Oriente**- no se encuentra representado ningún tipo de vegetación.

**Tabla 13.** Eje de variación SIM_ID.

| EJE Y | |
|---|---|
| SIM_ID | NÚMERO |
| SV | 44 |
| A-Mb/Chl | 43 |
| A-Mb/Ani | 42 |
| A | 41 |
| Pl-Ma/Ppa-Clm | 40 |
| Pl/Ppa | 39 |
| Vr-Py/Pyn | 38 |
| Vr-Mb/Mve | 37 |
| Vr-Mb/Hla | 36 |
| Vr-Mb/Ati | 35 |
| Vc-P-Rf/Cef | 34 |
| Vc-P/Cef | 33 |
| Rf/Esp | 32 |
| Rf/Eph | 31 |
| P-Rf/Cef | 30 |
| Rf-Baa2/Wto | 29 |
| Rf-Baa2/Eph-Bco | 28 |
| Rf-Baa2/Egr-Wto | 27 |
| Rf-Baa2/Bco | 26 |
| Rf-Baa2/Cef-Wto | 25 |
| P-Rf/Cef-Esp | 24 |
| P/Cef | 23 |
| Mb-Baa2/Bj-Wto | 22 |
| Mb/Mve | 21 |
| Mb/Emy | 20 |
| Mb/Ani | 19 |
| Mb-Rf/Eph-Ani | 18 |
| Mb-P-Rf/Cef-Ani | 17 |
| Mb/Chl | 16 |
| Mb/Bj | 15 |
| Ma/Wto | 14 |
| Ma/Chl | 13 |
| Ma/Bco | 12 |
| Bab1/Xsp-Dca | 11 |
| Baa2-Pl/Wto | 10 |
| Baa2-Mb/Wto-Ani | 9 |
| Baa2-Mb/Chl | 8 |
| Baa2-A/Wto | 7 |
| Baa2-A/Bco | 6 |
| Baa2/Chl | 5 |
| Baa1-Mb/Chl | 4 |
| Baa1/Msa | 3 |
| Baa1/Chl | 2 |
| Baa1/Bco | 1 |

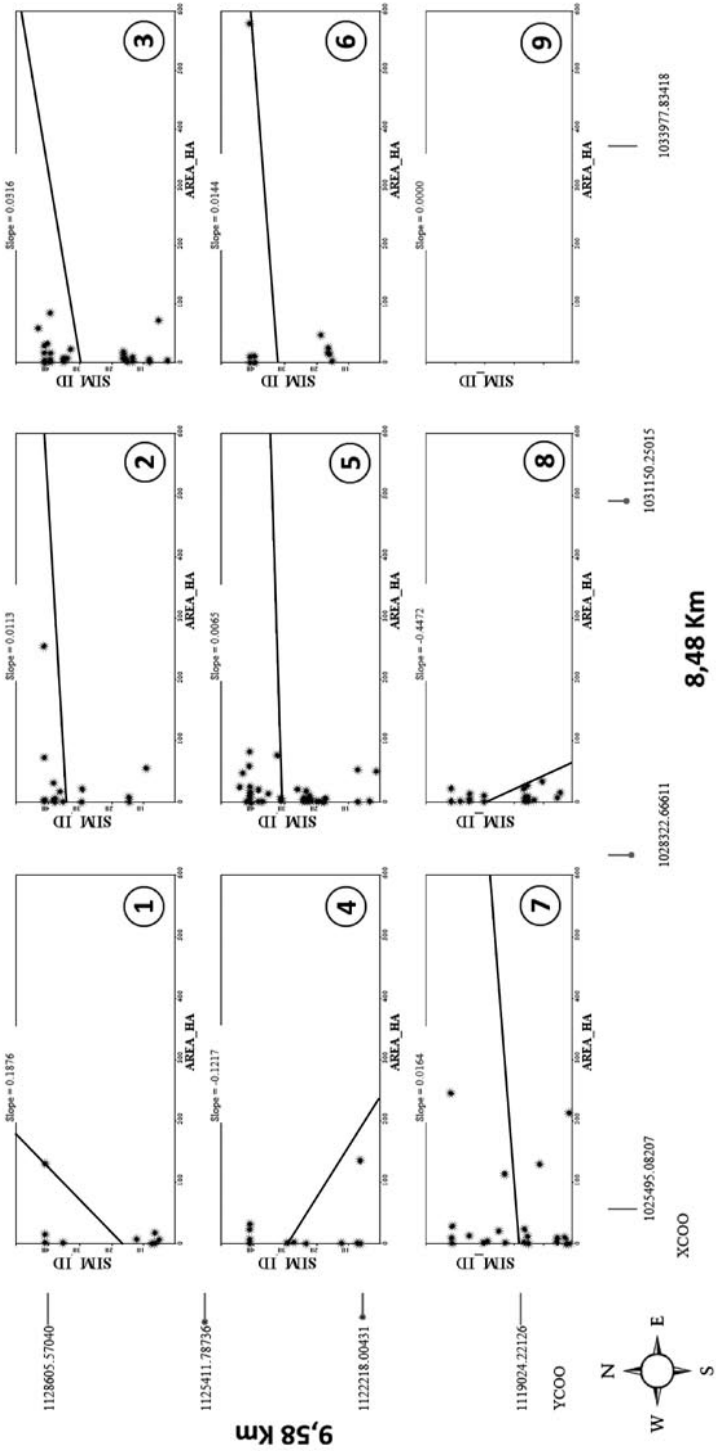

**Figura 6.** Distribución de la heterogeneidad en tipos de cobertura y área presentes en el páramo de Telecom.

## Páramo de Merchán

Se evaluó la heterogeneidad de coberturas a lo largo de 12,98 kilómetros en el eje **Y** (latitud) y 5,59 kilómetros en el eje **X** (longitud). Este páramo está altamente intervenido y presenta el menor número de tipos de vegetación y cobertura. El eje SIM-ID contiene 22 tipos de vegetación, distribuidos de la siguiente manera (tabla 14): el intervalo 0-10 representa dos tipos de bosques intervenidos y matorrales bajos en distintos arreglos florísticos. El intervalo 10-20 representa la escasa vegetación paramuna desde el tipo pajonal hasta los rosetales frailejonales de *Espeletia phaneractis*. La vegetación riparia se encuentra representada en cuatro tipos mientras que existen dos coberturas de origen antrópico. El intervalo 20-30 sólo contiene dos tipos de cobertura. La heterogeneidad de la zona se resume en la figura 7. **(1) Noroccidente-** en este sector predomina la cobertura agropecuaria en fragmentos inferiores a 100 hectáreas. Entre otros tipos de vegetación y con muy bajos valores de área estan los bosques intervenidos de *Weinmannia tomentosa* y los matorrales bajos de *Clethra fimbriata*. **(2) Norte-** en el norte se presentan once tipos de vegetación que van desde bosques intervenidos hasta rosetales frailejonales de *Espeletia phaneractis* que representan la mayor concentración de coberturas en el área de estudio; cabe mencionar la presencia de coberturas agropecuarias distribuidas en pequeñas extensiones. El área más grande se encuentra dominada por matorrales bajos entremezclados con vegetación a orillas de quebradas con *Ageratina tinifolia*. **(3) Nororiente-** esta región del área de estudio solo presenta los tipos de vegetación del intervalo 0-10 y la mayor expresión la adquieren los matorrales bajos de *Clethra fimbriata* (en fragmentos inferiores a 10 hectáreas). **(4) Occidente** –solamente presenta cuatro tipos de cobertura. Las áreas agropecuarias son dominantes aunque sus valores se encuentran por debajo de 100 hectáreas. **(5) Centro-** en el centro se concentra la mayor área agropecuaria presente en el área de estudio; esta ocupa alrededor de 2700 hectáreas. En los otros sectores la dominancia de la cobertura agropecuaria se mantiene lo cual hace que este páramo en relación a su tamaño sea el más intervenido.

**Tabla 14.** Eje de variación SIM_ID.

| EJE Y | |
|---|---|
| **SIM_ID** | **NÚMERO** |
| Pl-Baa2/Ppa-Clm | 22 |
| A-Rf/Eph | 21 |
| A-Ma/Chl | 20 |
| A | 19 |
| Vr-Pl/Ati-Ppa | 18 |
| Vr-Ma-Mb/Vst | 17 |
| Vr-Ma/Emy | 16 |
| Vr-Ma/Ati | 15 |
| Rf/Eph | 14 |
| Rf/Eph-Chl | 13 |
| Rf-Baa2/Chl | 12 |
| P/Cef | 11 |
| Mb-Vr/Ati | 10 |
| Mb-Baa2/Chl | 9 |
| Mb-Rf/Chl-Egr | 8 |
| Mb/Hla | 7 |
| Mb/Ani-Chl | 6 |
| Mb/Acl | 5 |
| Mb/Bj | 4 |
| Mb/Chl | 3 |
| Baa2/Wto | 2 |
| Baa2/Chl | 1 |

## Páramo Santuario y aledaños al complejo Carmen de Carupa

La región de páramo perteneciente al municipio de Carmen de Carupa se evaluó a lo largo de 35,66 kilómetros en el eje **Y** (latitud) y 18,27 kilómetros en el eje **X** (longitud). A pesar del fuerte impacto atribuido a la acción humana este páramo contiene fragmentos de vegetación natural importantes. El eje SIM-ID contiene 44 tipos de cobertura y están distribuidos en la figura ocho (8) de la siguiente manera: el intervalo 0-10 representa bosques semi conservados y bosques alterados o intervenidos. El intervalo 10-20 presenta bosques intervenidos y entremezclados con plantaciones forestales y dos formaciones de bosque andino bajo. La transición desde matorrales altos hasta matorrales bajos también se encuentran en este intervalo. El intervalo 20-30 contiene los tipos de vegetación más abundante en el páramo (matorrales bajos y pajonales), aunque cabe resaltar la inclusión de plantaciones forestales en este intervalo debido a su cercanía con los tipos de vegetación naturales. Se registran plantaciones de *Pinus patula* en áreas que deberían ser de páramo. El intervalo 30-40 presenta dos tipos de pajonales, rosetales frailejonales, vegetación casmófita y entremezclas donde la vegetación de páramo es la matriz en la cual se establecen algunos relictos de bosque andino alto. El intervalo 40-50 además de la vegetación riparia presenta dos tipos de cobertura de origen antrópico más algunos registros de sistemas hídricos (Tabla 15). La heterogeneidad de la zona se resume en la figura 8. **(1) Noroccidente**- no existe representación de ningún tipo de cobertura para el área de estudio. **(2) Norte**- todos los intervalos de cobertura se encuentran representados. La línea de tendencia indica un concentración de parches con cobertura agropecuaria cuya máxima área alcanza alrededor de 500 hectáreas. Otras coberturas importantes son los bosques semi conservados de *Weimannia tomentosa* y algunos tipos de matorral alto con áreas inferiores a 200 hectáreas. **(3) Nororiente**- presenta una tendencia similar a la anterior región en cuanto a las áreas agropecuarias. El registro máximo de superficie de esta cobertura alcanza 400 hectáreas. Para la zona se evidencia una ata concentración de formaciones con matorrales bajos en donde se destacan los matorrales bajos de *Aragoa cleefii* extendidos en superficies inferiores a 200 hectáreas. También cabe resaltar la presencia de rosetales frailejonales de *Espeletia grandiflora* entremezclada con relictos de bosque semi conservados de *Weinmannia tomentosa* en al menos 200 hectáreas de extensión. **(4) Occidente** –en el occidente de estos páramos a pesar de registrar una cobertura agropecuaria de al menos 800 hectáreas, la concentración de áreas se presenta al rededor del punto 30, lo cual indica un dominio en la región de pajonales con diferentes ensambles florísticos. Los pajonales de *Calamagrostis effusa* entremezclados con matorrales bajos de *Arcytophyllum nitidum* se presentan en áreas que no alcanzan las trescientas hectáreas. Las formaciones de bosque andino bajo dominadas por *Xylosma spiculiferum* y *Daphnopsis caracasana* se presentan en zonas de menos de 400 hectáreas. **(5) Centro**-el centro de la región carece de expresión de los tipos de cobertura del intervalo 30-40. La gran concentración de áreas agropecuarias es equilibrada según la línea de tendencia con algunos sectores representativos con matorrales altos de *Weinmannia tomentosa* (en áreas inferiores a 300 hectáreas) y bosques semi conservados entremezclados con rosetales frailejonales de *Espeletia phaneractis* (la superficie mayor está alrededor de 400 hectáreas). **(6) Oriente**-en este sector, las áreas agropecuarias alcanzan su mayor expresión (aproximadamente 1800 hectáreas) y sustituyen casi la totalidad de tipos de vegetación que deberían cubrir esta zona. Sólo se registra una formación de matorrales bajos con coberturas inferiores a 70 hectáreas. Los bosques semi conservados entremezclados con rosetales frailejonales de *Espeletia phaneractis* típicos de zonas que fueron alteradas se presentan significativamente en grupos con superficies inferiores a 350 hectáreas. **(7) Sur occidente**-presenta la mejor situación de la región ya que existe representación de la mayoría de coberturas naturales encontradas. Los matorrales altos alcanzan al menos superficies de 1800 hectáreas y los matorrales bajos de *Arcytophyllum nitidum* entremezclados con rosetales frailejonales de *Espeletia phaneractis* aproximadamente 700 hectáreas.**(8) Sur**-en esta región las plantaciones forestales alcanzan su mayor expresión (alrededor de 700 hectáreas) y las áreas agropecuarias presentan superficies que alcanzan 1600 hectáreas. Los bosques andinos altos intervenidos dominados por *Weinmannia tomentosa* se expresan en varios fragmentos inferiores a 400 hectáreas. Los matorrales bajos de *Arcytophyllum nitidum* se concentran en la región en sectores de al menos 200 hectáreas. **(9) Sur Oriente**- esta región solo se presentan áreas agropecuarias y su extensión es cercana a 400 hectáreas.

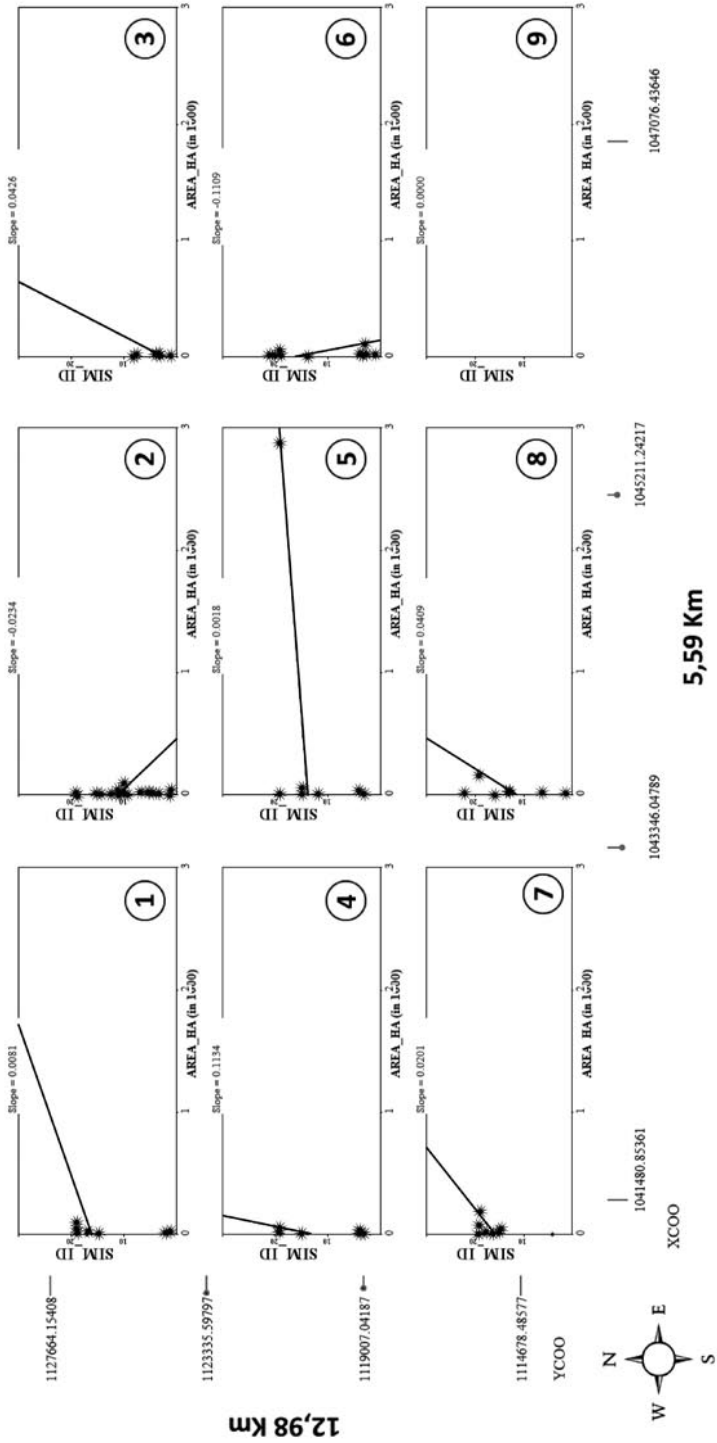

**Figura 7.** Distribución de la heterogeneidad en tipos de cobertura y área presentes en el páramo de Merchán.

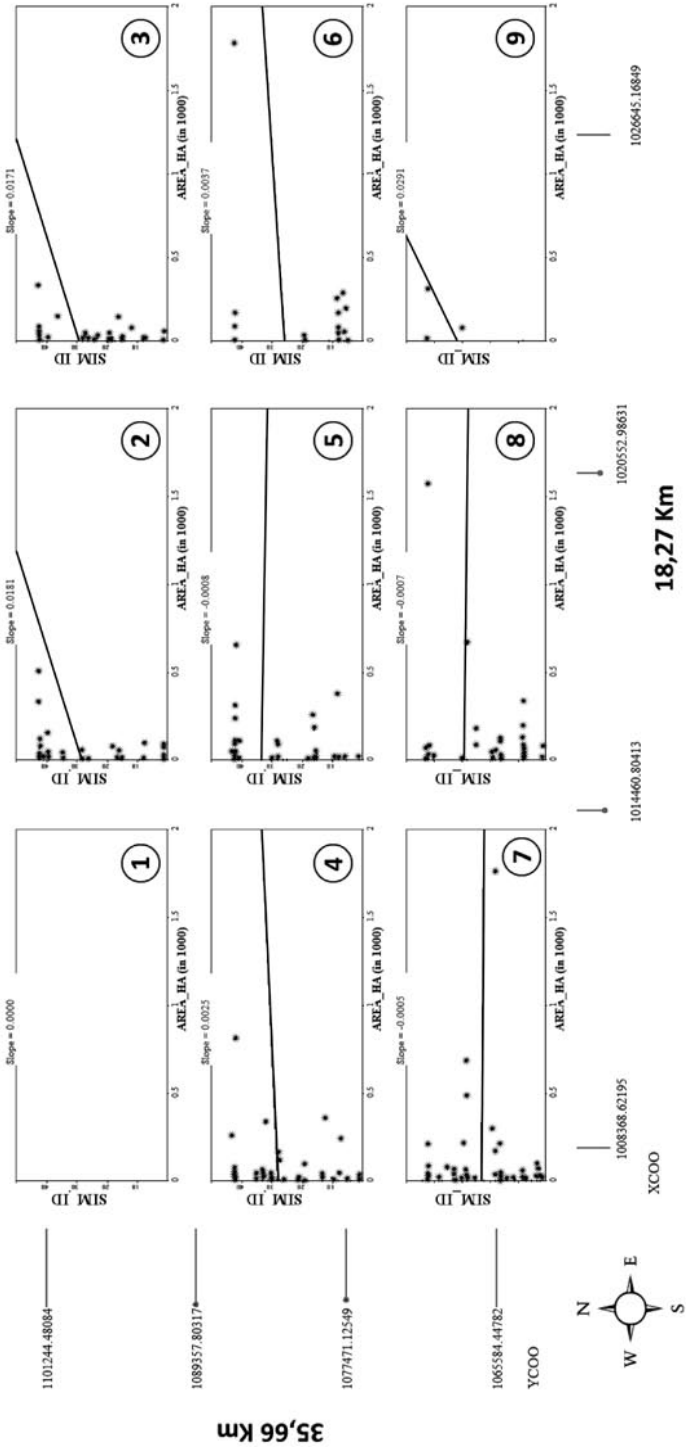

**Figura 8.** Distribución de la heterogeneidad en tipos de cobertura y área presentes en el páramo de Carmen de Carupa.

**Tabla 15.** Eje de variación SIM_ID.

| EJE Y | |
|---|---|
| **SIM_ID** | **NÚMERO** |
| H | 44 |
| A-Baa2/Wto | 43 |
| A | 42 |
| Vr-Py/Pyn | 41 |
| Vp/B-Rf-Baa2/Esp-Wto | 40 |
| Vp/B-Rf-Baa2/Egr-Wto | 39 |
| Vc-Mb/Ani-Cef | 38 |
| Rf-Mb/Pyn-Esp | 37 |
| Rf-Mb/Eph-Ani | 36 |
| Rf-Mb/Eph | 35 |
| Rf-Baa1/Egr-Wto | 34 |
| Rf/Eph | 33 |
| P-Mb/Cef-Ani | 32 |
| P-Mb/Cef | 31 |
| Pl/Ppa | 30 |
| P/Cef-Esp | 29 |
| P/Cef | 28 |
| Mb-Rf/Eph-Ani | 27 |
| Mb-P/Cef | 26 |
| Mb-P/Ani-Cef | 25 |
| Mb-Baa2/Bj-Wto | 24 |
| Mb-Baa2/Ani-Wto | 23 |
| Mb/Mve | 22 |
| Mb/Hsp | 21 |
| Mb/Emy | 20 |
| Mb/Bj | 19 |
| Mb/Ani | 18 |
| Mb/Acl | 17 |
| Ma1/Wto | 16 |
| Ma/Ati | 15 |
| B/Vp-Baa2-P-Rf/Wto-Cef-Esp | 14 |
| Bab2/Xsp-Dca | 13 |
| Bab1/Xsp-Dca | 12 |
| Baa2-Pl/Wto-Ppa | 11 |
| Baa2-P/Wto-Cef | 10 |
| Baa2-Ma/Wto | 9 |
| Baa2/Wto | 8 |
| Baa1-Rf/Wsp-Eph | 7 |
| Baa1-P/Wto-Cef | 6 |
| Baa1-P/Wsp-Cef | 5 |
| Baa1-Mb/Wsp-Hju | 4 |
| Baa1-Mb/Wsp-Ani | 3 |
| Baa1-Ch/Wto-Cht | 2 |
| Baa1/Wto | 1 |

## Páramo El Tablazo

Para el páramo del Tablazo se evaluó la heterogeneidad de 45 tipos de vegetación a lo largo de 23,86 kilómetros en el eje **Y** (latitud) y 16,76 kilómetros en el eje **X** (longitud). El eje SIM-ID contiene 45 tipos de cobertura y están distribuidos en la figura nueve (9) de la siguiente manera: el intervalo 0-10 representa bosques semi conservados y bosques alterados o intervenidos. El intervalo 10-20 presenta bosques intervenidos y entremezclados con plantaciones forestales y dos formaciones de bosque andino bajo y unos matorrales altos. El intervalo 20-30 incluye vegetación de páramo, como matorrales bajos, chuscales y un tipo de rosetal frailejonal dominado por *Espeletia corymbosa* entremezclado con bosque andino alto intervenido. El intervalo 30-40 presenta el resto de vegetación de páramo además de las coberturas de origen antrópico. El intervalo 40-50 presenta otros tipos de cobertura (tabla 16). La heterogeneidad de la zona se resume en la figura 9 **(1) Noroccidente**-solamente presenta dos tipos de cobertura se destacan los matorrales bajos de *Aragoa cleefii* que se extienden en al menos 100 hectáreas.**(2) Norte**- para el norte de la región de estudio se presentan nueve tipos de cobertura y la línea de tendencia indica una concentración de coberturas del tipo bosque intervenido dominado por *Weinmannia tomentosa* entremezclada con matorrales bajos de *Arcytophyllum nitidum* sobre superficies que alcanzan 250 hectáreas. Los sectores de producción agropecuaria alcanzan las mismas dimensiones. **(3) Nororiente**-para el nororiente de la región, el intervalo 30-40 presenta una gran concentración de coberturas que indican un alto grado de intervención. Las áreas agropecuarias alcanzan 1600 hectáreas mientras las plantaciones forestales se encuentran en menos de 100 hectáreas. De las coberturas naturales se destacan los matorrales bajos dominados por *Arcytophyllum nitidum* en al menos 750 hectáreas y los bosques intervenidos dominados por *Weinmannia tomentosa* en superficies inferiores a 100 hectáreas. **(4) Occidente** –la región occidental carece de la expresión de la mayoría de los tipos de vegetación del intervalo 10-20. Las zonas de producción agropecuaria alcanzan 1000 hectáreas. De las coberturas naturales se destaca la concentración en el sector de los matorrales bajos de *Bejaria resinosa* entremezclados con bosques intervenidos de *Weinmannia tomentosa* en al menos 100 hectáreas. Los intervalos 0-10 y 30-40 se encuentran bien representados pero sus coberturas individuales no sobrepasan 200 hectáreas. **(5) Centro-** se presentan nueve tipos de cobertura. El intervalo 40-50 carece de expresión en la zona y las coberturas más extensas son la agropecuaria con superficies que alcanzan 400 hectáreas y los matorrales altos de *Weinmannia tomentosa* con tamaños inferiores a 250 hectáreas **(6) Oriente**-. la tendencia indica una concentración de superficies en los patrones de vegetación con bosques semi conservados e intervenidos y matorrales altos. La expresión de estos tipos nunca sobrepasa 250 hectáreas. **(7) Sur occidente-** sólo se presentan matorrales altos de *Weinmannia tomentosa* entremezclados con rosetales frailejonales de *Espeletia phaneractis* en menos de 10 hectáreas y áreas de producción agropecuaria en al menos 100 hectáreas. **(8) Sur-** el intervalo 40-50 alcanza su mayor expresión sin embargo los tipos de vegetación con bosques y matorrales altos dominan en área cubierta. Los bosques andinos bajos semi-conservados dominados por *Xylosma spiculiferum* y *Daphnopsis caracasana* son los más extensos con cerca de 300 hectáreas en su máxima expresión. Las zonas de producción agropecuaria se extienden alrededor de 150 hectáreas. **(9) Sur Oriente-** sólo se presenta la cobertura agropecuaria en menos de 50 hectáreas.

**Tabla 16.** Eje de variación SIM_ID.

| EJE Y | |
|-------|-----|
| SIM_ID | NÚMERO |
| SVC | 45 |
| SV | 44 |
| Q | 43 |
| H | 42 |
| Pl/Ppa | 41 |
| A-Baa2/Wto | 40 |
| A | 39 |
| Vr-Mb/Ani | 38 |
| Vr-Ma-Mb/Vst | 37 |
| Vr-Ma/Ati | 36 |
| Rf-Mb/Eph-Ani | 35 |
| Rf/Esp | 34 |
| Rf/Egr | 33 |
| P/Cef | 32 |
| Mb/Vfl | 31 |
| Mb/Hla | 30 |
| Mb/Ani | 29 |
| Baa2/Ecy-Wto | 28 |
| Ch/Cht | 27 |
| P-Ma/Wto | 26 |
| Mb-Vc/Bj | 25 |
| Mb-Rf/Ani-Egr | 24 |
| Mb-Baa2/Bj-Wto | 23 |
| Mb-P/Ani-Cef | 22 |
| Mb/Bj | 21 |
| Mb/Acl | 20 |
| Ma-Rf/Wto-Eph | 19 |
| Ma/Xsp | 18 |
| Ma/Wto-Ani | 17 |
| Ma/Wto-Acl | 16 |
| Ma/Wto | 15 |
| Ma/Mru | 14 |
| Bab2-Pl/Xsp-Dca-Ppa | 13 |
| Bab2/Xsp-Dca | 12 |
| Bab1/Xsp-Dca | 11 |
| Baa2-Pl/Wto | 10 |
| Baa2-Pl/Qhu | 9 |
| Baa2-Mb/Wto-Ani | 8 |
| Baa2/Wto | 7 |
| Baa2/Vst | 6 |
| Baa2/Qhu | 5 |
| Baa1/Wto | 4 |
| Baa1/Qhu | 3 |
| Baa1-Rf/Wto-Esp | 2 |
| Baa2-P-Rf/Wto-Cef-Esp | 1 |

## <u>Páramos de la jurisdicción CORPO-GUAVIO</u>

En los páramos de la jurisdicción de CORPOGUAVIO se diferenciaron seis unidades de tamaño considerable y otras pequeñas alrededor de estas. Debido a esta distribución, las áreas grandes se encuentran representadas en varios cuadrantes mientras las superficies pequeñas se encuentran en su totalidad dentro de un cuadrante. Por otra parte, debido a los límites geográficos establecidos durante el desarrollo del proyecto (el cual dependió del dominio y la administración del territorio por diferentes entidades del estado) se excluye en el análisis la jurisdicción del páramo de Chingaza, contigua a las áreas de páramo de los municipios de Fómeque y Guasca y que en conjunto forman una unidad continua. Se evaluaron 60 tipos de cobertura que se observan en la tabla 17. Las áreas con páramo de esta jurisdicción se desarrollan a través en 61,01 kilómetros en el eje **Y** (latitud) y 53,65

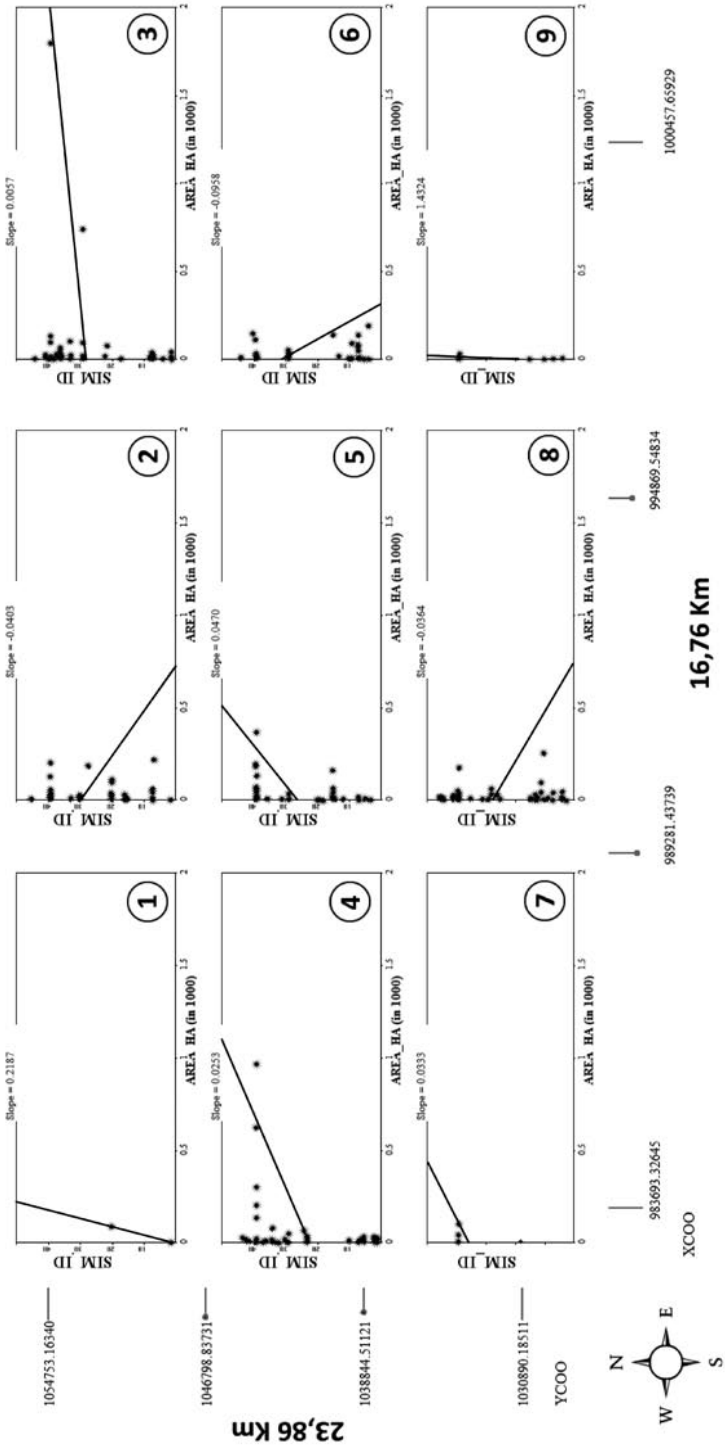

**Figura 9.** Distribución de la heterogenidad en tipos de cobertura y área presentes en el páramo El Tablazo.

kilómetros en el eje **X** (longitud). En el eje SIM-ID se observan los siguientes intervalos: 0-10 representa formaciones de bosque andino alto y de matorrales altos conservados. En el intervalo 10-20 se presentan matorrales altos en diferentes arreglos florísticos, entremezclados con otros tipos de vegetación. El intervalo 20-30 es muy diverso en matorrales bajos y dos tipos de pajonales. En el intervalo 30-40 se registran pajonales frailejonales, chuscales y vegetación propia de pantano. El intervalo 40-50 presenta siete tipos de vegetación de pantano y los rosetales frailejonales dominantes. Dentro del intervalo 50-60 se registran las turberas, la vegetación ribereña y los otros tipos de cobertura encontrados. La heterogeneidad de la zona se resume en la figura 10. **(1) Noroccidente**- contiene un elevado número de tipos de cobertura. El intervalo 10-40 presenta una concentración elevada de chuscales de *Chusquea tessellata* hacia el sector del municipio de Guasca. Las zonas con este tipo de vegetación presenta extensas áreas que en algunos casos alcanzan 700 hectáreas. Los pajonales de *Calamagrostis effusa* entremezclados con rosetales frailejonales de *Espeletia grandiflora* se expresan en zonas hasta de 400 hectáreas. Otros tipos de cobertura natural importantes en la zona, son los bosques conservados de *Drimys granadensis* y de *Weinmannia microphylla* expresados en la zona en superficies hasta de 400 hectáreas. Las zonas de producción agropecuaria se presentan en áreas inferiores a 800 hectáreas. **(2) Norte**- los pajonales de *Calamagrostis effusa* entremezclados con matorrales bajos de *Arcytophyllum nitidum* se presentan en al menos 700 hectáreas. El tipo de bosque alto andino conservado dominado por *Weinmannia microphylla* ocupa superficies inferiores a 200 hectáreas. **(3) Nororiente**- no se expresan los tipos de cobertura del intervalo 40-50. Al igual que en la zona anterior, una de las mayores concentraciones de cobertura la muestran los bosques andinos altos dominados por *Weinmannia microphylla* que se expresan hasta 350 hectáreas. Algunos

**Tabla 17.** Eje de variación SIM_ID.

| EJE Y | |
|---|---|
| SIM_ID | NÚMERO |
| X | 60 |
| Q | 59 |
| Pl | 58 |
| Ap | 57 |
| Ac | 56 |
| H | 55 |
| E | 54 |
| Vr | 53 |
| Vt/Rhp | 52 |
| Vt/Ove | 51 |
| Rf-P/Es-Cef | 50 |
| Rf-P/Egr | 49 |
| Rf/Egr | 48 |
| Rf/Ea | 47 |
| Pt/Pls-Crv-Els | 46 |
| Pt/Hol | 45 |
| Pt/Est | 44 |
| Pt/Cve | 43 |
| Pt/Cru | 42 |
| Pt/Cli | 41 |
| Pt/Cja | 40 |
| Pt/Cac | 39 |
| Ch-Rf/Cht-Egr | 38 |
| Ch-P/Cht-Cbo | 37 |
| Ch/Cht | 36 |
| P-Rf-Ch/Cef-Egr | 35 |
| P-Rf/Cef-Egr | 34 |
| P-Rf/Cef-Ear | 33 |
| P-Rf/Cef | 32 |
| P-Mb/Cef | 31 |
| P-Mb/An-Cef | 30 |
| P/Fdo | 29 |
| P/Cef | 28 |
| P/Cbo | 27 |
| Mb-P/Cef | 26 |
| Mb-P/Ani-Cef | 25 |
| Mb-Ch/Hgo-Cht | 24 |
| Mb/Vfl | 23 |
| Mb/Hgo | 22 |
| Mb/Gra | 21 |
| Mb/Ani | 20 |
| Mb/Aab | 19 |
| Ma-P/Wmi-Cef | 18 |
| Ma-P/Cef | 17 |
| Ma-Ch/Msp-Cs | 16 |
| Ma-Ch/Bgl-Cht | 15 |
| Ma/Wsp | 14 |
| Ma/Wmi | 13 |
| Ma/Vfl | 12 |
| Ma/Ppp | 11 |
| Ma/Pga | 10 |
| Ma/Mni | 9 |
| Ma/Dri | 8 |
| Ma/Bgl | 7 |
| Ma/Ati | 6 |
| Baa1-Ma/Wmi | 5 |
| Baa1-Ch/Wmi-Chs | 4 |
| Baa1/Wsp | 3 |
| Baa1/Wmi | 2 |
| Baa1/Dri | 1 |

pajonales también están presentes aunque nunca sobrepasan 100 hectáreas de superficie. **(4) Occidente** –se expresan un gran número de tipos de cobertura. Como ya se mencionó previamente, esto se debe a que se incluyen varios sectores por encima de 2800 metros en el análisis. Los chuscales de *Chusquea tessellata* entremezclados con pajonales de *Calamagrostis bogotensis* dominan y se expresan hasta en 750 hectáreas. Los tipos de cobertura del intervalo 20-40 y que incluyen desde matorrales bajos hasta vegetación de pantano se concentran en fragmentos hasta de 350 hectáreas de superficie. Las zonas agropecuarias se extienden hasta en 200 hectáreas al igual que los dos primeros tipos de bosques conservados. **(5) Centro**- se presenta una gran expresión de las coberturas dominadas por pajonales de *Calamagrostis effusa* entremezcladas con rosetales frailejonales de *Espeletia grandiflora*, chuscales de *Chusquea tessellata* (aproximadamente 1900 hectáreas) y pajonales de *Calamagrostis effusa* entremezclados con matorrales bajos de *Arcytophyllum nitidum* (aproximadamente 1750 hectáreas). La gran concentración de bosques conservados es evidente y su máxima expresión alcanza 800 hectáreas de superficie. Alrededor del punto veinte se concentran un gran número de tipos de matorrales altos y bajos con coberturas inferiores a 100 hectáreas. Las plantaciones forestales también se encuentran presentes en la zona en al menos 200 hectáreas. **(6) Oriente**- sólo presenta dos tipos de cobertura. Los bosques conservados de *Weinmannia microphylla* que se extienden por alrededor de 250 hectáreas. **(7) Sur occidente**- los bosques conservados dominados por *Drimys granadensis* se extienden en fragmentos hasta de 1200 hectáreas. Los matorrales bajos de *Arcytophyllum nitidum* se expresan en la zona en alrededor de 400 hectáreas y algunos pajonales se concentran en tamaños inferiores a 350 hectáreas.**(8) Sur**-sólo se presentan cinco coberturas y de estas se destacan los bosques conservados de *Drimys granadensis* en fragmentos inferiores a 400 hectáreas. **(9) Sur**

**Oriente**-en el sur oriente del área de estudio solamente se registra un pequeño sector con bosques conservados.

## Área de estudio perteneciente al páramo de Sumapaz

Para el área de estudio perteneciente al páramo de Sumapaz se evaluaron 46 tipos de cobertura a lo largo de 74,81 kilómetros en el eje Y (latitud) y 44,18 kilómetros en el eje X (longitud). En el eje SIM-ID se muestran los siguientes intervalos: de 0-10 bosques conservados e intervenidos en diferentes expresiones florísticas. En el intervalo 10-20 se presentan los bosques intervenidos y entremezclados con vegetación de páramo así como algunos matorrales altos. El intervalo 20-30 presenta toda la variedad típica de ambientes paramunos y algunas mezclas con vegetación de mayor porte. Para el intervalo 30-40 se registran pajonales, frailejonales, chuscales entremezclados con vegetación boscosa además de dos coberturas de origen antrópico. En el intervalo 40-50 se presentan otros tipos de cobertura (tabla 18). La heterogeneidad de la zona se resume en la figura 11. **(1) Noroccidente**- carece de expresión de los tipos de cobertura encontrados. **(2) Norte**-presenta el mayor registro de coberturas agropecuarias, cerca de 2000 hectáreas. Los bosques conservados entremezclados con intervenidos dominados por *Clusia multiflora* alcanzan en la zona alrededor de 1000 hectáreas **(3) Nororiente**- presenta una concentración de áreas en el intervalo 20-30 típico de vegetación de páramo. Los pajonales de *Calamagrostis effusa* entremezclados con rosetales frailejonales de diferentes especies de *Espeletia* alcanzan una de las más grandes expresiones (aproximadamente 12500 hectáreas). Los bosques de *Weinmannia tomentosa* entremezclados con chuscales de *Chusquea tessellata* se presentan en fragmentos inferiores a 1000 hectáreas. Las áreas de producción agropecuaria se cubren con parches menores a 500 hectáreas. **(4)**

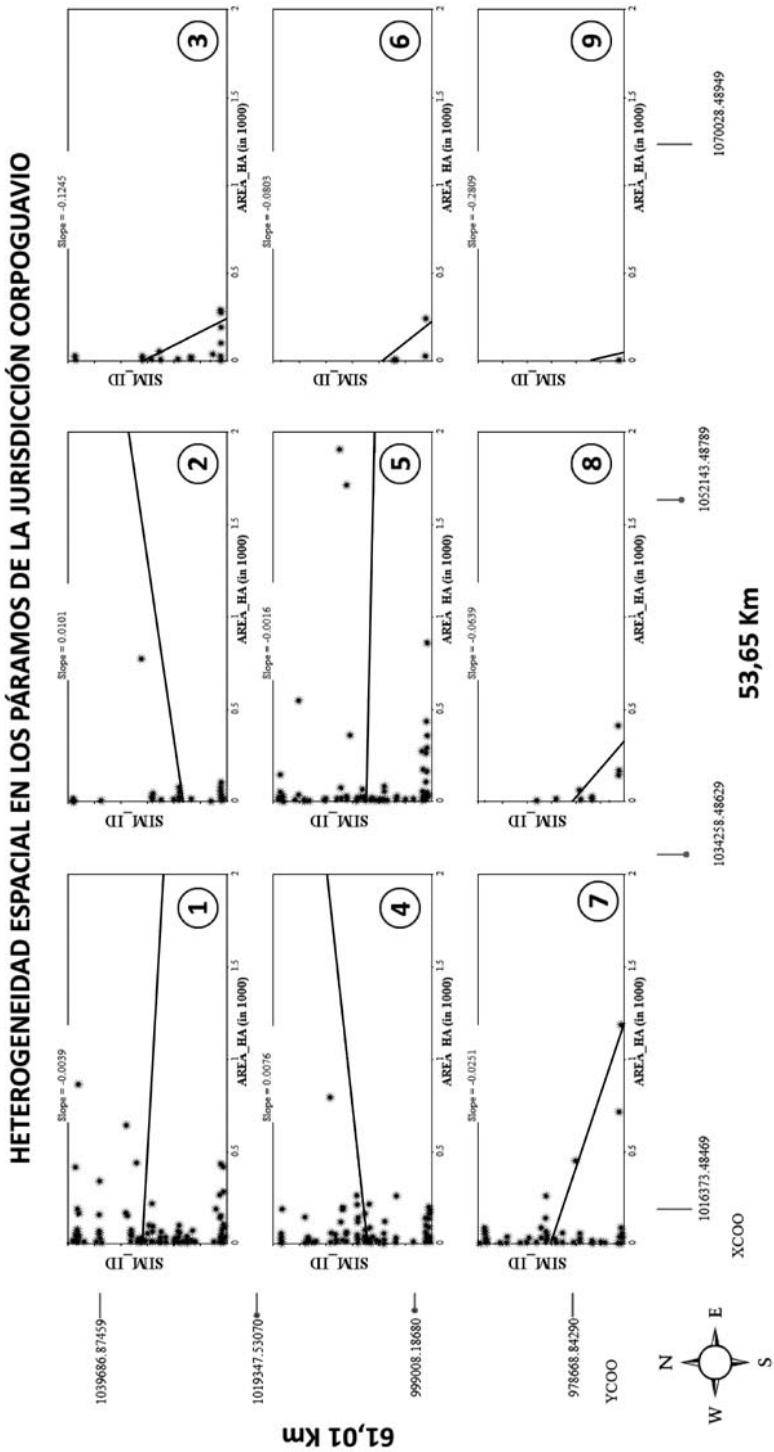

**Figura 10.** Distribución de la heterogeneidad en tipos de cobertura y área presentes en los páramos de la jurisdicción CORPOGUAVIO.

**Occidente-** sólo se presentan tres tipos de bosque conservado con coberturas hasta de 2500 hectáreas. **(5) Centro-** la línea de tendencia indica una concentración del área hacia los tipos de cobertura boscosa. Alrededor del punto 10 de bosques intervenidos, las concentraciones no superan 1000 hectáreas de cobertura, mientras que los bosques conservados de *Weinmannia rollotii* pueden alcanzar hasta 1500 hectáreas. Los matorrales bajos de diferentes composiciones florísticas entremezclados con pajonales de *Calamagrostis effusa* se presentan es superficies hasta de 1000 hectáreas y los pajonales de *Calamagrostis effusa* entremezclados con frailejonales con distintas especies de *Espeletia* alcanzan alrededor de 2500 hectáreas. Las áreas de producción agropecuaria se presentan en superficies de hasta 2000 hectáreas. **(6) Oriente-** para la región oriental del páramo de Sumapaz, los pajonales de *Calamagrostis effusa* entremezclados con rosetales frailejonales dominados por diferentes especies de *Espeletia* son los que cubren mayor superficie con fragmentos que alcanzan 7000 hectáreas. Los bosques conservados de *Weinmannia rollotii* junto a pajonales y rosetales frailejonales entremezclados con elementos de bosques intervenidos se distribuyen en parches que alcanzan 1500 hectáreas. Las áreas agropecuarias se presentan en al menos 1000 hectáreas. **(7) Sur occidente-** se presentan dos concentraciones de áreas en los intervalos 0-10 y 10-20, en las localidades pertenecientes a bosques intervenidos y a vegetación típica de páramo. En el primer intervalo los bosques intervenidos dominados por *Clusia multiflora* alcanzan superficies de hasta 2000 hectáreas, mientras que los bosques intervenidos de *Weinmannia tomentosa* alcanzan hasta 1000 hectáreas. La vegetación riparia con matorrales altos de *Ageratina tinifolia* se expresan en superficies hasta de 2000 hectáreas. Las áreas de producción agropecuaria se presentan al menos en 500 hectáreas. **(8) Sur-** se presenta una concentración de fragmentos de vegetación boscosa intervenida con algunos fragmentos de bosques conservados. La superficie mayor la alcanzan los bosques intervenidos dominados por *Clusia multiflora* en aproximadamente 1500 hectáreas. Los pajonales de *Calamagrostis effusa* entremezclados con rosetales frailejonales dominados por diferentes especies de *Espeletia* alcanzan superficies hasta de 2000 hectáreas al igual que las áreas agropecuarias. **(9) Sur Oriente-** el suroriente de páramo de Sumapaz no se incluye en el análisis.

**Tabla 18.** Eje de variación SIM_ID.

| EJE Y | |
|---|---|
| SIM_ID | NÚMERO |
| Baa1/Clm | 1 |
| Baa1/Msa | 2 |
| Baa1/Wro | 3 |
| Baa1-Baa2/Clm | 4 |
| Baa1-Baa2/Wro | 5 |
| Baa1-P/Wto-Cef | 6 |
| Baa2/Clm | 7 |
| Baa2/Msa | 8 |
| Baa2/Wro | 9 |
| Baa2/Wto | 10 |
| Baa2-A/Wro | 11 |
| Baa2-A/Wto | 12 |
| Baa2-P/Wto-Cef | 13 |
| Bab1/Xsp-Dca | 14 |
| B/Vp-Baa2-Ch/Wto-Cht | 15 |
| B/Vp-Baa2-P-Rf/Wto-Cef-Esp | 16 |
| B/Vp-Baa-Ch/Wto-Cht | 17 |
| Ma/Mru | 18 |
| Ma/Wro | 19 |
| Mb/Ani | 20 |
| Mb/Bj | 21 |
| Mb/Vfl | 22 |
| Mb-Ch/Cht | 23 |
| Mb-P/Cef | 24 |
| P-Rf/Cef-Esp | 25 |
| Rf/Egr | 26 |
| Ch/Cht | 27 |
| Vc-Mb/Aab | 28 |
| Vp/B-Ch-Baa2/Cht-Wto | 29 |
| Vp/B-Mb-Baa2/Aab-Clm | 30 |
| Vp/B-P-Rf-Baa2/Cef-Wto | 31 |
| Vp/B-Rf-Baa2/Egr-Msa | 32 |
| Vp/B-Rf-Baa2/Egr-Wto | 33 |
| Vp/B-Rf-Baa2/Msa | 34 |
| Vp/B-Rf-Baa2/Wro | 35 |
| Vp-Vr-P-Ma/Cef-Clm | 36 |
| Vr-Ma/Ati | 37 |
| Vr-Py/Pyn | 38 |
| A | 39 |
| A-Ma/Clm | 40 |
| A-Ma/Wro | 41 |
| A-Pl/Egl | 42 |
| H | 43 |
| Si | 44 |
| SV | 45 |
| U | 46 |

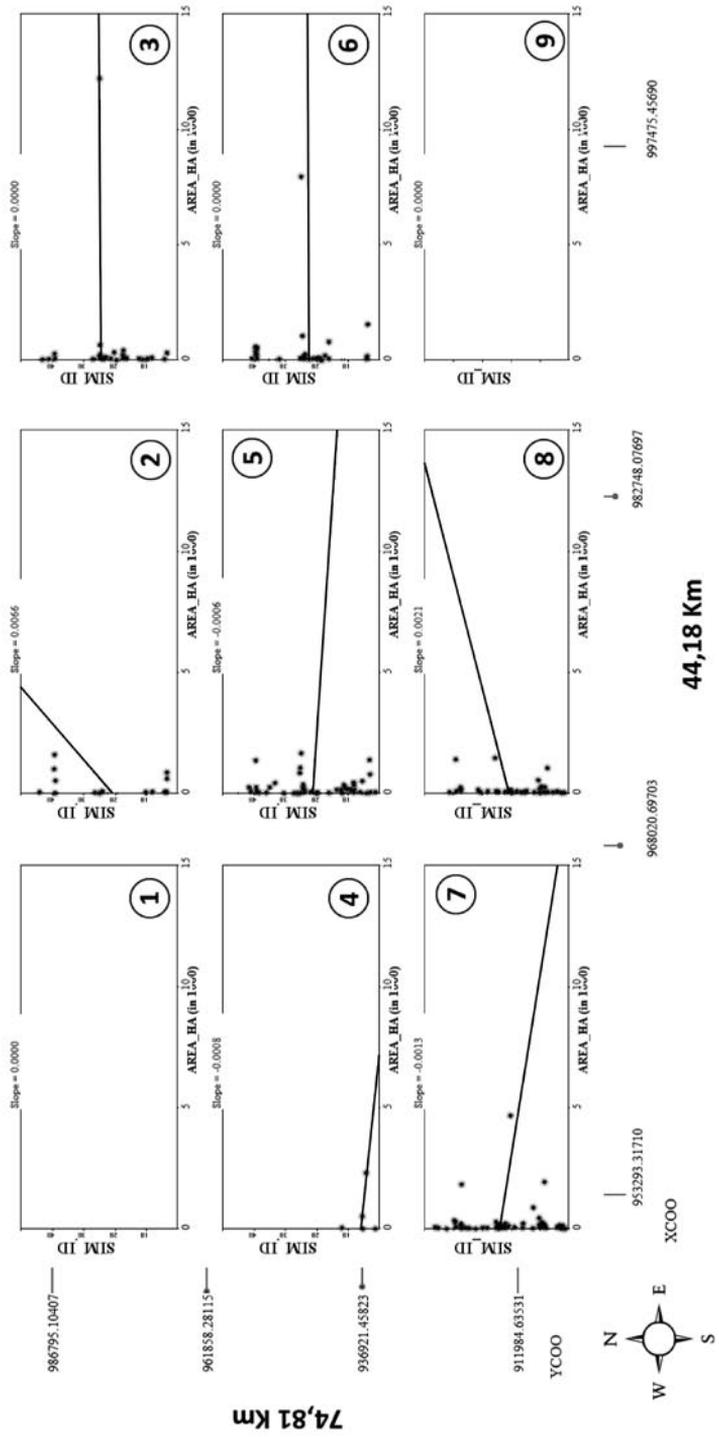

**Figura 11.** Distribución de la heterogeneidad en tipos de cobertura y área presentes en el área de estudio del páramo de Sumapaz.

## Páramo del Parque Nacional Natural Los Nevados

Para el área de estudio perteneciente al área de páramo del Parque Nacional Natural Los Nevados –cordillera Central, se evaluaron 36 tipos de cobertura a lo largo de 37,49 kilómetros en el eje **Y** (latitud) y 24,78 kilómetros en el eje **X** (longitud). Los patrones de distribución de áreas encontrados en esta región difieren de los encontrados para el resto de páramos estudiados. En el eje SIM-ID (Tabla 19) se muestran los siguientes intervalos: de 0-10 se presentan los bosques conservados, todos los tipos de matorral y un tipo de pajonal. En el intervalo 10-20 se concentran los prados, pajonales y pastizales. En el intervalo 20-30 se presentan los cojines de plantas vasculares y un tipo de rosetal dominado por *Senecio isabelis*. Para el intervalo 30-40 se registran tres tipos de vegetación natural y otras coberturas. La heterogeneidad de la zona se resume en la figura 12. **(1) Noroccidente**- se registraron cuatro tipos de vegetación con valores inferiores a 50 hectáreas. la tendencia indica tamaños superiores de bosques conservados y matorrales altos **(2) Norte**- Se presenta una máxima concentración de áreas en el intervalo 10-20. Los prados de *Lachemilla orbiculata* con parches de bosques conservados de *Hesperomeles ferruginea* y cojines de plantas vasculares de *Plantago rigida* presentan los mayores fragmentos que alcanzan hasta 400 hectáreas. Otros tipos de vegetación que se destacan son los pajonales de *Calamagrostis effusa* con pastizales de *Agrostis perennans* y cojines de plantas vasculares con *Plantago rigida* en tamaños hasta de 200 hectáreas y los pastizales de *Agrostis perennans* en fragmentos menores a 200 hectáreas. El intervalo 0-10 sólo presenta un tipo de matorral bajo en áreas menores a las 100 hectáreas. **(3) Nororiente**-.para esta región aparte de la zonas cubierta por nieve (la más extensas), el tipo de cobertura más importante la constituye el suelo desnudo con zonas de herbazales dominadas por *Pentacalia gelida, Baccharis caespitosa* y *Stereocaulon vesuvianum* en superficies que alcanzan 1750 hectáreas. En zonas hasta de 750 hectáreas se presentan pajonales de *Calamagrostis effusa* entremezclados con pastizales de *Agrostis perennans* y algunos tipos de vegetación presentes en el intervalo 10-20, que se extienden hasta en 500 hectáreas. Los bosques conservados de *Weinmannia* spp. entremezclados con prados de *Lachemilla orbiculata* y con algunas áreas agropecuarias son los más extensos en la zona y sus superficies alcanzan alrededor de 400 hectáreas. **(4) Occidente** –las áreas se concentran entre los tipos de vegetación con prados, pajonales y pastizales en diferentes arreglos florísticos. Estos se encuentran en general alrededor de 100 hectáreas y entre ellos la mayor superficie la alcanzan los pastizales de *Agrostis perennans* entremezclados con pajonales de *Calamagrostis effusa* en zonas hasta de 700 hectáreas. La mayor superficie en el grupo de cojines de plantas vasculares dominadas por *Plantago rigida,* la presenta el tipo entremezclado con pajonales de *Calamagrostis recta* y *Calamagrostis effusa* y con pastizales de *Agrostis perennans* hasta en alrededor de 150 hectáreas. Los bosques más representativos de la región son los dominados por especies de *Weinmannia,* entremezclados con prados de *Lachemilla orbiculata* (aproximadamente en 900 hectáreas). **(5) Centro-** se observan tres concentraciones de áreas. Cabe mencionar que los bosques no se expresan dentro de la zona. En el intervalo 0-10 los matorrales bajos dominados por *Loricaria colombiana* se presentan el al menos 200 hectáreas. En el intervalo 10-20, los pajonales de *Calamagrostis recta* y *Calamagrostis effusa* se distribuyen en parches hasta de 1100 hectáreas y la concentración de pastizales y pajonales hacia el final del intervalo se presenta en superficies inferiores a 250 hectáreas. Otras coberturas con superficies significativas son los cojines de plantas vasculares dominados por *Plantago rigida* y *Distichia muscoides* entremezclados con pajonales de *Calamagrostis recta* en áreas superiores a 500 hectáreas y las zonas con suelo desnudo y herbazales dominados por *Pentacalia gelida, Baccharis caespitosa* y *Stereocaulon vesuvianum* en sectores hasta de 350 hectáreas. **(6) Oriente-** presenta expresión de todos los intervalos. Los bosques conservados son más frecuentes que en la región central y en algunos casos alcanzan cerca de 400 hectáreas. Las dos coberturas con mayor extensión corresponden a pajonales de *Calamagrostis effusa* entremezclados con pastizales de *Agrostis perennans* en fragmentos que alcanzan más de 800 hectáreas y algunos arreglos florísticos de plantas vasculares dominadas por *Plantago rigida* en superficies de hasta 700 hectáreas. **(7) Sur occidente-** presenta la mayor expresión, cerca de 1800 hectáreas de bosques conservados entremezclados con prados. Para el intervalo 10-20, los pastizales de *Agrostis perennans* en combinación con pajonales de *Calamagrostis effusa* y *C. recta* se presentan en sectores hasta de 350 hectáreas. **(8) Sur-** presenta concentración de los tipos de vegetación del intervalo 10-20. Al igual que en la zona anterior, los pastizales de *Agrostis perennans* en combinación con pajonales de *Calamagrostis effusa* y *C. recta* se expresan en gran proporción con respecto a los otros tipos de cobertura y alcanzan más de 750 hectáreas. **(9) Sur Oriente-** la tendencia de los sectores se mantiene aunque se presenta un incremento de aproximadamente tres cuartas partes en el tamaño máximo de los fragmentos de los tipos de vegetación presentes en el intervalo 10-20 con respecto al anterior.

**Tabla 19.** Eje de variación SIM_ID.

| EJE Y | |
|---|---|
| SIM_ID | NÚMERO |
| H | 36 |
| N2 | 35 |
| N1 | 34 |
| Hz/Bca-Lni | 33 |
| Cj/Dmu | 32 |
| Sd-Hz/Pge-Bca-Sve | 31 |
| R/Sis-Lal | 30 |
| Pv-P-Ps/Pri-Cre-Cef-Ahe | 29 |
| Pv-P-Ps/Pri-Cef-Ahe | 28 |
| Pv-P/Pri-Dmu-Cre-Cef | 27 |
| Pv-P/Pri-Dmu-Cre | 26 |
| Pv-P/Pri-Dmu-Cef | 25 |
| Pv-P/Pri-Cre-Cef | 24 |
| Pv-P/Pri-Cre | 23 |
| Pv/Pri | 22 |
| Ps-Pv/Ahe-Pri | 21 |
| Ps-P/Ahe-Cre-Cef | 20 |
| Ps-P/Ahe-Cef | 19 |
| Ps/Ahe | 18 |
| P-Pv/Cre-Cef-Pri | 17 |
| P-Ps-Pv/Cef-Ahe-Pri | 16 |
| P-Ps/Cef-Ahe | 15 |
| Pd-Pv/Lor-Pri | 14 |
| Pd-Baa1-Pv/Lor-Hfe-Pri | 13 |
| Pd/Lor | 12 |
| P/Cre-Cef | 11 |
| P/Cre | 10 |
| Mb-Baa1/Pve-Cre-Pse | 9 |
| Mb/Pve | 8 |
| Mb/Lco | 7 |
| Mb/Hla | 6 |
| Ma-P/Emy-Cef | 5 |
| Baa1-Pd-A/Wsp-Lor | 4 |
| Baa1-Pd/Wsp-Lor | 3 |
| Baa1-Pd/Hfe-Lor | 2 |
| Baa1/Hfe | 1 |

### Páramo de Guanacas, área del volcán Puracé

Para el área de estudio perteneciente los páramos Guanacas y Las Delicias se evaluaron 38 tipos de cobertura a lo largo de 18,85 kilómetros en el eje **Y** (latitud) y 18,25 kilómetros en el eje **X** (longitud). El eje SIM-ID (tabla 20) presenta los siguientes intervalos: de 0-10 se encuentran bosques conservados e intervenidos entremezclados con áreas agropecuarias y algunas mezclas con vegetación de páramo. En el intervalo 10-20 se concentran los matorrales altos entremezclados con vegetación de páramo y sistemas agropecuarios. En el intervalo 20-30 se presentan matorrales bajos pajonales y rosetales frailejonales en diferentes arreglos florísticos. Para el intervalo 30-40 se registran los chuscales, las turberas y otras coberturas como las zonas agropecuarias. La heterogeneidad de la zona se resume en la figura 13. En general debido a la simetría de la región y a su forma redondeada la

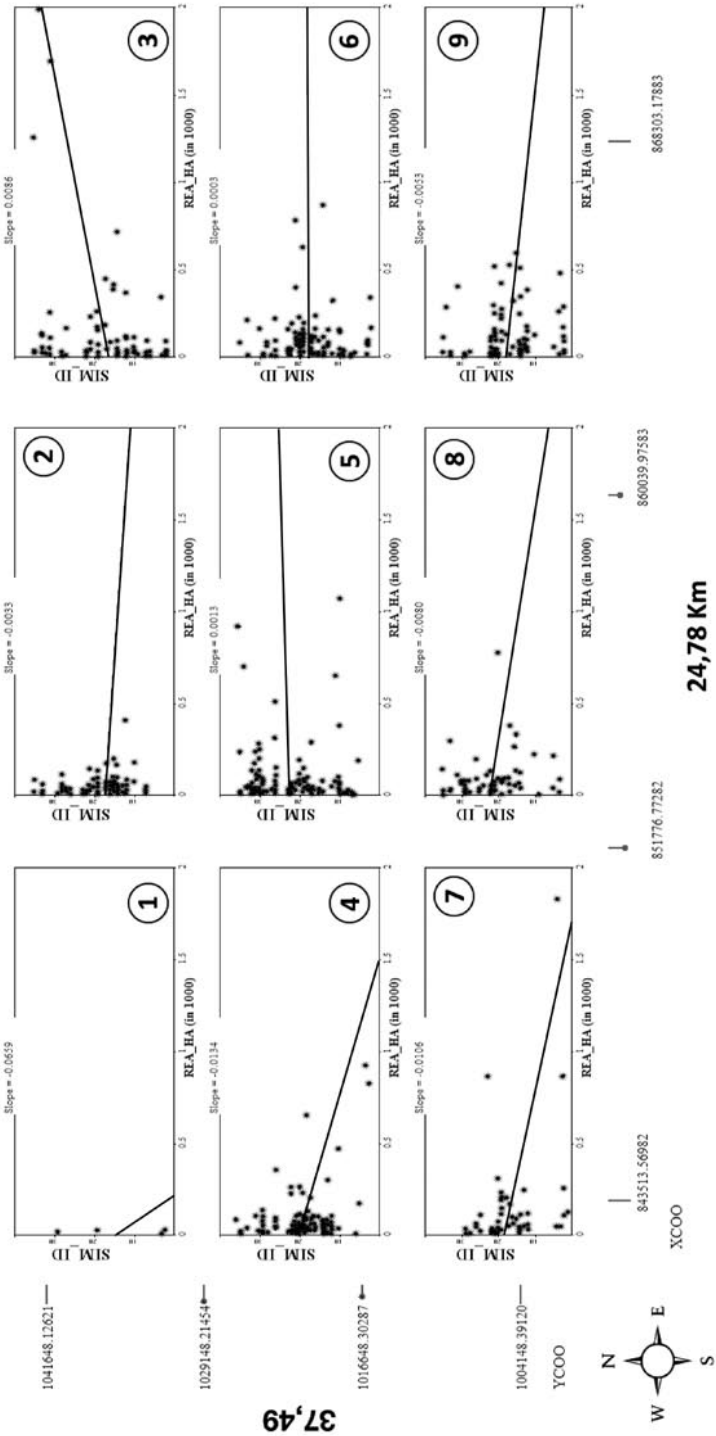

**Figura 12** Distribución de la heterogeneidad en tipos de cobertura y área presentes en el área de estudio perteneciente a páramo en el Parque Nacional Natural Los Nevados.

concentración de las coberturas se presenta en la zona central de la región **(1) Noroccidente**- se presentan cuatro tipos de cobertura con valores muy bajos de superficie. **(2) Norte**- se registran 11 tipos de cobertura con una tendencia de acumulación de área hacia la región de rosetales frailejonales de *Espeletia hartwegiana* entremezclados con pajonales de *Calamagrostis effusa* en alrededor de 1000 hectáreas. **(3) Nororiente**- se presentan siete tipos de cobertura en áreas reducidas. **(4) Occidente** – se presentan tres registros con bajos valores. **(5) Centro**- en la región central se observa la mayor concentración de coberturas los bosques conservados dominados por *Clethra rugosa*,

**Tabla 20.** Eje de variación SIM_ID.

| EJE Y | |
|---|---|
| SIM_ID | NÚMERO |
| A | 38 |
| H | 37 |
| T/ Ope-Cyp | 36 |
| Ch-Rf-P/ Cht-Eha-Cef | 35 |
| Ch-Rf-Mb/ Cht-Eha-Dci | 34 |
| Ch-Rf-Baa1/ Cht-Eha | 33 |
| Ch-Rf-A/ Cht-Eha | 32 |
| Ch-Rf/ Cht-Eha | 31 |
| Ch/ Cht-Dem | 30 |
| Rf-P-Mb/ Eha-Cef-Dci-Blo | 29 |
| Rf-P-Ma/ Eha-Cef-Bca | 28 |
| Rf-P-Baa1/ Eha-Cef-Bca-Wpu | 27 |
| Rf-P/ Eha-Cef | 26 |
| Rf-Ch/ Eha-Cht | 25 |
| P/Cef | 24 |
| Mb-Ma/ Dci-Cht-Wbr-Msa | 23 |
| Mb/ Hla | 22 |
| Mb/ Dci-Cht | 21 |
| Ma-Ps/ Cru-Dry-Emy-Asp | 20 |
| Ma-Ch/ Bca-Wpu-Cht | 19 |
| Ma-Ch/ Ati-Cht | 18 |
| Ma-A/ Wbr-Msa | 17 |
| Ma-A/ Cru-Dry-Emy | 16 |
| Ma-A/ Bca | 15 |
| Ma/ Wbr-Msa | 14 |
| Ma/ Cru-Dry-Emy | 13 |
| Ma/Ast- Bca-Wpu | 12 |
| Ma/ Ati | 11 |
| Baa1-Mb/ Bca-Wbr-Dci-Blo | 10 |
| Baa1-Ch/ Cru-Dry-Emy-Cht | 9 |
| Baa1-Ch/ Bca-Wpu-Cht | 8 |
| Baa2-A/ Wbr-Msa | 7 |
| Baa2-A/ Cru-Dry-Emy | 6 |
| Baa2-A/ Bca-Wpu | 5 |
| Baa1/ Wbr-Msa | 4 |
| Baa2/ Cru-Dry-Emy | 3 |
| Baa1/ Cru-Dry-Emy | 2 |
| Baa1/ Bca-Wpu | 1 |

*Drimys granadensis* y *Escallonia myrtilloides* alcanzan 11000 hectáreas y los chuscales de *Chusquea tessellata* entremezclados con rosetales frailejonales de *Espeletia hartwegiana* se presentan en zonas hasta de 2000 hectáreas. **(6) Oriente**- se registran 11 tipos de cobertura con concentración de hasta 1000 hectáreas en los tipos de vegetación cerrada. **(7) Sur occidente**- presenta la mayor concentración de cobertura agropecuaria y se extiende aproximadamente en 2000 hectáreas. Otros tipos de cobertura boscosa también se registran para la zona aunque es evidente su fuerte estado de intervención. **(8) Sur**- el sector sur presenta 11 tipos de cobertura, en los cuales dominan los rosetales frailejonales en distintos arreglos florísticos que se extienden hasta en zonas con 1000 hectáreas de superficie. **(9) Sur Oriente**- sólo se presentan dos coberturas los matorrales altos de *Ageratina tinifolia* mezclados con chuscales de *Chusquea tessellata* en una pequeña área y chuscales de *Chusquea tessellata* mezclados con rosetales frailejonales de *Espeletia hartwegiana* y bosque altoanandino conservado en aproximadamente 900 hectáreas.

## DEPENDENCIA ESPACIAL

Como se mencionó en la metodología, los diagramas de dispersión de Moran (Moran, 1948) pretenden comparar la contiguidad en las regiones de un mismo tipo de vegetación por medio de los pesos espaciales. Para cada región de páramo se presenta la dispersión de primer y segundo grado con un umbral entre 1000 y 2000 metros. Para el primer caso la dispersión de los pesos indican que tan cerca se encuentra un siguiente parche de vegetación del mismo tipo y para el segundo si existe un parche del mismo tipo posterior al siguiente más cercano de otro tipo. En un diagrama de Moran típico, cada valor en el eje de las abscisas se distribuye según sus pesos para un solo tipo de variable, es decir que en el sentido estricto se evalúa solamente la expresión de un tipo (por ejemplo número de votos por un candidato, el PIB de

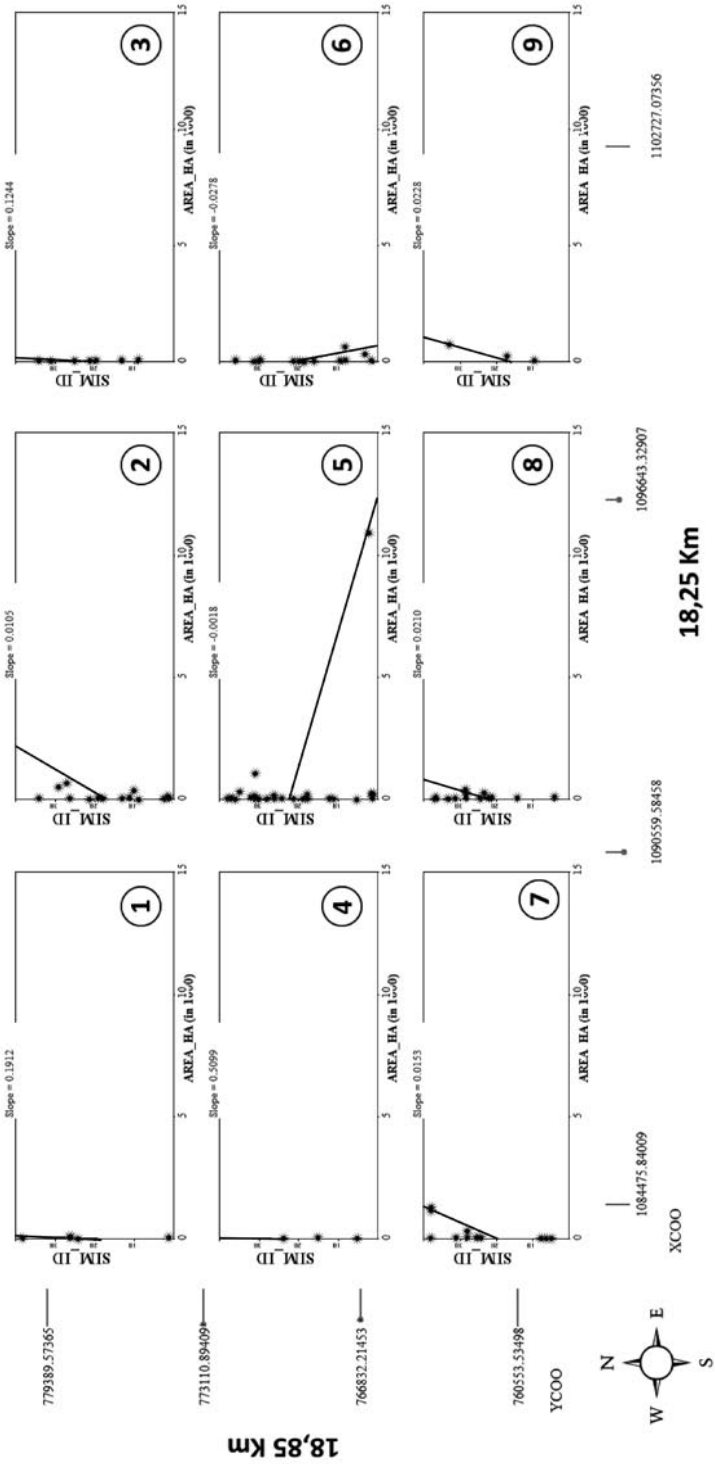

**Figura 13.** Distribución de la heterogeneidad en tipos de cobertura y área presentes en el área de estudio perteneciente a los páramos de Guanacas y Las Delicias –área del volcán Puracé.

397

cada región o la cobertura de determinado tipo de vegetación), los valores positivos indicarían una contigüidad y dependencia positiva y los valores negativos lo contrario. Al utilizar un variable categórica consecutiva y entera, la variable se distribuye de menor a mayor en el eje de las abscisas y cada número indica un tipo de vegetación diferente y a diferencia del sentido clásico del análisis, los extremos de los pesos indican vegetación separada y fragmentada y la cercanía vegetación conectada. Es importante recalcar que la distribución natural de algunos tipos de vegetación de páramo es fragmentada en sí; sin embargo, la conectividad entre los parches y sus interacciones depende de una distancia mínima. El eje de variación SIM-ID es similar al presentado en el aparte sobre heterogeneidad. En la figura 14 se observa la distribución de los tipos de cobertura según el peso (contigüidad) de los parches que la componen y representa un ejemplo de la forma de interpretarlas.

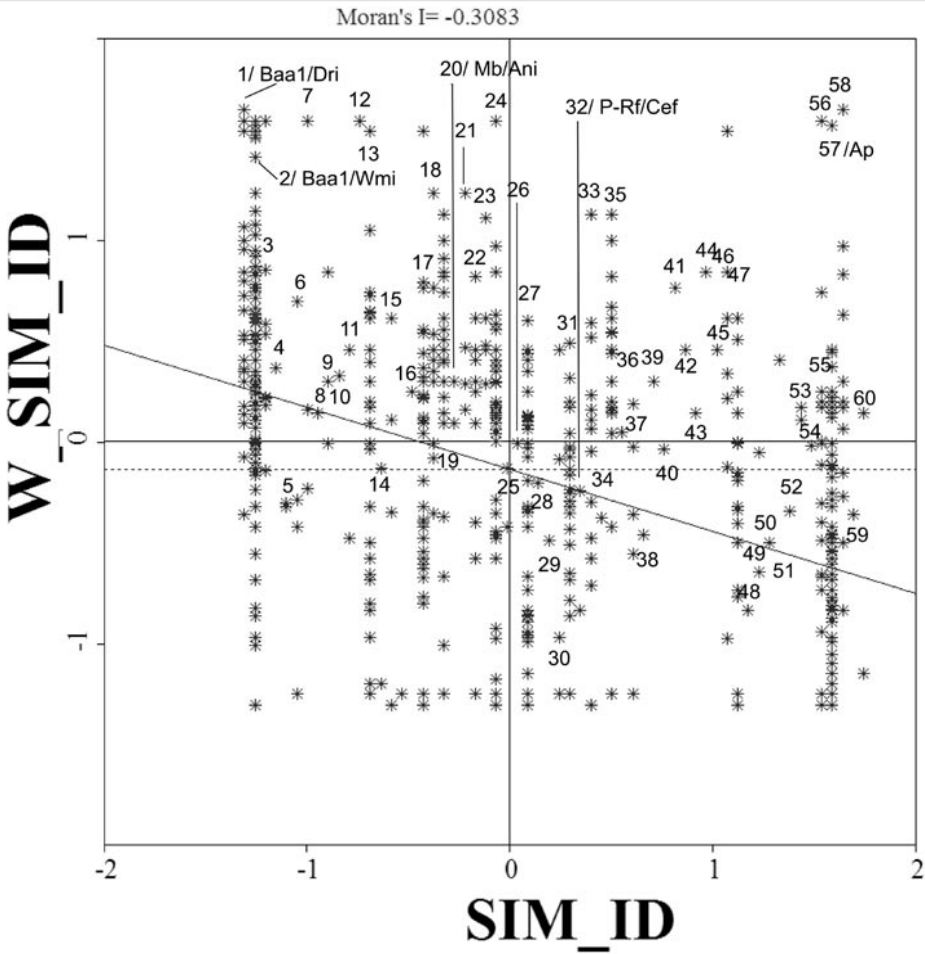

**Figura 14.** Distribución de los pesos de dependencia espacial (contigüidad grado 1 –umbral 1000 metros) para los parches de cobertura presentes en el área de páramo de la jurisdicción CORPOGUAVIO. Observese que las mejores conectividades las presentan los bosques conservados uno (1) y (2), así como las coberturas de origen antrópico con dominancia de pastos (57).

## Alta montaña de la serranía de Perijá (Figura 15 A y B)

En general la serranía presenta patrones importantes en la distribución de los tipos de vegetación páramuna, que indican una buena salud de las coberturas de la zona. Se observa que en el intervalo 0-20, los bosques conservados e intervenidos presentan los valores de contigüidad más bajos de la región tanto de primer como de segundo grado. Esta situación responde a la entrada de la actividad humana, principalmente con cultivos ilícitos, tala, quemas y ganadería. Cabe resaltar que los bosques presentan patrones de continuidad en segundo grado. En el intervalo 20-40 los tipos de vegetación se distribuyen dejando espacios considerables, la mayor agrupación se presenta en los matorrales bajos con diferentes arreglos florísticos. Otras dos concentraciones con algún grado de conectividad se presentan en el intervalo 40-60 alrededor de los tipos de vegetación dominados por pajonales entremezclados con matorrales altos y en los tipos de vegetación dominados por pajonales y frailejonales entremezclados con otros tipos de vegetación. Los herbazales, así como los rosetales frailejonales altos, se presentan muy dispersos y bajo condiciones típicas como sitios rocosos y expuestos a los fuertes vientos presentes en las cuchillas o partes más altas.

## Páramos: La Rusia, Belén y Guantiva (Figura 15 C y D)

Según el eje de variación SIM-ID que se muestra en los dos diagramas de dispersión, los tipos de vegetación que presentan mayor continuidad en el espacio son los pajonales entremezclados con matorrales bajos (cerca al centroide de las figuras). La gran zona de intervención antrópica, al parecer, no presenta una alta conectividad de primer grado, lo cual indica la alta concentración del disturbio. No obstante en segundo grado, estos pesos aumentan y representan la mayor contigüidad en la zona en este nivel. Esta situación es indicadora del mecanismo de acción humana sobre las áreas que opera mediante pequeñas zonas de clareo que se extienden hasta formar zonas extensas como las presentes en este páramo. Los chuscales y los frailejonales del intervalo 20-30 presentan concentraciones de pesos importantes tanto en primero como en segundo grado de contigüidad.

## Páramo de Telecom (Figura 15 E y F)

En el páramo de Telecon los pesos del el eje SIM-ID se distribuyen de manera dispersa para los intervalos con relictos de cobertura natural e indican la total pérdida de conectividad en la región. Valores de contigüidades de segundo grado se presentan hacia el sector del centroide del diagrama para algunos tipos de matorrales bajos. Las zonas agropecuarias dominan y presentan alta contigüidad tanto de primero como en segundo grado.

## Páramo de Merchán (Figura 16 G y H)

El páramo de Merchán presenta el estado más crítico de todas las regiones estudiadas. El diagrama de dispersión de Moran prácticamente sólo muestra conectividad de las áreas agropecuarias. El segundo grado de contigüidad es nulo para los pocos relictos de cobertura natural existentes. Sólo las áreas agropecuarias se presentan en primero y segundo grado de contigüidad y cabe mencionar que esta puede ser la suerte de muchas regiones paramunas de Colombia si continúa la presión antrópica y no se toman medidas urgentes para mitigarla.

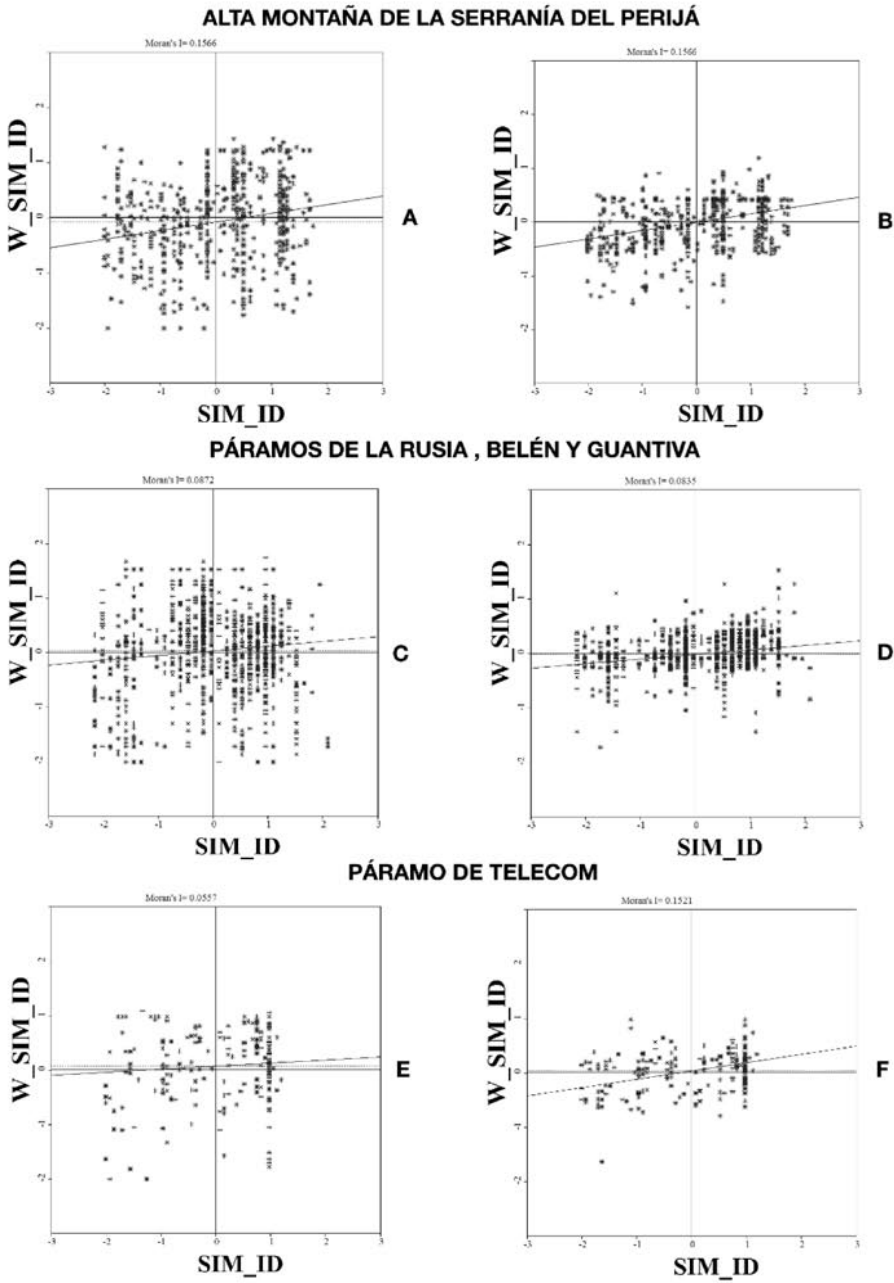

**Figura 15.** Distribución de los pesos de contigüidad por identificador de cobertura SIM-ID para los páramos de la alta montaña de la serranía de Perijá y los páramos de la Rusia, Belén y Guantiva y para el páramo de Telecom.

**Páramo Santuario y aledaños a Carmen de Carupa y páramo El Tablazo (Figura 16 I, J, K y L)**

Estos páramos representan el estadio anterior al presentado para los páramos de Telecom y Merchán. Las coberturas naturales aún presentan algún grado de contigüidad, aunque esta se concentra en tipos abundantes y no se distribuye de manera homogénea en los distintos patrones como sucede en las áreas conservadas de la serranía de Perijá, ciertas zonas de los páramos en la Jurisdicción CORPOGUAVIO o el PNN Los Nevados. Al igual que en los casos anteriores las mayores contigüidades en los pesos se presentan para la cobertura agropecuaria. Las coberturas con matorrales bajos y algunos frailejonales tienen conectividades que se evidencian especialmente en segundo grado.

**Páramos de la jurisdicción CORPOGUAVIO (Figura 17 M y N)**

Para los páramos de la jurisdicción CORPOGUAVIO la distribución de los pesos está ligada a la separación de las grandes unidades de páramo. Esta es la razón por la cual los valores de los pesos se encuentran amplificados a lo largo del eje de las ordenadas W-SIM-ID. A pesar de lo anterior es posibles distinguir contigüidades de primer y segundo orden en las coberturas naturales. Los valores mayores de contigüidad se presentan tanto en primer como en segundo grado para los tipos de vegetación boscosa e indican un importante estado de conservación en la zona. Muchas áreas con coberturas naturales también presentan esta particularidad. Como se evidenció en la sección de fragmentación, las coberturas de origen antrópico también se presentan y la separación de los valores en el diagrama de dispersión indican que el disturbio es continuo y se presenta en toda la zona de estudio.

**Área de estudio perteneciente al páramo de Sumapaz (Figura 17 O y P)**

Para esta región el tipo de cobertura con valores mayores de contigüidad tanto de primer y segundo orden es la agropecuaria. Los bosques de esta región desafortunadamente no presentan mucha conectividad entre sus parches lo cual genera una desprotección de los tipos paramunos ubicados al interior. La distribución de los pesos en todos los intervalos de variación SIM-ID (en primer y segundo grado) indican un grado de conservación importante de las coberturas páramunas existentes. Los tipos de cobertura entremezclados son los que presentan menores conectividades en la zona.

**Páramo del Parque Nacional Natural Los Nevados (Figura 17 Q y R)**

Esta región representa una distribución óptima y equilibrada de patrones en donde se observa que la mayoría de patrones tienen una dependencia espacial con los patrones de su tipo tanto en primer grado como en segundo. Las mayores conectividades se presentan hacia el intervalo central especialmente en las áreas dominadas con pajonales, pastizales y cojines de plantas vasculares. La conectividad de los tipos boscosos no se evidencia debido al corte arbitrario del área de referencia.

**Páramo de Guanacas área del volcán Puracé (Figura 18 S y T)**

El área de los páramos de Guanacas y Las Delicias presenta una muy baja conectividad de sus patrones de vegetación y cobertura debido a las grandes extensiones presentes. No obstante, se presentan algunos indicios de contigüidad en las regiones con bosques conservados. Este caso representa un estado de conservación con grandes extensiones de vegetación por tal motivo no se evidencian conectividades importantes.

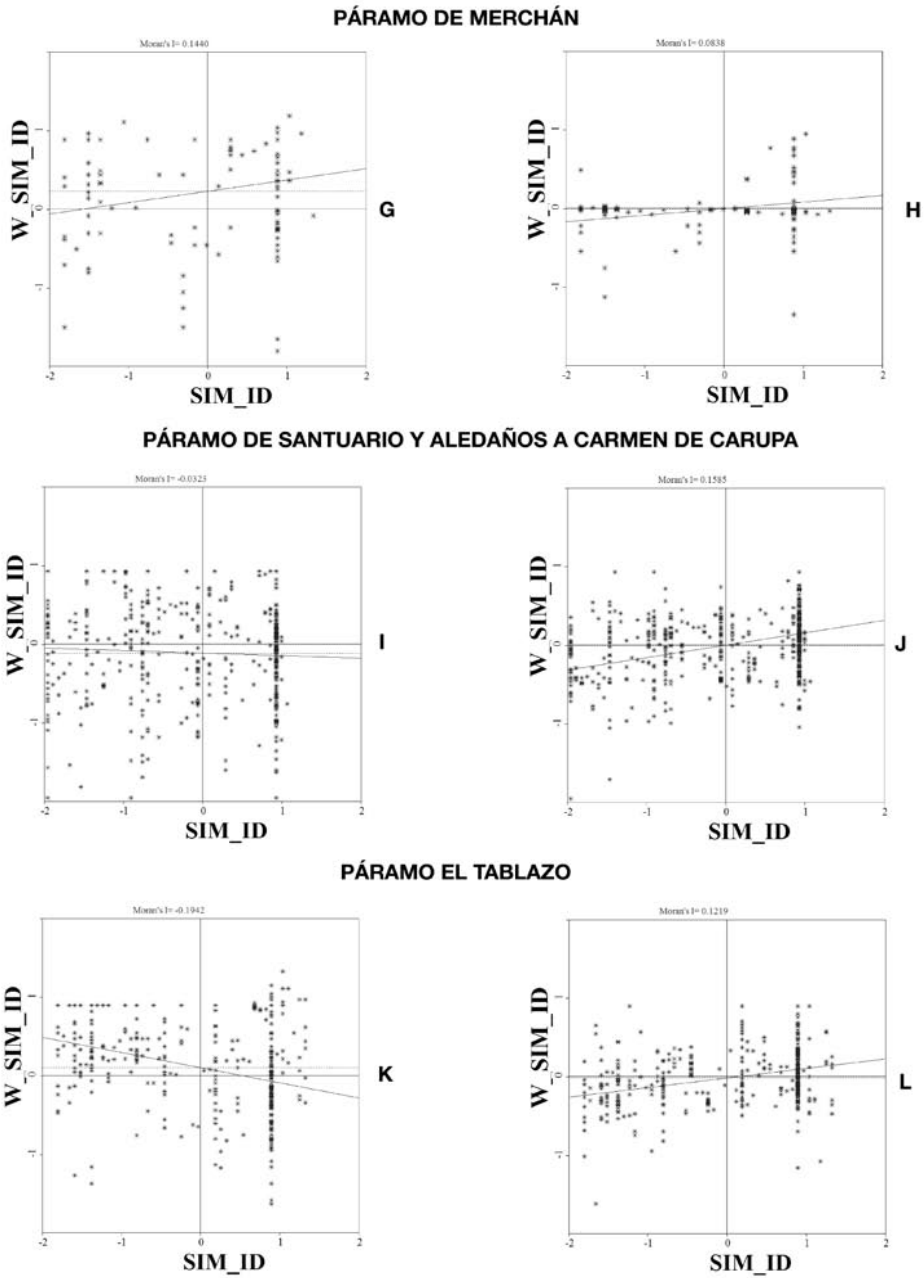

**Figura 16.** Distribución de los pesos de contigüidad por identificador de cobertura SIM-ID para el páramo de Merchán, el páramo de Santuario y aledaños al municipio de Carmen de Carupa y el páramo El Tablazo.

## PÁRAMOS JURISDICCIÓN CORPOGUAVIO

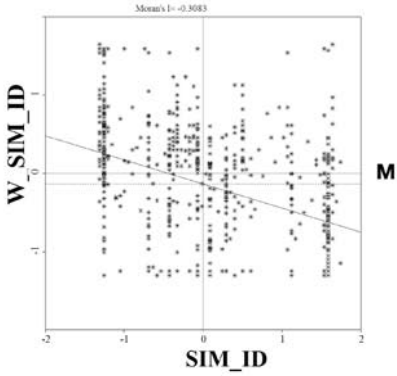

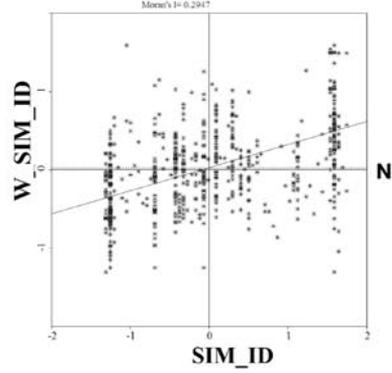

## ÁREA DE ESTUDIO PERTENECIENTE AL PÁRAMO DE SUMAPAZ

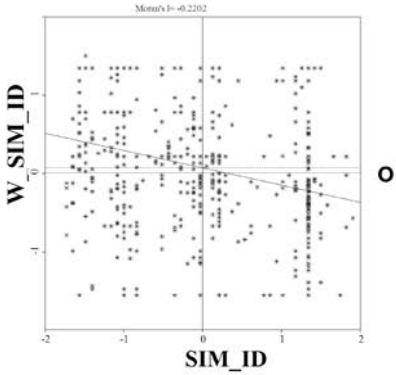

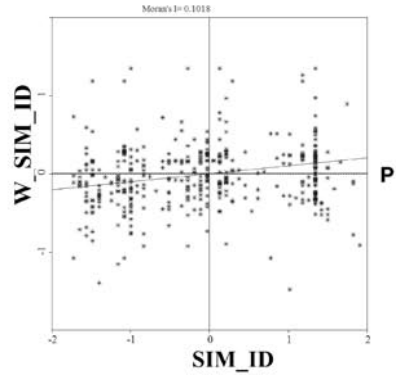

## ÁREA DE PÁRAMO DEL PNN LOS NEVADOS

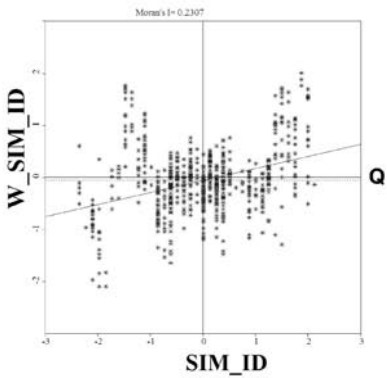

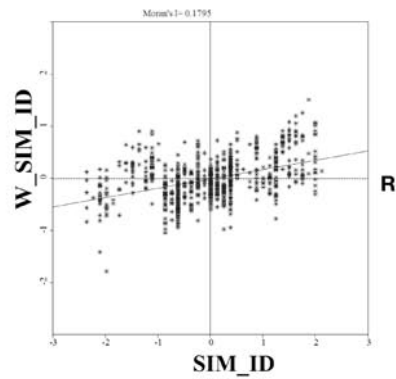

**Figura 17.** Distribución de los pesos de contigüidad por identificador de cobertura SIM-ID para los páramos de la jurisdicción CORPOGUAVIO, el área de estudio perteneciente al páramo de Sumapaz y el área de páramo del PNN Los Nevados.

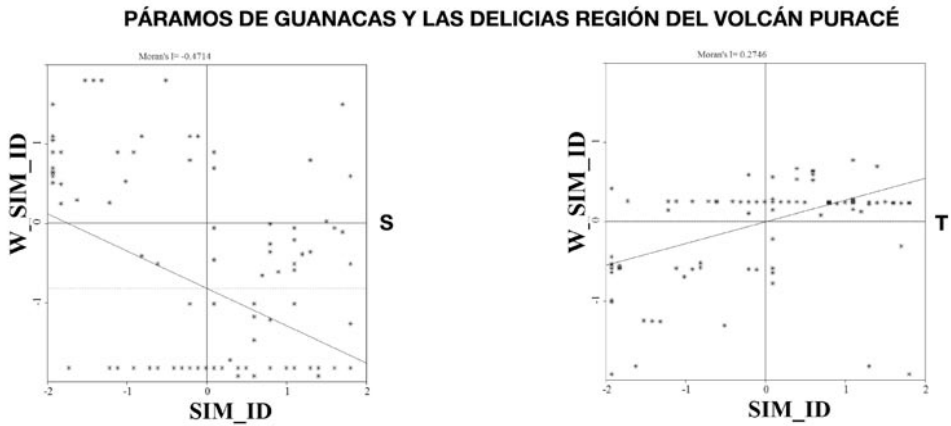

**Figura 18.** Distribución de los pesos de contigüidad por identificador de cobertura SIM-ID para los páramos de Guanacas y las Delicias, región del volcán Puracé.

En la tabla 21. Se resaltan los tipos de vegetación con mayor y menor grado de conectividad grado 1 en cada una de las áreas de páramo estudiadas.

## CONSIDERACIONES FINALES

En general, en la alta montaña de la serranía de Perijá en los páramos ubicados en el Parque Nacional Natural Los Nevados y en gran parte del área pertenecientes a los páramos de la jurisdicción CORPOGUAVIO los tipos de vegetación en las áreas estudiadas muestran un buen estado de conservación y de conectividad. La intervención de origen antrópico, el aumento del cultivo de papa, el cambio en el uso del suelo y la presencia de batallones de alta montaña, principalmete en los páramos de la cordillera Oriental, han ocasionado que la estructura de distribución natural de los tipos de vegetación se vea afectada hasta tal punto que la pérdida de conectividad hace inviable cualquier proceso de revegetalización y recuperación. Esta problemática se presenta en gran magnitud en los páramos La Rusia, Belén, Guantiva (hacia la vertiente oriental), Telecom, Merchán, Tablazo, sector nororiental del páramo de Sumapaz y la vertiente Oriental de los páramos de Carmen de Carupa y sectores aledaños. Para los páramos de Guanacas y Las Delicias (área de influencia del vocán Puracé) la problemática está concentrada hacia el sector noroccidental de la formación.

Según las áreas paramunas evaluadas, se presentan las siguientes condiciones:

### Serranía de Perijá

La alta montaña de la serranía de Perijá presenta 64 tipos de cobertura distribuidos alrededor 57 kilómetros en el eje **Y** (latitud) y 21 kilómetros en el eje **X** (longitud).

El noroccidente la región- carece de coberturas por encima de los 2800 m. En el norte prevalece la vegetación boscosa, dominan los bosques intervenidos. Las mayores áreas no superan 800 hectáreas. En el nororiente de la serranía existen coberturas de la mayoría de los tipos muestreados. Las áreas más grandes las ocupan los bosques conservados. Existe un área de chuscales de aproximadamente 800 hectáreas. Los bosques conservados de *Hesperomeles ferruginea* (SIM-VEG 1) presentan áreas por

**Tabla 21.** Tipos de vegetación con elevados y bajos valores de dependencia espacial (conectividad-grado 1-1000 metros).

| SIM_ID | SÍMBOLO | TIPO DE VEGETACIÓN | ESPECIES DOMINANTES | DEPENDENCIA-CONECTIVIDAD GRADO 1 | TAMAÑO DE LOS PARCHES |
|---|---|---|---|---|---|
| colspan=6 | **ALTA MONTAÑA DE LA Serranía de Perijá** |||||
| 53 | Rfb/Epe-Ale | Frailejonales bajos | *Espeletia perijaensis y Aa leucantha* | ELEVADA | MEDIANOS Y GRANDES |
| 52 | Rfa-P/Epe-Ldi-Cef | Frailejonales arborescentes entremezclados con pajonales | *Espeletia perijaensis, Libanothamnus divisoriensis y Calamagrostis effusa* | | |
| 42 | Mb-P/Hba-Cef | Matorrales bajos entremezclados con pajonales | *Hypericum baccharoides y Calamagrostis effusa* | | |
| 39 | Mb/Lst-Hph | Matorrales bajos | *Lourtegia stoechadifolia* e *Hypericum magdalenicum* | | |
| 31 | Ma2/Tme-Cmu | Matorrales altos | *Ternstroemia meridionalis, Clusia multiflora, Weinmannia pinnata y Prumnopitys montana* | | |
| 55 | Rfb-Mb-P/Epe-Lst-Cef-Cin | Frailejonales bajos entremezclados con matorrales bajos y pajonales | *Espeletia perijaensis, Lourtegia stoechadifolia, Stevia lucida, Calamagrostis effusa y Calamagrostis intermedia* | MEDIA | GRANDES |
| 6 | Baa1/Wpi-Rro | Bosque alto andino conservado | *Weinmannia pinnata y Decussocarpus rospigliosii* | MUY BAJA O AUSENTE | MEDIANOS |
| 5 | Baa1/Wpi-Bin | Bosque alto andino conservado | *Weinmannia pinnata, Brunellia integrifolia y Roupala pseudocordata.* | | |
| colspan=6 | **PÁRAMOS DE LA RUSIA, BELEN Y GUANTIVA** |||||
| 24 | Co/Ocl-Cco | Cortaderal | *Oreobolus cleefii, Cortaderia colombiana y Diplostephium colombianum.* | ELEVADA | PEQUEÑOS |
| 23 | Ch/Cht | Chuscales | *Espeletia incana, Monnina salicifolia, Sphagnum sancto-josephense, Geranium sibbaldioides y Chusquea tessellata* | | |
| 15 | P-Mb/Cef-Hla | Pajonales con matorrales | *Calamagrostis effusa, Hypericum laricifolium y Ageratina tinifolia.* | | |
| 3 | Baa1/Wmi-Iku | Bosque andino alto | *Weinmannia microphylla, Ilex kunthiana y Miconia* sp. | MEDIA | MEDIANOS |
| 1 | Baa1/Pqj | | *Polylepis quadrijuga, Diplostephium tenuifolium y Escallonia myrtilloides* | | |
| 22 | Rf-P/Esp-Cef | Rosetales-frailejonales-pajonales | *Espeletia congestiflora, Calamagrostis effusa, Pentacalia vaccinioides, Paramiflos glandulosa y Arcytophyllum nitidum* | BAJA | MEDIANOS |
| 19 | Rf/Eph | Rosetales-frailejonales | *Espeletia phaneractis, Calamagrostis effusa.* | | |
| 9 | Baa1-Mb/Wmi-Hla | Bosque andino alto entremezclado con vegetación de páramo | *Weinmannia microphylla, Ilex kunthiana* e *Hypericum laricifolium* | MUY BAJA O AUSENTE | MUY BAJA O AUSENTE |
| colspan=6 | **PÁRAMO DE TELECOM** |||||
| 41 | A | Áreas de uso agropecuario | | ELEVADA | GRANDES |
| 38 | Vr-Py/Pyn | Rosetales | *Puya nitida* | MEDIA | MUY PEQUEÑOS |
| 15 | Mb/Bj | Matorrales bajos | *Bejaria resinosa* | | |
| colspan=6 | **PÁRAMO DE MERCHÁN** |||||
| 19 | A | Áreas de uso agropecuario | | ELEVADA | GRANDES |
| 15 | Vr-Ma/Ati | vegetación riparia con matorrales altos | *Ageratina tinifolia* | MEDIA | MUY PEQUEÑOS |
| 3 | Mb/Chl | Matorrales bajos | *Clethra fimbriata* | | |
| colspan=6 | **PÁRAMOS DE CARMEN DE CARUPA Y SECTORES ALEDAÑOS** |||||
| 42 | A | Áreas de uso agropecuario | | ELEVADA | GRANDES |
| 29 | P/Cef-Esp | Pajonales | *Calamagrostis effusa y Espeletia* sp. | MEDIA | MEDIANOS |
| 18 | Mb/Ani | Matorrales bajos | *Ageratina tinifolia* | | |
| 8 | Baa2/Wto | Bosque andino alto inter-venido | *Weinmannia tomentosa* | | |
| 1 | Baa1/Wto | Bosque andino alto | *Weinmannia tomentosa* | | |

## Continuación tabla 21.

| | | ALTA MONTAÑA DE LA Serranía de Perijá | | | |
|---|---|---|---|---|---|
| **SIM_ID** | **SÍMBOLO** | **TIPO DE VEGETACIÓN** | **ESPECIES DOMINANTES** | **DEPENDENCIA-CONECTIVIDAD GRADO 1** | **TAMAÑO DE LOS PARCHES** |
| 5 | **Baa1-P/Wsp-Cef** | Bosque andino alto y pajonales | *Weinmannia* sp. y *Calamagrostis effusa* | **MUY BAJA O AUSENTE** | **MEDIANOS** |
| 4 | **Baa1-Mb/Wsp-Hju** | Bosque andino alto y matorrales bajos | *Weinmannia* sp. y *Hypericum juniperinum* | | |
| | | **PÁRAMO DEL TABLAZO** | | | |
| 39 | **A** | Áreas de uso agropecuario | | **ELEVADA** | **GRANDES** |
| 29 | **Mb/Ani** | Matorrales bajos | *Ageratina tinifolia* | **MEDIA** | **MEDIANOS** |
| 20 | **Mb/Acl** | Matorrales bajos | *Aragoa cleefii* | | |
| 7 | **Baa2/Wto** | Bosque andino alto intervenido | *Weinmannia tomentosa* | | |
| 30 | **Mb/Hla** | Matorrales bajos | *Hypericum laricifolium* | **MUY BAJA O AUSENTE** | **MEDIANOS** |
| 24 | **Mb-Rf/Ani-Egr** | Matorrales bajos con frailejonales | *Arcytophyllum nitidum* y *Espeletia grandiflora* | | |
| 10 | **Baa2-Pl/Wto** | Bosque andino alto intervenido entremezclado con plantaciones forestales | *Weinmannia tomentosa* | | |
| | | **PÁRAMOS PERTENECIENTES A LA JUSISDICCIÓN CORPOGUAVIO** | | | |
| 2 | **Baa1/Wmi** | Bosque Alto Andino conservado o semi conservado | *Weinmannia microphylla* | **ELEVADA** | **GRANDES** |
| 1 | **Baa1/Dri** | Bosque Alto Andino conservado o semi conservado | *Drimys granadensis* | | |
| 27 | **P/Cbo** | Pajonales | *Calamagrostis bogotensis* | **MEDIA** | **MEDIANOS** |
| 24 | **Mb-Ch/Hgo-Cht** | Matorrales bajos entremezclados con chuscales | *Hypericum goyanesii* y *Chusquea tessellata* | | |
| 17 | **Ma-P/Cef** | | *Calmagrostis effussa* | | |
| 43 | **Pt/Cve** | Vegetación de Pantano | *Crassula venezuelensis* | **MUY BAJA O AUSENTE** | **MEDIANOS** |
| 42 | **Pt/Cru** | | *Cyperus* aff. *rufus* | | |
| 40 | **Pt/Cja** | | *Carex jamesonii* var. *chordalis* | | |
| | | **PÁRAMO DE SUMAPAZ** | | | |
| 39 | **A** | Áreas de uso agropecuario | | **ELEVADA** | **MEDIANOS Y GRANDES** |
| 25 | **P-Rf/Cef-Esp** | Pajonale con rostetales frailejonales | *Calamagrostis effusa - Espeletia* sp. | **MEDIA** | **GRANDES** |
| 24 | **Mb-P/Cef** | Matorrales bajos y pajonales | *Calamagrostis effusa* | | |
| 22 | **Mb/Vfl** | Matorrales bajos | *Vaccinium floribundum* | | |
| 36 | **Vp-Vr-P-Ma/Cef-Clm** | Vegetación riparia, pajonales y matorrales | *Calamagrostis effusa - Clusia multiflora* | **MUY BAJA O AUSENTE** | **GRANDES** |
| 35 | **Vp/B-Rf-Baa2/Wro** | Vegetación de páramo entremezclada con bosques intervenidos | *Weinmannia rollotii* | | |
| 34 | **Vp/B-Rf-Baa2/Msa** | | *Miconia salicifolia* | | |
| 33 | **Vp/B-Rf-Baa2/Egr-Wto** | | *Espeletia grandiflora - Weinmannia rollotii* | | |
| 32 | **Vp/B-Rf-Baa2/Egr-Msa** | Vegetación de páramo (Rosetal-frailejonal) dominando sobre bosque altoandino | *Espeletia grandiflora - Miconia salicifolia* | | |
| 31 | **Vp/B-P-Rf-Baa2/Cef-Wto** | Vegetación de páramo (Pajonal-rosetal-frailejonal) dominando sobre bosque altoandino | *Weinmannia tomentosa - Calamagrostis effusa* | | |
| 30 | **Vp/B-Mb-Baa2/Aab-Clm** | Vegetación de páramo (Matorrales bajos) dominando sobre bosque altoandino | *Aragoa abietina - Clusia multiflora* | | |
| 29 | **Vp/B-Ch-Baa2/Cht-Wto** | Vegetación de páramo (Chuscal) dominando sobre bosque altoandino | *Chusquea tessellata - Weinmannia tomentosa* | | |
| 28 | **Vc-Mb/Aab** | Vegetación casmófita principalmente matorrales bajos | *Aragoa abietina* | | |
| 2 | **Baa1/Msa** | Bosque andino alto | *Miconia salicifolia* | | |
| 1 | **Baa1/Clm** | | *Clusia multiflora* | | |

**Continuación tabla 21.**

| SIM_ID | SÍMBOLO | TIPO DE VEGETACIÓN | ESPECIES DOMINANTES | DEPENDENCIA-CONECTIVIDAD GRADO 1 | TAMAÑO DE LOS PARCHES |
|---|---|---|---|---|---|
| **ALTA MONTAÑA DE LA Serranía de Perijá** | | | | | |
| **PÁRAMO DEL PARQUE NACIONAL NATURAL LOS NEVADOS** | | | | | |
| 24 | Pv-P/Pri-Cre-Cef | Cojines de plantas vasculares | *Plantago rigida-Distichia muscoides-Calamagrostis effusa* | ELEVADA | PEQUEÑOS |
| 23 | Pv-P/Pri-Cre | | *Plantago rigida-Calamagrostis recta* | | |
| 22 | Pv/Pri | | *Plantago rigida* | | |
| 21 | Ps-Pv/Ahe-Pri | Pastizales y cojines de plantas vasculares | *Agrostis perennans-Plantago rigida* | | |
| 20 | Ps-P/Ahe-Cre-Cef | Pastizales y pajonales | *Agrostis perennans-Calamagrostis recta-Calamagrostis effusa* | | |
| 19 | Ps-P/Ahe-Cef | | *Agrostis perennans-Calamagrostis effusa* | | |
| 18 | Ps/Ahe | Pastizales | *Agrostis perennans* | | |
| 17 | P-Pv/Cre-Cef-Pri | Pajonales y cojines de plantas vasculares | *Calamagrostis recta-Calamagrostis effusa-Plantago rigida* | | |
| 16 | P-Ps-Pv/Cef-Ahe-Pri | Pajonales con pastizales y cojines de plantas vasculares | | | |
| 15 | P-Ps/Cef-Ahe | Pajonales y pastizales | *Calamagrostis effusa-Agrostis perennans* | | |
| 14 | Pd-Pv/Lor-Pri | Prados y cojines de plantas vasculares | *Lachemilla orbiculata-Plantago rigida* | | |
| 13 | Pd-Baa1-Pv/Lor-Hfe-Pri | Prados con parches de bosques y cojines de plantas vasculares | *Lachemilla orbiculata-Hesperomeles ferruginea-Plantago rigida* | | |
| **PÁRAMO DE GUANACAS SECTORES ALEDAÑOS AL VOCÁN PURACÉ** | | | | | |
| 1 | Baa1/ Bca-Wpu | Bosque andino alto entremezclado con áreas agropecuarias | *Brunellia cayambensis* y *Weinmannia pubescens* | ELEVADA | GRANDES |

encima de cien hectáreas. En el occidente del área de estudio existen tres tipos de bosques conservados y uno intervenido que presentan áreas superiores a 200 hectáreas. Existen numerosos matorrales altos pero con valores bajos de cobertura y apenas dos tipos de chuscales registran alrededor de 400 hectáreas. En el centro del área de estudio se presentan las mayores densidades de coberturas entre los tipos de bosques conservados entremezclados con herbazales y matorrales altos, bosques intervenidos, matorrales bajos entremezclados con pajonales y rosetales frailejonales arborescentes de *Libanothamnus divisioriensis* y frailejonales de *Espeletia perijaensis*. Los bosques conservados entremezclados con matorrales y herbazales presentan los mayores registros aproximadamente 600 hectáreas. Hacia el Oriente de la serranía domina la vegetación paramuna con una concentración mayor a 4700 hectáreas y pertenecientes al tipo rosetales frailejonales entremezclados con matorrales bajos y pajonales. Otras combinaciones de rosetal frailejonal también son importantes, sin embargo, se encuentran distribuidas por debajo de 300 hectáreas de superficie. Para el sur occidente de la región de estudio se presenta solo tipo de cobertura dentro de los matorrales altos con cobertura inferior a 50 hectáreas. En el sur de la serranía se presenta mayor concentración de área en formaciones paramunas con respecto a los otros tipos de vegetación. Las formaciones de rosetales frailejonales y pastizales entremezclados con matorrales bajos y en algunos casos bosques no intervenidos se registran con un área aproximada a 500 hectáreas. Otro registro alrededor de 500 hectáreas se ubica hacia los matorrales bajos. Para el sur oriente de la serranía se presentan los menores registros de vegetación. Al menos dos tipos de vegetación ubicados dentro de los bosques conservados entremezclados con herbazales y algunos chuscales tienen áreas alrededor de 300 hectáreas.

La alta montaña del Perijá presenta muy pocas regiones de producción agropecuaria, aunque los bosques intervenidos y zonas de cultivos ilícitos están presentes en menor proporción.

**Páramos de La Rusia, Belén y Guantiva**

Los páramos de esta zona de estudio forman un continuo a lo largo de 55,9 kilómetros en el eje Y (latitud) y con una variación longitudinal 33,17 kilómetros (X ). Debido al sentido nororiental de la cordillera el sector Norte presenta apenas seis tipos de cobertura. Las áreas más extensas son de bosques intervenidos de *Quercus humboldtii* y alcanzan alrededor de mil hectáreas. Para el nororiente de la región se presenta al igual que hacia el centro, el mayor número coberturas en donde se destacan cuatro mil hectáreas de bosque conservado y 5000 hectáreas de bosque intervenido entremezclado con áreas agropecuarias. Hacia el occidente del área de estudio se encuentran doce tipos de cobertura. Las coberturas más extensas pertenecen a formaciones de tipo boscoso. Los bosques intervenidos entremezclados con áreas agropecuarias se presentan en aproximadamente 2000 hectáreas, mientras los bosques conservados de *Weinmannia microphylla* entremezclados con chuscales de *Chusquea tessellata* se presentan en más de 3000 hectáreas. La región central presenta casi todos lo tipos de cobertura diferenciados en el estudio. La tendencia general de la región central muestra un predominio de la formación de bosques conservados dominados por *Weinmannia microphylla*, *Ilex kunthiana* y *Miconia* sp. Es importante mencionar que aunque la región en general presenta características de conservación ya se evidencia la llegada de la intervención antrópica con dominio de áreas agropecuarias entremezcladas con bosques conservados de *Weinmannia microphylla* en aproximadamente 1000 hectáreas. Las áreas con bosques conservados entremezclados con pequeñas áreas de producción agropecuaria se distribuyen en aproximadamente 2000 hectáreas. Los matorrales altos de *Polylepis quadrijuga* se presentan concentrados en la región central aunque los valores de cobertura de sus parches no superan las 500 hectáreas. Otra cobertura con gran expresión en la zona central son los pajonales de *Calamagrostis effusa* entremezclados con distintos tipos de matorrales bajos en alrededor de 1800 hectáreas. Hacia el sector oriental se presenta una concentración de pajonales de *Calamagrostis effusa* entremezclados con distintos tipos de matorrales bajos cuyos valores individuales de cobertura no superan 1500 hectáreas. Los rosetales frailejonales entremezclados con pajonales también están presentes al oriente con coberturas que no superan las mil hectáreas.

Al suroccidente se presenta una disminución drástica de los tipos de vegetación boscosa a pesar de que se registra un área cercana a 3500 hectáreas de bosques conservados dominados por *Polylepis quadrijuga*. Los rosetales frailejonales dominados por *Espeletiopsis guacharaca*, *Arcytophyllum nitidum* y *Bejaria resinosa* entremezclados con matorrales bajos de *Arcytophyllum nitidum* se concentran con áreas individuales menores a 1000 hectáreas.

Para el sector sur se presenta un gran bosque de vegetación boscosa intervenida entremezclada con áreas agropecuarias en alrededor de 4000 hectáreas aunque la mayor cobertura pertenece a áreas agropecuarias con alrededor de 27000 hectáreas.

Para el suroriente los mayores valores que se registran no superan 500 hectáreas para los bosques entremezclados con áreas agropecuarias.

La alta complejidad en la forma de los bosques indica un grado de conservación o intervención intermedia pero importante mientras la alta complejidad presente en las zonas agropecuarias que se extienden por más de 30000 hectáreas indica que la zona hacia ese sector es irrecuperable por lo menos con la tecnología actual. Las zonas agropecuarias cuando están ingresando a un sistema presentan valores bajos de dimensión fractal ya que se manifiestan como clareos angulares o redondeados.

La vegetación de páramo presenta valores de complejidad altos que indican un estado de conservación importante con relación al área (se presentan áreas con vegetación de páramo muy extensas en comparación con otros sitios en Colombia); sin embargo, el aspecto general de la zona muestra igualmente una zona de intervención de las más grandes del país ubicadas en cercanía a un páramo que revela una sobreexplotación de recursos ofrecidos por el sistema. Sin duda de las zonas paramunas colombianas ésta debe ocupar un lugar de interés, ya que se encuentra amenazada por los sistemas agropecuarios del flanco oriental.

**Páramos de la jurisdicción CAR**

El alarmante estado de transformación, especialmente en el páramo de Merchán, reclama con urgencia que se implementen las medidas de manejo y zonificación ambiental, es necesario concertar con los propietarios de las tierras, la extensión del límite del cultivo de papa en los escasos sitios que aún mantienen condiciones originales. Es preocupante el estado de transformación de las rondas de las quebradas y riachuelos que surten a los acueductos municipales, entre ellos el de Merchán.

En el páramo de Telecom debe entrar a acordarse con las fuerzas militares nuevas transformaciones del medio natural, aunque la presencia de estas fuerzas del orden es disuasiva para evitar colonizaciones e invasiones, las construcciones y la proyección futura -si la hay, obviamente- tendrán impacto, acción que no es deseable ya que en términos de representatividad para la zona de Saboyá, este es el páramo con mejor estado de conservación que se observó.

En el complejo de páramos de Carmen de Carupa, el de menor extensión es el de Santuario donde prácticamente solamente se encuentran pequeñas islas de páramo rodeadas por cultivos de papa, por tanto es urgente entrar a reglamentar la protección de estas pequeñas islas representativas.

**Páramos de la jurisdicción CORPO-GUAVIO**

La mayoría de área perteneciente a los páramos de la jurisdicción CORPOGUAVIO presenta un buen estado en cuanto a la conservación de sus coberturas. Los páramos que presentan mayor intervención son los ubicados hacia los municipios de Gama, Ubalá, Guasca y Gachalá mientras los ubicados hacia el sector de Junín presentan una alta diversidad en tipos de vegetación y áreas no muy intervenidas por la acción humana. La mayor área de superficie páramuna se encuentra entre Guasca y Fómeque y la intervención disminuye a medida que las formaciones se acercan al PNN Chingaza o cuando se incrementa la cota altitudinal.

En la región de páramo del municipio de Gama existen aún relictos de bosque con alguna importancia en la zona aunque la fuerte acción de las áreas con cultivos forestales desplazó por completo las formaciones típicas de páramo. Caso similar se presenta en la región de páramo del municipio de Ubalá que contrasta con una extensión considerable de páramo en el municipio de Gachetá, donde la dimensión fractal es alta para sus tipos de cobertura natural.

**Área de páramo perteneciente al Parque Nacional Natural Los Nevados**

Esta región representa una distribución de patrones óptima y equilibrada, en donde se observa que la mayoría de patrones tienen una dependencia espacial con los patrones de su tipo tanto en primer grado como en segundo. Las mayores conectividades se presentan en las áreas dominadas con pajonales, pastizales y cojines de plantas vasculares. La conectividad de los tipos boscosos no se evidencia debido al corte arbitrario del área de referencia.

**Área de estudio perteneciente a los páramos de Guanacas y Las Delicias**

Presentan una muy baja conectividad de sus patrones de vegetación y cobertura debido a la existencia de grandes bloques de vegetación, lo cual reduce la fragmentación existente. No obstante, en la franja altoandina se presentan algunos indicios de contigüidad en las localidades con bosques conservados rodeadas de algún grado de intervención. Este caso representa un estado de conservación con grandes extensiones de vegetación, por tal motivo no se evidencian conectividades importantes.

**AGRADECIMIENTOS**

A los doctores Luis Norberto Parra y Luis Jairo Silva de la Universidad Nacional de Colombia, sede Medellín por la revisión detallada y sus valiosos comentarios.

**LITERATURA CITADA**